\documentclass[reprint,aps,prb,twocolumn,superscriptaddress]{revtex4}
\usepackage{amsmath,amsfonts,amssymb}
\usepackage{graphicx,float,calc}
\usepackage{color,bm}
\usepackage{ulem}
\usepackage{braket}
\usepackage[colorlinks,urlcolor=blue,citecolor=blue,linkcolor=blue]{hyperref}
\usepackage{hyperref}
\usepackage{chngcntr}
\begin{document}

\title{Dynamics of asymmetrically deformed skyrmion driven by internal forces and strain force in a flower-shaped magnetic nanostructure}

\author{Zhen-Yu Tan\footnote{These authors contributed equally to this work.}}
\affiliation{School of Physics and Materials Science and Research Center for Advanced Information Materials, Guangzhou University, Guangzhou 510006, China}
\author{Ji-Pei Chen$^*$\footnote{chenjp@gzhu.edu.cn}}
\affiliation{School of Physics and Materials Science and Research Center for Advanced Information Materials, Guangzhou University, Guangzhou 510006, China}
\author{Yu-Ke Shi}
\affiliation{School of Physics and Materials Science and Research Center for Advanced Information Materials, Guangzhou University, Guangzhou 510006, China}
\author{Yuan Chen}
\affiliation{School of Physics and Materials Science and Research Center for Advanced Information Materials, Guangzhou University, Guangzhou 510006, China}
\author{Ming-Hui Qin}
\affiliation{Institute for Advanced Materials, South China Normal University, Guangzhou 510006, China}
\author{Xing-Sen Gao\footnote{xingsengao@scnu.edu.cn}}
\affiliation{Institute for Advanced Materials, South China Normal University, Guangzhou 510006, China}
\author{Jun-Ming Liu}
\affiliation{Laboratory of Solid State Microstructures and Innovative Center of Advanced Microstructures, Nanjing University, Nanjing 210093, China}

\begin{abstract}
Magnetic skyrmions emerge as promising quasi-particles for encoding information in next-generation spintronic devices. Their innate flexibility in shape is essential for the applications although they were often ideally treated as rigid particles. In this work, we investigated the voltage-controlled uniform strain mediated dynamics of deformed skyrmions in heterostructures with a flower-shaped magnetic nanostructure, using micromagnetic simulations. The simulated results revealed the possible states of isolated skyrmion nucleated in the nanostructure, which can be mutually switched by applying suitable in-plane strain pulses. In addition, it was found that the skyrmion motions are driven by the emerging internal forces and strain force, which originate from the asymmetric deformation of skyrmion structures. Furthermore, an analytical model of deformed skyrmions was proposed to interpret the dependences of internal forces and strain force on the asymmetric deformation of skyrmion, with some formulae derived for these forces in a semi-analytical approach. Further calculations based on these formulae verified the forces appearing in the skyrmion motion, with the resulting forces showing consistence with the simulated data. This suggested that our semi-analytical model successfully captures the main physics responsible for the motion of deformed skyrmion in the nanostructure. Our work extends the understanding of the mechanics emerging in deformed skyrmion, and provides an effective approach for deterministic manipulation of deformed skyrmion motion via strain forces and internal forces, which may be instructive to design of skyrmion-based spintronic devices.

\end{abstract}

\date{\today}

\maketitle
\section{introduction}
Magnetic skyrmions, topologically stable spin textures, are often observed in magnetic materials that lack inversion symmetry. The nanoscale skyrmions exhibit particle-like behaviors including their creation, annihilation and controllable mobility. These novel characteristics make them promising candidate as information carriers for future spintronic devices \cite{nagaosa2013topological,fert2017magnetic,wiesendanger2016nanoscale,jiang2015sge,kang2016skyrmion,zhang2020skyrmion}, which permit deterministic manipulation of skyrmions by means of injected spin-polarized currents, external electric-field, and so on \cite{romming2013writing,sampaio2013nucleation,fert2013v,iwasaki2013current,hsu2017electric,jiang2015blowing,wang2020electric,
ba2021electric,geirhos2020macroscopic,chen2016exotic,lin2022manipulation,yao2022vector}. For the application of skyrmions in spintronic devices, many feasible designs have been proposed for confining the skyrmions in geometric nanostructures, such as nanostripes and nanodisks, which may allow the precise control of individual skyrmions \cite{sampaio2013nucleation,fert2013v,iwasaki2013current}. These achievements underpin a wide variety of emerging skyrmion-based spintronic devices, ranging from racetrack memories, magnetic random access memories, logic circuits to neuromorphic devices, etc. \cite{nagaosa2013topological,fert2017magnetic,wiesendanger2016nanoscale,jiang2015sge,kang2016skyrmion,zhang2020skyrmion,
romming2013writing,sampaio2013nucleation,fert2013v,iwasaki2013current}, which have great benefits of all-electrical control and energy-efficiency fashion \cite{fert2017magnetic,wiesendanger2016nanoscale}.\par
Generally, the skyrmions are ideally regarded as rigid circular-shaped particles with no distortions under external stimuli, and so far the majority of studies on skyrmion motion are conducted based on this assumption \cite{schutte2014inertia,lin2013particle,yasin2022real,liu2022flexoresponses,liu2023emergent}. However, it was found that skyrmions with elliptical shape or asymmetric deformation may appear in the presence of applied current with even small current density\cite{yasin2022real}, uniaxial in-plane magnetic anisotropy\cite{cheng2021elliptical}, strain
gradients\cite{liu2022flexoresponses} or defects in sample\cite{liu2023emergent}, exhibiting their flexibility rather than the rigidity. In addition, direct experimental observations and micromagnetic simulations verified that these distortions may in turn affect the skyrmion dynamics, as manifested especially by the modification of skyrmion Hall angle\cite{yasin2022real,cheng2021elliptical,liu2022flexoresponses,liu2023emergent}. It is known that the skyrmion motion driven by external stimuli can be understood in the frame of Thiele equation \cite{iwasaki2013current,thiele1973steady}, which describes phenomenologically the dynamic behaviors of skyrmion particles under the influence of various forces including Magnus force, dissipative force as well as the internal and external reversible forces [see Eqs. (\ref{eq3})-(\ref{eq5})]. It was revealed that the dynamic behaviors of deformed skyrmion can be well explained by a modified Thiele equation, in which a corrective gyromagnetic coupling vector $\textbf{\textit{G}}$ in Magnus force term and/or a corrective asymmetric dissipative force tensor ${\cal D}$ in dissipative force term were introduced to adapt the skyrmion deformation\cite{yasin2022real, liu2022flexoresponses,liu2023emergent}. Moreover, some studies pointed out that the internal forces stemmed from internal distortions of the skyrmions exert essential impact on their motion, as evidenced only by the micromagnetic simulation data \cite{koshibae2016berry,koshibae2017theory}. Nevertheless, the underlying mechanisms for the deformation induced internal force were not clearly uncovered in these studies. For this issue, one may refer to the pioneering works by Thiele \cite{thiele1973steady}, in which the internal reversible force $\textbf{\textit{F}}_{\mathrm{in}}$ originates from the gradient of internal energy $\it E_{\mathrm{in}}$ of the skyrmion configurations as $\textbf{\textit{F}}_{\mathrm{in}}=-\nabla {\it E}_{\mathrm{in}}$. On the assumption of rigid circular-shaped skyrmion particles, the internal force term was always ignored since it is identically zero due to the rotational and translational invariance of ${\it E}_{\mathrm{in}}$ for the skyrmion \cite{leutner2022skyrmion}. Thus, these enlightened us to investigate the possible internal forces arising from asymmetric distributions of $E_{\mathrm{in}}$ in deformed skyrmion structures, and also their effects on the skyrmion motions.\par
For developing energy efficient spintronic devices, the feasible schemes to distort the skyrmion using voltage-controlled in-plane strain are currently being investigated, with some impressive progress has been made on the strain gradients driven skyrmion motions in multiferroic heterostructures \cite{liu2022flexoresponses,yanes2019skyrmion,du2023strain}. In these schemes, the strain gradient locally breaks inversion symmetry, and it naturally creates an emerging force that drives the skyrmions moving along the strain gradient directions. On the contrary, the homogeneous strain alone does not break inversion \cite{du2023strain}, and there are few reports on the alone application of a spatially uniform uniaxial strain for driving skyrmion motion. Although the voltage-induced uniform uniaxial strain in multiferroic devices can distort the skyrmion, whether it may activate the skyrmion motion alone remains unclear. In this regard, the automotion of domain wall associated with energy efficient applications might provide access to this problem \cite{liu2022flexoresponses,nikonov2014automotion,mawass2017switching,yershov2018geometry,gao2016dynamic}. It was demonstrated that distorted domain wall structures can be moved with the assistance of the internal driving forces coming from internal energy (i.e., demagnetization and magnetic anisotropy) in the system, but not any external driving force from external stimuli (i.e., external field or  spin-polarized current). These provoked our interest in deformed skyrmions for the urge to explore their motion under the combined assistance of uniform uniaxial strain and internal driving forces, and also for their potential spintronic applications.\par
In this work, we aimed at studying the motion of deformed skyrmions mediated by voltage-controlled uniform strain in multiferroic heterostructures with a flower-shaped magnetic nanostructure (see Fig. \ref{fig1}), by using micromagnetic simulations. We first demonstrated the dynamic behaviors of isolated skyrmion confined in this nanostructure by applying suitable in-plane strain pulses. It was found that the skyrmion motions are driven by internal forces and strain force, which naturally originate from the skyrmion with asymmetric deformation. These simulation results were further confirmed by an analytical model of deformed skyrmions with a set of analytical formulae derived for these forces. Our present study extends the understanding of mechanics emerging in deformed skyrmion, and provides an effective approach for controlling the dynamic behaviors of asymmetrically deformed skyrmion via strain forces and internal forces, which may guide the design of skyrmion-based spintronic devices.
\section{Model and Simulation Methods}
\begin{figure}[t]
\centering
\includegraphics[width=0.45\textwidth]{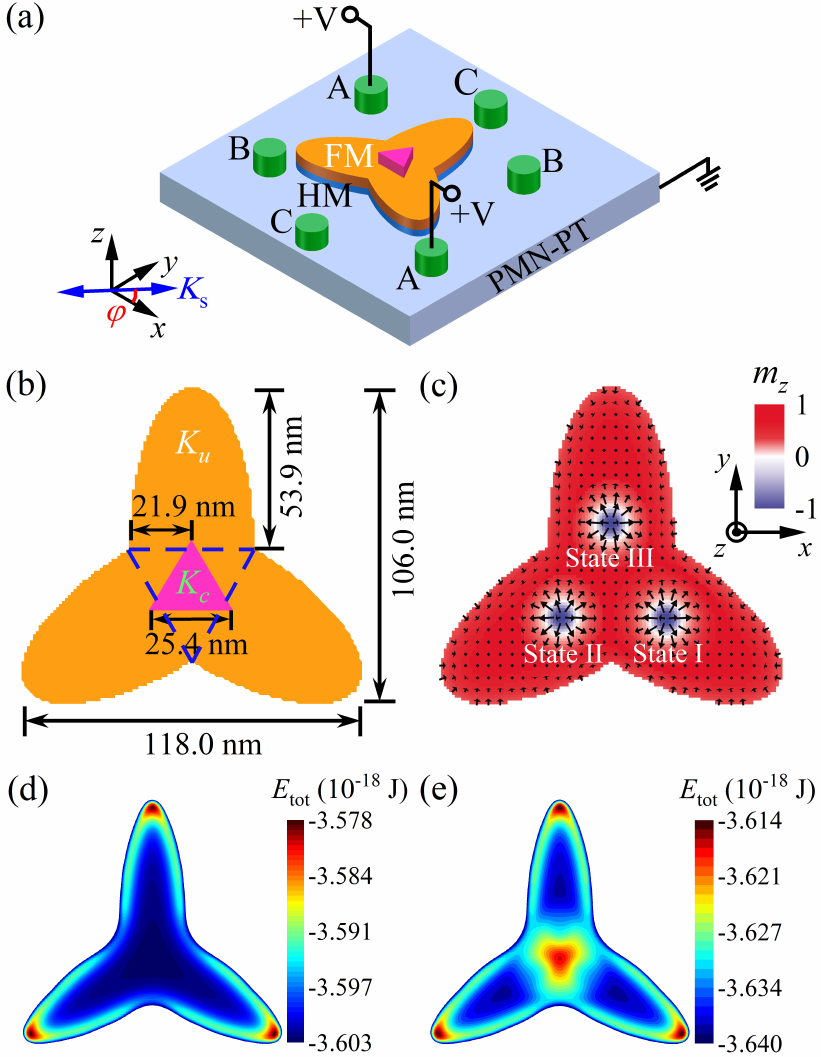}
\caption{(Color online) (a) Sketch of multiferroic heterostructure consisting of tripetalous flower-shaped FM/HM layers on PMN-PT film, with a capping nanoisland fabricated on the top surface of FM/HM layers, and three pairs of upper electrodes ${\mathrm{A}}\mbox{-}{\mathrm{A}}$, ${\mathrm{B}}\mbox{-}{\mathrm{B}}$ and ${\mathrm{C}}\mbox{-}{\mathrm{C}}$ built on (001)-oriented PMN-PT film substrates. (b) The schematic view of the flower-shaped FM/HM layers with the inhomogeneous PMA distribution on the $xy$-plane, in which the PMA strength $K_{\mathrm{c}}$ in capping island and $K_{\mathrm{u}}$ in the outer ring of the FM/HM film are different. The flower-shaped nanostructure is composed of three semi-ellipses and a central equilateral triangle, as marked by blue dashed lines. The half-length of the major-axis and minor-axis of the semi-ellipses here are 54.0 nm and 22.0 nm, respectively. The side-length of equilateral trianglar capping island is 25.4 nm. (c) Three possible equilibrium skyrmion states in the flower-shaped nanostructure, in which the isolated skyrmion may be nucleated at one of the three semi-ellipses, denoted as states I, II and III. (d) and (e) Plots of the potential energy profiles for the isolated skyrmion located at different positions in the nanostructures in the cases of (d)$K_{\mathrm{c}}=K_{\mathrm{u}}= 0.80\ \mathrm{MJ/m^{3}}$ and (e) $K_{\mathrm{c}}=0.93\ \mathrm{MJ/m^{3}}$, $K_{\mathrm{u}}= 0.80\ \mathrm{MJ/m^{3}}$. }
\label{fig1}
\end{figure}

In this work, we studied the voltage control of magnetic skyrmion on an ultrathin flower-shaped ferromagnetic/heavy-metal (FM/HM) nanostructure such as the Co/Pt layers, taking into account the strain-mediated magnetoelectric coupling in multiferroic heterostructures. We proposed a capping nanoisland/FM/HM multilayers fabricated on the top surface of a $ \mathrm{PbMg_{1/3}Nb_{2/3}O_{3}\mbox{-}PbTiO_{3}}$ (PMN-PT) piezoelectric substrate, which is a room-temperature single-phase multiferroic material and shows good merit in piezoelectric properties for the large piezostrain, as schematically shown in Fig.~\ref{fig1}(a). The tripetalous flower-shaped FM/HM nanostructure considered here is composed of three semi-ellipses and a central equilateral triangle [see Fig.~\ref{fig1}(b)], and it is surrounded by three pairs of top electrodes (${\mathrm{A}}\mbox{-}{\mathrm{A}}$, ${\mathrm{B}}\mbox{-}{\mathrm{B}}$ and ${\mathrm{C}}\mbox{-}{\mathrm{C}}$) with each pair lying on the major axis of the semi-ellipses, which can be afforded by the recent advances in sophisticated micro-device processing techniques \cite{gao2016dynamic,yao2018electrically,cui2013method,hu2009electric,song2022strain}. The top equilateral triangular capping layer is expected to enhance the perpendicular magnetic anisotropy (PMA) of the system, which can be grown by choosing a proper capping material with modified material parameters e.g. strain and sample thickness. Therefore, the PMA strength in and out of the capping regions ($K_{\mathrm{c}}$ and $K_{\mathrm{u}}$) are different, which induces an inhomogeneous PMA distribution on the flower-shaped nanostructure, as presented in Fig.~\ref{fig1}(b).

The magnetic state in the Co/Pt film is usually dependent on the total free energy ($\it E$), including Heisenberg exchange energy, Dzyaloshinskii-Moriya (DM) interaction energy, PMA energy, and demagnetization energy, which is written as \cite{sampaio2013nucleation,chen2016exotic,cui2013method,hu2009electric,song2022strain,cui2015generation}:
\begin{equation}\begin{split}\label{eq1}
E =& \iiint \left\{A_{\mathrm{ex}}\left(\nabla\cdot\textbf{\textit{m}}\right)^{2}
+D\left[m_{z}\left(\nabla\cdot\textbf{\textit{m}}\right)-\left(\textbf{\textit{m}}\cdot\nabla\right) m_{z} \right] \right.\\
&\left.-K_{z}\left(\textbf{\textit{m}}\cdot\textbf{\textit{e}}_{z}\right)^{2}
-\frac{\mu_{0}}{2}\mathrm{M}_{\mathrm{s}}\textbf{\textit{m}}\cdot\textbf{\textit{H}}_{\mathrm{d}} -K_{\mathrm{s}}\left(\textbf{\textit{m}}\cdot\textbf{\textit{e}}_{\mathrm{s}}\right)^{2}
\right\}{\mathrm{d}}V,\\
\end{split}\end{equation}
where $A_{\mathrm{ex}}$, $D$ and $K_{z}$ are the ferromagnetic exchange, DM interaction and PMA constants, respectively. $\textbf{\textit{H}}_{\mathrm{d}}$ is the magnetostatic self-interaction fields. $\textbf{\textit{m}}=\bf{M}$$/\mathrm{M_{\mathrm{s}}}$ is the normalized magnetization vector with $\mathrm{M_{\mathrm{s}}}$ the saturation magnetization. The PMA easy-axis is along the $\pm z$-axis.
For experimental scheme, an application of positive voltage on the electrodes ${\mathrm{A}}\mbox{-}{\mathrm{A}}$, ${\mathrm{B}}\mbox{-}{\mathrm{B}}$ or ${\mathrm{C}}\mbox{-}{\mathrm{C}}$ (the bottom electrode is grounded) may induce an in-plane uniaxial strain along the axis connecting the electrodes, with the corresponding magnetoelastic energy written as an effective in-plane uniaxial magnetic anisotropy $-K_{\mathrm{s}}\left(\textbf{\textit{m}}\cdot\textbf{\textit{e}}_{\mathrm{s}}\right)^{2}$, with the strength of anisotropy constant $K_{\mathrm{s}}$ and the easy axis of uniaxial anisotropy $\textbf{\textit{e}}_{\mathrm{s}}$. Therefore, the application of voltage to the ${\mathrm{A}}\mbox{-}{\mathrm{A}}$, ${\mathrm{B}}\mbox{-}{\mathrm{B}}$ or ${\mathrm{C}}\mbox{-}{\mathrm{C}}$  electrodes generate strain at the angle of $\varphi = 150^{\circ}$, $30^{\circ}$ or $270^{\circ}$ with respect to the $x$-axis \cite{cui2015generation,chen2020voltage}. In addition, previous study demonstrated that the strength and the direction of in-plane uniaxial strains can be deterministically tuned by controlling the voltage applied on two different pairs of electrodes, according to the voltage control law \cite{chen2020voltage}. For example, to obtain a desired strain with strength $K_{\mathrm{s}}$ and direction along the angle of $\varphi = 120^{\circ}$ that lies between the connecting lines of the electrodes pairs ${\mathrm{A}}\mbox{-}{\mathrm{A}}$ and ${\mathrm{C}}\mbox{-}{\mathrm{C}}$, one may choose the suitable magnitude of voltage $\mathrm{V}_{\mathrm{A}}$ and $\mathrm{V}_{\mathrm{B}}$ applied on the electrodes pairs ${\mathrm{A}}\mbox{-}{\mathrm{A}}$ and ${\mathrm{C}}\mbox{-}{\mathrm{C}}$.\par
To investigate the dynamics of the magnetic structures driven by the in-plane uniaxial strains, the simulations were conducted by employing the software package Mumax3 \cite{vansteenkiste2014design}, in which time-dependent magnetization dynamics was computed by solving the Landau-Lifshitz-Gilbert (LLG) equation \cite{landau1992theory,gilbert1955lagrangian}:
\begin{equation}\begin{split}\label{eq2}
\frac{\mathrm{d}\textbf{\textit{m}}}{\mathrm{d}t} =-\gamma\textbf{\textit{m}} \times \textbf{\textit{H}}_{\mathrm{eff}}+\alpha\textbf{\textit{m}} \times\frac{\mathrm{d}\textbf{\textit{m}}}{\mathrm{d}t},\\
\end{split}\end{equation}
where the first and second terms on the right side of the equation describe the gyromagnetic precession and the Gilbert damping, respectively. $\textbf{\textit{H}}_{\mathrm{eff}} =-(1/\mu_{0} \mathrm{M_{\mathrm{s}}})\partial E/\partial \textbf{\textit{m}}$ is the effective field, $\gamma$ is the Gilbert gyromagnetic ratio, and $\alpha$ is the damping coefficient.
In the simulations, we considered that the Co/Pt bilayers contain a 1-nm-thick cobalt film. The typical parameters for studying the Co/Pt system were adopted as \cite{sampaio2013nucleation}: the saturation magnetization $\mathrm{M_{\mathrm{s}}}=580\ \mathrm{KA/m}$, exchange constant $A_{\mathrm{ex}}=15\ \mathrm{pJ/m}$, DM interaction constant $D = 3\ \mathrm{mJ}/\mathrm{m}^{2}$, gyromagnetic ratio $\gamma  = 2.211 \times 10^{5}\ \mathrm{m/(A\cdot s)}$. The PMA strength for the Co/Pt film is fixed at
$K_{\mathrm{u}}= 0.80\ \mathrm{MJ/m^{3}}$. In most calculations, the PMA strength for the capping island was chosen as $K_{\mathrm{c}}=0.93\ \mathrm{MJ/m^{3}}$, or stated elsewhere. The nanomagnets were divided into unit cells with cell size of $1 \times 1 \times 1 \,\mathrm{nm^{3}}$ for the simulations.\par
In the widely studied Pt/Co systems, the Gilbert damping constant was usually chosen as $\alpha = 0.3$ for a 0.4-nm-thick Co layer\cite{sampaio2013nucleation}. However, some experiments and first-principles calculations on Pt/Co multilayer thin films indicated that the value of $\alpha$ decreases significantly with increasing the thickness of Co layer, and $\alpha = 0.1$ is adaptive for a 1.0-nm-thick Co film \cite{neilinger2018ferromagnetic,mizukami2010gilbert,barman2007ultrafast,barati2014gilbert}. As suggested by these experimental findings, some recent studies on skyrmion dynamics in Pt/Co film are conducted under a low damping constant of $\alpha = 0.1$ \cite{chauwin2019skyrmion,zhang2020stochastic,fattouhi2021logic,purnama2015guided}. Therefore, we carried out the simulations by adopting $\alpha = 0.1$ for studying Pt/Co thin film with a 1.0-nm-thick Co layer in the present work.

\section{Results and Discussion}
\subsection{Nucleation of skyrmion in the flower-shaped nanostructure with inhomogeneous PMA}
Fig.~\ref{fig1}(c) presents three possible isolated skyrmion structures nucleated in the flower-shaped nanostructure with inhomogeneous PMA. It should be noted that only the isolated skyrmions with down-core polarity were considered here, which may be generated randomly at one of the three semi-ellipses. To further elucidate the role of capping island on the nucleation of skyrmion, we produced isolated skyrmion at different positions in the flower-shaped nanostructure, and then calculated the total energy $E$ over the whole nanostructure for skyrmion at different positions, as the energy profiles plotted for the cases of $K_{\mathrm{u}}= K_{\mathrm{c}}= 0.80\ \mathrm{MJ/m^{3}}$ [Fig.~\ref{fig1}(d)] and $K_{\mathrm{u}}= 0.80\ \mathrm{MJ/m^{3}}, K_{\mathrm{c}}=0.93\ \mathrm{MJ/m^{3}}$ [Fig.~\ref{fig1}(e)]. As a comparison, one may see in Figs.~\ref{fig1}(d) and ~\ref{fig1}(e) that the energy valley locates at center of nanostructure for the case of $K_{\mathrm{u}}= K_{\mathrm{c}}= 0.80\ \mathrm{MJ/m^{3}}$, while there are three energy valleys distributing in the three semi-ellipses and an energy peak appearing at the center of nanostructure for the case of $K_{\mathrm{u}}= 0.80\ \mathrm{MJ/m^{3}}$ and $K_{\mathrm{c}}=0.93\ \mathrm{MJ/m^{3}}$. In this sense, the spatial total energy profile of skyrmion over the nanostructure here may be understand to be the potential energy landscape in the nanostructure, which can be directly modified by the nonuniform PMA \cite{lin2022manipulation,hao2021skyrmion}. Therefore, the nanoisland serves as a potential barrier for $K_{\mathrm{c}}> K_{\mathrm{u}}$ in this model, and the isolated skyrmion becomes stable in the potential wells of the three semi-ellipses. The following discussions will focus on the case of $K_{\mathrm{u}}= 0.80\ \mathrm{MJ/m^{3}}$ and $K_{\mathrm{c}}=0.93\ \mathrm{MJ/m^{3}}$, wherein three skyrmion states are identified as state I, II, or III according to their positions on the nanostructure.
\subsection{Skyrmion dynamics in the flower-shaped nanostructure induced by in-plane strain pulses}
\begin{figure}[t]
\centering
\includegraphics[width=0.48\textwidth]{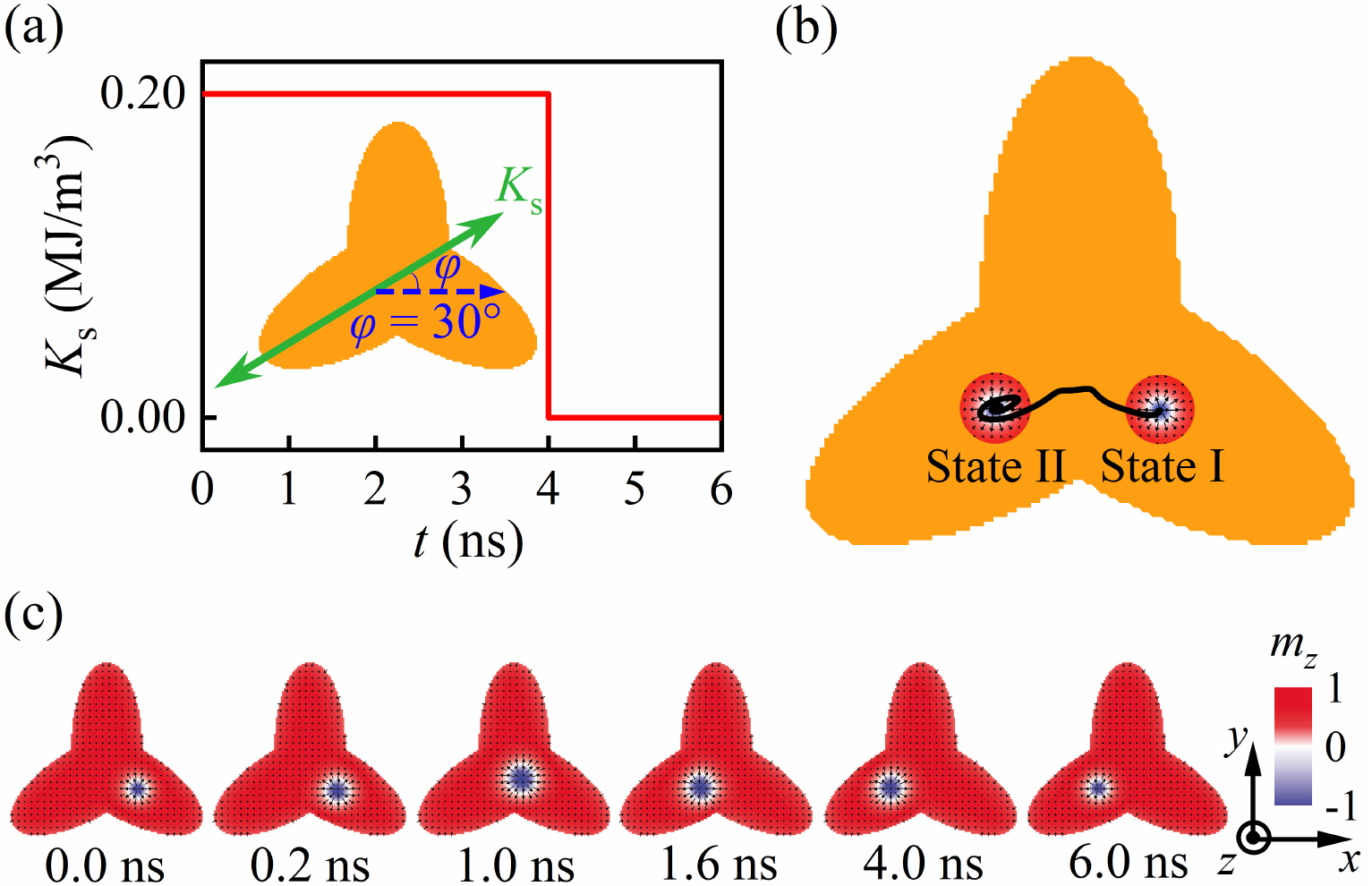}
\caption{ (Color online) The motion of skyrmion in the flower-shaped nanostructure induced by an in-plane strain pulse. (a) An applied in-plane strain pulse with strength $K_{\mathrm{s}}=0.20\ \mathrm{MJ/m^{3}}$ and duration of 4.0 ns. The insert shows that strain was applied at the angle of $\varphi = 30^{\circ}$ with respect to the $x$-axis. (b) Trajectory of the skyrmion center driven by the strain pulse, from the initial skyrmion state (State I) to the final skyrmion state (State II). Here the skyrmion center $(X, Y)$ is defined as $X = \iint xq \mathrm{d}x\mathrm{d}y/\iint q \mathrm{d}x\mathrm{d}y$ and $Y = \iint yq \mathrm{d}x\mathrm{d}y/\iint q \mathrm{d}x\mathrm{d}y$, where $q = \textbf{\textit{m}} \cdot (\partial_{x}\textbf{\textit{m}}\times \partial_{y}\textbf{\textit{m}}) /4 \pi$ is the topological charge density, and the integrals are performed over the coordinate $xy$-plane \cite{lin2013driven}. (c) Snapshots show the process of the skyrmion movement driven by the strain pulse.}
\label{fig2}
\end{figure}

\begin{figure}[t]
\centering
\includegraphics[width=0.45\textwidth]{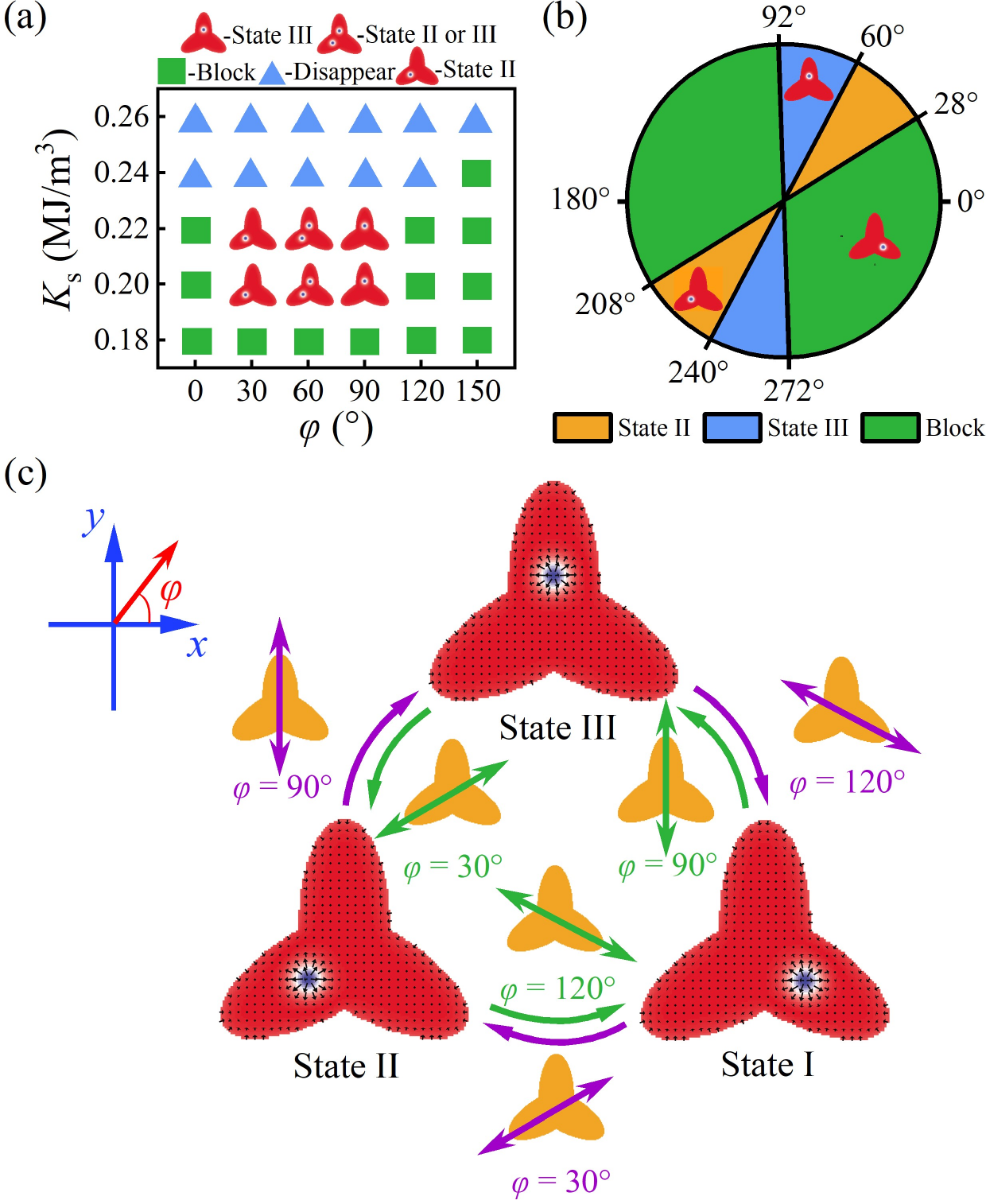}
\caption{(Color online) (a) The state diagram illustrates different final states for skyrmion motion from the initial skyrmion state (State I), under various strain strength $K_{\mathrm{s}}$ and applied angle $\varphi$ of the strain pulse. (b) A pie chart shows the effect of the strain angle $\varphi$ on the skyrmion motion from initial State I, under a fixed $K_{\mathrm{s}}=0.20\ \mathrm{MJ/m^{3}}$. The sections colored green, orange and blue represent the angle conditions for motion blocked state, final State II and State III, respectively. Here the pie chart is generated from a set of data points for $\varphi = 2^{\circ}, 4^{\circ}, \cdots , 360^{\circ}$ at intervals of $2^{\circ}$. (c) Schematic illustrations of switching sequences for the skyrmion states (State I, II and III) driven by strain pulse with a fixed strength $K_{\mathrm{s}}=0.20\ \mathrm{MJ/m^{3}}$. Here we used the straight bi-arrows to mark the direction of applied strain, and the curved pink and green arrows to denote the clockwise and counterclockwise switching sequences for skyrmion states.}
\label{fig3}
\end{figure}

In this section, we studied the skyrmion dynamics in the flower-shaped nanostructure induced by in-plane strain pulses. Here, the direction of the applied in-plane strain is defined by the angle $\varphi$ with respect to the $x$-axis [see Fig.~\ref{fig2}(a)].
The simulations start from the initial skyrmion state (State I) at $t = 0.0\ \mathrm{ns}$. We first tested the effect of strain pulses on the skyrmion dynamics, using a strain pulse with moderate amplitude of $K_{\mathrm{s}}=0.20\ \mathrm{MJ/m^{3}}$ and duration of 4.0 ns. As seen in Figs.~\ref{fig2}(b) and ~\ref{fig2}(c), the simulation results showed that the skyrmion first moves toward the capping region at the beginning of the strain pulse (see snapshots at $t = 0.0\ \mathrm{ns} \sim 1.0\ \mathrm{ns}$ ), and then it turns the trajectory to run into the left-down semi-ellipse (see snapshots at $t = 1.6\ \mathrm{ns} \sim 4.0\ \mathrm{ns}$). Subsequently, the strain was turned off and the skyrmion evolves into an equilibrated state, i.e., State II at 6.0 ns.
Further simulations demonstrated that the skyrmion motion in the nanostructure is sensitive to the strain direction (angle $\varphi$) and strain amplitude $K_{\mathrm{s}}$, with the final states for skyrmion motion summarized in Fig.~\ref{fig3}(a). In this diagram, the ``Disappear" states occupy the region of large $K_{\mathrm{s}}$, while the ``Block" states dominate the region of small $K_{\mathrm{s}}$. For these results, one may understand that the large $K_{\mathrm{s}}$ deforms the skyrmion structures largely and finally make it disappear, while the small $K_{\mathrm{s}}$ are not sufficient to activate the skyrmion movement for escape from the potential well. For the moderate $K_{\mathrm{s}}$ region in the diagram, it was found that the skyrmion may be deterministically switched to State II or State III by adopting certain suitable strain directions, while its motion was blocked at other strain directions. To further demonstrate the effect of strain direction on skyrmion motion, a pie chart in Fig.~\ref{fig3}(b) showed the angle conditions for different final states (i.e., ``Block" state, State II and State III) of skyrmion motion at a fixed moderate strain strength $K_{\mathrm{s}}=0.20\ \mathrm{MJ/m^{3}}$. These results are reasonable by considering that the strain with $60^{\circ}<\varphi<92^{\circ}$ and $28^{\circ}<\varphi<60^{\circ}$ orients toward the upper and left-down semi-ellipses, which makes the skyrmion tends to switch to State II and State III, respectively. Additionally, we found in the simulations that skyrmion may run to the upper or left-down semi-ellipse at the particular angle $\varphi=60^{\circ}$, relying on the nuance of the initial skyrmion structure. This is mainly because that the strain with $\varphi=60^{\circ}$ just points along the minor-axis of the right-down ellipse, making the possible switching of skyrmion state from initial State I to State II or III due to the symmetry of the nanostructure. In this sense, the skyrmion states of State I, II and I can be deterministically and mutually switching by applying strain pulse with moderate strength $K_{\mathrm{s}}$ and suitable angle $\varphi$, as the switching sequences shown schematically in Fig.~\ref{fig3}(c).
\subsection{Skyrmion motion driven by internal forces and strain force in the flower-shaped nanostructure}
\begin{figure}[t]
\centering
\includegraphics[width=0.50\textwidth]{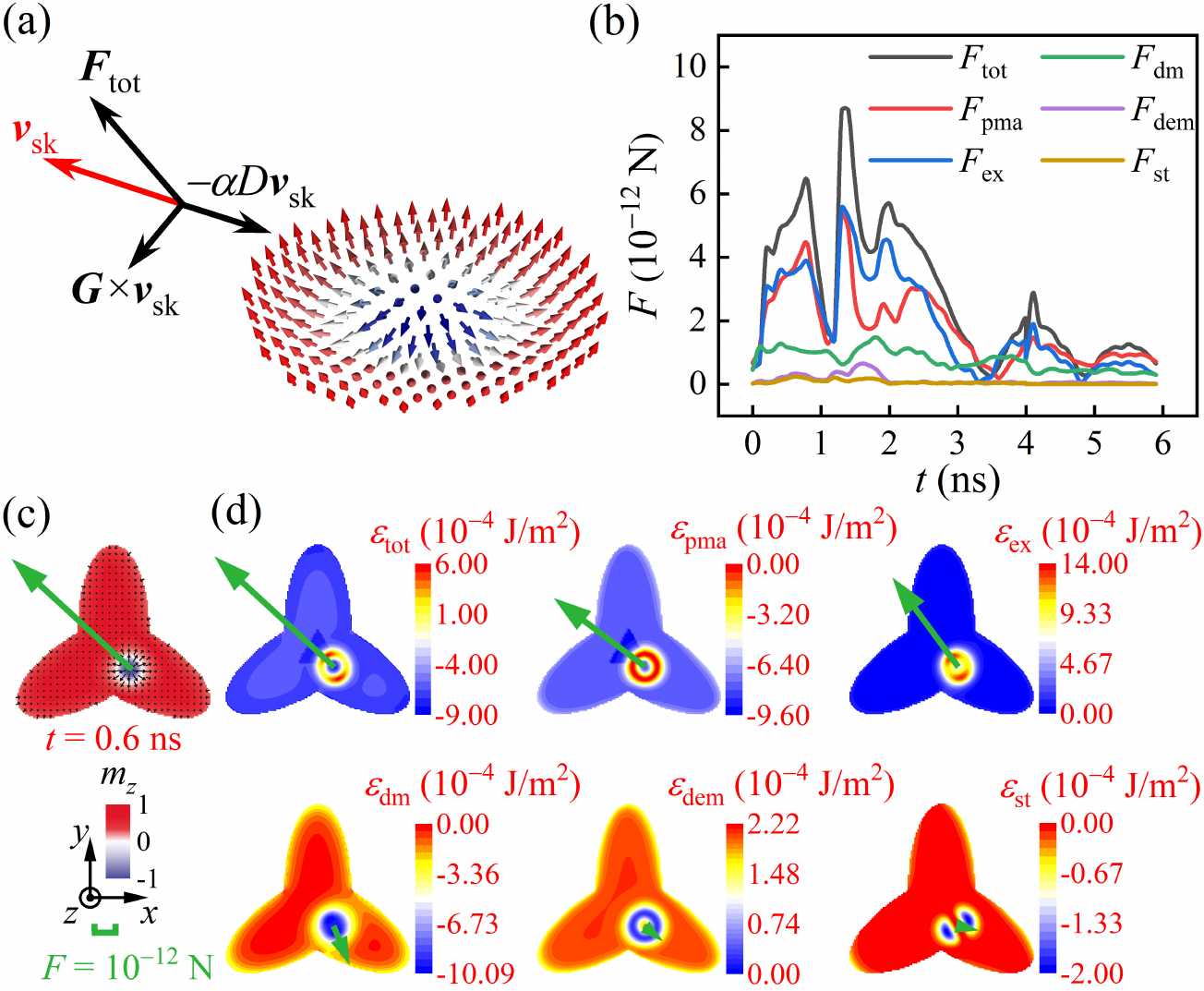}
\caption{(Color online) (a) Schematic illustrations of skyrmion motion driven by internal forces and strain force in the flower-shaped nanostructure. (b) Plots of the total force $F_\mathrm{tot}$ and various component forces (i.e., $F_\mathrm{ex}$, $F_\mathrm{dm}$, $F_\mathrm{pma}$, $F_\mathrm{dem}$, and $F_\mathrm{st}$) acting on the skyrmion as a function of time $t$ for the skyrmion motion presented in Fig.~\ref{fig2}. (c) Snapshot shows the magnetic state at $t = 0.6$ ns in the process of skyrmion movement depicted in Fig.~\ref{fig2}. (d) Illustrations of the distribution of the total areal energy density $\varepsilon_\mathrm{tot}$, various components (i.e., $\varepsilon_\mathrm{ex}$, $\varepsilon_\mathrm{dm}$, $\varepsilon_\mathrm{pma}$, $\varepsilon_\mathrm{dem}$ and $\varepsilon_\mathrm{st}$) on the $xy$-plane and the corresponding emerging forces for the skyrmion state shown in (c). Here we used the green arrows to represent the forces, and the inserted green measure gauge to mark the force with magnitude of $10^{-12}$ N.}
\label{fig4}
\end{figure}

In this section, we paid our attention to understand the mechanism of skyrmion dynamics triggered by the in-plane strain pulse. For this issue, we investigated the skyrmion motion using the Thiele equation \cite{iwasaki2013current,thiele1973steady}:
\begin{equation}\begin{split}\label{eq3}
\textbf{\textit{G}}\times\textbf{\textit{v}}_{\mathrm{sk}}+\alpha \mathcal{D}\textbf{\textit{v}}_{\mathrm{sk}}
-\textbf{\textit{F}}_\mathrm{tot}=0,
\end{split}\end{equation}
where $\textbf{\textit{v}}_{\mathrm{sk}}$ is the drift velocity of skyrmion, and $\mathcal{D}$ is a dissipative force tensor. The first term in the left-hand side of equation represents the Magnus force with gyromagnetic coupling vector $\textbf{\textit{G}}$. The second and the third terms describe the dissipative force and the total reversible force $\textbf{\textit{F}}_\mathrm{tot}$ acting on the skyrmion, respectively. The skyrmion motion driven by all forces was schematically illustrated in Fig.~\ref{fig4}(a).
The total reversible force $\textbf{\textit{F}}_\mathrm{tot}$ was divide into two terms
\begin{equation}\begin{split}\label{eq4}
\textbf{\textit{F}}_\mathrm{tot}=\textbf{\textit{F}}_\mathrm{in}+\textbf{\textit{F}}_\mathrm{st}.
\end{split}\end{equation}
The external strain force term $\textbf{\textit{F}}_\mathrm{st}$ relates to the strain energy, and the internal force term $\textbf{\textit{F}}_\mathrm{in}$ contains the forces arising from the internal exchange energy, DM interaction energy, PMA energy, and demagnetization energy as
\begin{equation}\begin{split}\label{eq5}
\textbf{\textit{F}}_\mathrm{in}=\textbf{\textit{F}}_\mathrm{ex}+\textbf{\textit{F}}_\mathrm{dm}
+\textbf{\textit{F}}_\mathrm{pma}+\textbf{\textit{F}}_\mathrm{dem}.
\end{split}\end{equation}
The reversible forces acting on the skyrmion is defined as
\begin{subequations}\label{eq6}
\begin{align}
&\textbf{\textit{F}}_\mathrm{tot}=\iint\textbf{\textit{f}}_\mathrm{tot} \mathrm{d}S,\label{eq6a}
\\
&\textbf{\textit{f}}_\mathrm{tot}=\textbf{\textit{f}}_\mathrm{ex}+\textbf{\textit{f}}_\mathrm{dm}
+\textbf{\textit{f}}_\mathrm{pma}+\textbf{\textit{f}}_\mathrm{dem}+\textbf{\textit{f}}_\mathrm{st},\label{eq6b}
\\
&\textbf{\textit{F}}_{\mu}=\iint\textbf{\textit{f}}_{\mu} \mathrm{d}S,\mu= \mathrm{ex}, \mathrm{dm}, \mathrm{pma}, \mathrm{dem}\ \mathrm{or}\ \mathrm{st},\label{eq6c}
\\
&\textbf{\textit{f}}_{\mu}=-\nabla \varepsilon_{\mu},\label{eq6d}
\end{align}
\end{subequations}
where $\textbf{\textit{f}}_{\mu}$ and $\varepsilon_{\mu}$ are the areal force density and areal energy density on the coordinate $xy$-plane for various energy components, and the integrations in Eqs. (\ref{eq6a}) and (\ref{eq6c}) are restricted to the region of skyrmion configuration.

To proceed, we studied the reversible forces arising in the skyrmion motion exhibited in Fig.~\ref{fig2}. Here we calculated the reversible forces by adopting a numerical procedure. First, we obtained a set of areal energy density $\varepsilon_{\mu}$ from the simulated data of magnetic states. Second, we numerically calculated the force density $\textbf{\textit{f}}_{\mu}$ on the $xy$-plane according to the definition Eq. (\ref{eq6d}), and then made a summation of $\textbf{\textit{f}}_{\mu}$ over the skyrmion region to get the force components $\textbf{\textit{F}}_{\mu}$ and total force $\textbf{\textit{F}}_\mathrm{tot}$. The results plotted in Fig.~\ref{fig4}(b) shows the evolutions of $\textbf{\textit{F}}_\mathrm{tot}$ and $\textbf{\textit{f}}_{\mu}$ as a function of time $t$, which reveals the role of the reversible forces on skyrmion motions in this process. It was seen clearly that the total reversible force $\textbf{\textit{F}}_\mathrm{tot}$ is contributed mainly by the force components $\textbf{\textit{F}}_\mathrm{pma}$ and $\textbf{\textit{F}}_\mathrm{ex}$ (the maximum is $~ 5.7 \times 10^{-12}$ N) and some less by $\textbf{\textit{F}}_\mathrm{dm}$, while scarcely by $\textbf{\textit{F}}_\mathrm{dem}$ and $\textbf{\textit{F}}_\mathrm{st}$ as the maximum is less than $1.0\times 10^{-12}$ N.

To further identify the sources of driving forces appearing in the nanostructure, we focused on studying the distributions of energy density $\varepsilon_{\mu}$ on the nanostructure. In Fig.~\ref{fig4}(d), some $\varepsilon_{\mu}$ profiles and the relevant forces were shown for a typical magnetic state at $t = 0.6$ ns [see Fig.~\ref{fig4}(c)] in the skyrmion motion. It was first noticed that the pattern of $\varepsilon_\mathrm{ex}$ lacks rotational symmetry which indicates the obvious internal deformation of skyrmion structure, because the $\varepsilon_\mathrm{ex}$ pattern for the circular skyrmion is always rotationally symmetric. In addition, the appearance of $\textbf{\textit{F}}_\mathrm{pma}$ is mainly due to the nonuniform PMA around the capping region, as we will discuss in Sec. III(D) below. Moreover, we found that some ripple-like lines exist in the profiles of $\varepsilon_\mathrm{dm}$ and $\varepsilon_\mathrm{dem}$, as the shape similar to the edge of the nanostructure. This manifests the geometric confinement and boundary effect of nanostructure on the energy distribution, since both the demagnetization energy and DM energy may induce shape anisotropy in the geometrically confined systems \cite{streubel2016magnetism,zheng2017characteristics,cubukcu2016dzyaloshinskii,chen2021control}.

According to the above analysis, all three factors (i.e., deformation of skyrmion structure, nonuniform PMA, and the nanostructure geometry and size) synthetically bring strong impact on energy density distributions in the system, leading to emergence of the total reversible force $\textbf{\textit{F}}_\mathrm{tot}$. In the following of this work, we will focus on analyzing the roles of these factors in the emergence of the forces $\textbf{\textit{F}}_\mathrm{ex}$, $\textbf{\textit{F}}_\mathrm{dm}$, $\textbf{\textit{F}}_\mathrm{st}$, $\textbf{\textit{F}}_\mathrm{pma}$, and $\textbf{\textit{F}}_\mathrm{dem}$.

\subsection{Internal forces and strain force in deformed skyrmion structures}
The above findings have given rise to a puzzling question: why and how the internal forces and strain force appear in the deformed skyrmion structure and the nanostructure with nonuniform PMA? To answer this question, we deduced a set of analytical formulae for the relation between emerging forces and deformed skyrmion structures, nanostructure geometry and size, or the nonuniform PMA in this section.

The areal energy density functions for the internal exchange, DM interaction, PMA energy and the strain energy on the $xy$-plane of the nanostructure are
\begin{equation}\begin{split}\label{eq7}
&\varepsilon_\mathrm{ex}=Ad(\nabla \textbf{\textit{m}})^{2},\\
&\varepsilon_\mathrm{dm}=Dd\textbf{\textit{m}}\cdot[(\hat{z} \times \nabla)\times \textbf{\textit{m}}],\\
&\varepsilon_\mathrm{pma}=-K_{z}dm_{z}^{ 2},\\
&\varepsilon_\mathrm{st}=-K_{\mathrm{s}}d(\textbf{\textit{m}}\cdot \textbf{\textit{e}}_{\mathrm{s}})^{2},\\
\end{split}\end{equation}
with the film thickness $d$.

\begin{figure}[t]
\centering
\includegraphics[width=0.45\textwidth]{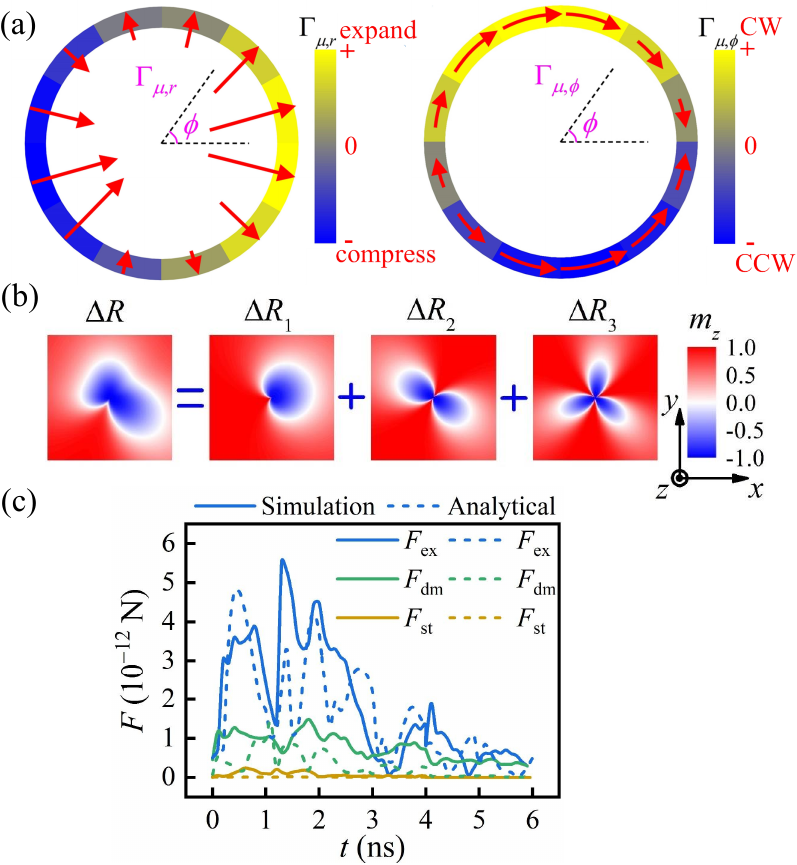}
\caption{(Color online) (a) Schematic illustrations of effect of the radial and
tangential components (i.e., $\Gamma_{\mu,r}$ and $\Gamma_{\mu,\phi}$) of angular density of force $\bf{\Gamma}_{\mu}$ on different polar angles segments of the asymmetrically deformed skyrmion texture. The inward and outward straight arrows represent the compression ($\Gamma_{\mu,r}<0$) and expansion ($\Gamma_{\mu,r}>0$), respectively. The CW and CCW curve arrows denote the CW ($\Gamma_{\mu,\phi}>0$) and CCW ($\Gamma_{\mu,\phi}<0$) rotation, respectively. The asymmetric distributions of $\bf{\Gamma}_{\mu}$ lead to nonzero force $\textbf{\textit{F}}_{\mu}$. (b) Schematic illustrations of deformation $\Delta R$ in the deformed skyrmion [Fig.~\ref{fig4}(c)] constituted of three main modes of $\Delta R_1$, $\Delta R_2$ and $\Delta R_3$. (c) Comparison between the analytical and simulation results of the time evolutions of forces $F_\mathrm{dm}$, $F_\mathrm{ex}$ and $F_\mathrm{st}$ for the skyrmion motion displayed in Fig.~\ref{fig2}.}
\label{fig5}
\end{figure}

We considered a deformed skyrmion configuration with down-core polarity as \cite{nagaosa2013topological,liu2022flexoresponses,wang2018theory,romming2015field}
\begin{equation}\begin{split}\label{eq8}
\textbf{\textit{m}}(\textbf{\textit{r}})=(\sin\Theta\cos\phi, \sin\Theta\sin\phi, \cos\Theta),\\
\end{split}\end{equation}
where the magnetization at position $\textbf{\textit{r}}(r, \phi)$ is described by azimuthal angle $\Theta(r, \phi)$ and polar angle $\phi$. The cross section of a skyrmion is approximately expressed by a standard $360^{\circ}$ domain wall profile \cite{liu2022flexoresponses,wang2018theory,romming2015field}:
\begin{equation}\begin{split}\label{eq9}
\Theta(r, \phi)=-2\arctan \bigg[ \frac{\sinh(R/w)}{\sinh(r/w)} \bigg],\\
\end{split}\end{equation}
with boundary conditions $\Theta(r=0)=-\pi$ and $\Theta(r=\infty)=0$. Here $w$ is a constant representing the domain wall width, and $R=R(\phi)$ is a function of polar angle denoting the distance between skyrmion core and points at various polar angle $\phi$ on the $m_{z}=0$ contour of skyrmion configuration. Thus the deformation of skyrmion may be characterized by the function $R(\phi)$, which can be divided into different deformation modes using the Fourier expansion \cite{liu2022flexoresponses}
\begin{equation}\begin{split}\label{eq10}
R(\phi)=R_{0}+\sum_{n}a_{n}\cos(n\phi)+b_{n}\sin(n\phi),\\
\end{split}\end{equation}
with the integer $n\,(n=1, 2, 3,\cdots)$ denoting the rotational periodicity.
By substituting Eq. (\ref{eq8}) into Eq. (\ref{eq7}), the areal energy density functions can be rewritten in terms of $\Theta$ as
\begin{subequations}\label{eq11}
\begin{align}
&\varepsilon_\mathrm{ex}=Ad\bigg[ \bigg(\frac{\partial \Theta}{\partial r}\bigg)^{2}+\frac{1}{r^{2}}\bigg(\frac{\partial \Theta}{\partial \phi}\bigg)^{2}+\frac{\sin^{2}\Theta}{r^{2}} \bigg],\label{eq11a}\\
&\varepsilon_\mathrm{dm}=-Dd\bigg(\frac{\partial \Theta}{\partial r}+\frac{\sin2\Theta}{2r}\bigg),\label{eq11b}\\
&\varepsilon_\mathrm{st}=-K_{\mathrm{s}}d\sin^{2}\Theta\cos^{2}\beta,\label{eq11c}
\end{align}
\end{subequations}
and $\beta$ is the included angle between the directions of strain and polar angle $\phi$ of local magnetization $\textbf{\textit{m}}(\textbf{\textit{r}})$. Substituting the ansatz $\Theta(r, \phi)$ [Eq. (\ref{eq9})] into Eqs. (\ref{eq11a})-(\ref{eq11c}), one may further get the expressions for areal energy density in terms of $r$ and $\phi$. Subsequently, the areal forces density functions $\textbf{\textit{f}}_\mathrm{\mu}$ can be solved by using the relation
\begin{equation}\begin{split}\label{eq12}
&\textbf{\textit{f}}_\mathrm{\mu}=-\nabla \varepsilon_\mathrm{\mu}=f_{\mu,r}\textbf{\textit{e}}_{r}
+f_{\mu,\phi}\textbf{\textit{e}}_{\phi},\\
&f_{\mu,r}=-\frac{\partial \varepsilon_{\mu}}{\partial r},\\
&f_{\mu,\phi}=-\frac{1}{r}\frac{\partial \varepsilon_{\mu}}{\partial \phi},\\
\end{split}\end{equation}
and various force components $\textbf{\textit{F}}_\mathrm{\mu}$ are obtained after the integral
\begin{equation}\begin{split}\label{eq13}
\textbf{\textit{F}}_{\mu}&=F_{\mu,x}\textbf{\textit{e}}_{x}+F_{\mu,y}\textbf{\textit{e}}_{y},\\
\textbf{\textit{f}}_{\mu}&=f_{\mu,x}\textbf{\textit{e}}_{x}+f_{\mu,y}\textbf{\textit{e}}_{y},\\
F_{\mu,x}&=\int^{360^{\circ}}_{0^{\circ}}\mathrm{d}\phi \int^{\infty}_{0}f_{\mu,x}r\mathrm{d}r\\
&=\int^{360^{\circ}}_{0^{\circ}}\mathrm{d}\phi \int^{\infty}_{0}(f_{\mu,r}\cos\phi-f_{\mu,\phi}\sin\phi)r\mathrm{d}r,\\
F_{\mu,y}&=\int^{360^{\circ}}_{0}\mathrm{d}\phi \int^{\infty}_{0}f_{\mu,y}r\mathrm{d}r\\
&=\int^{360^{\circ}}_{0^{\circ}}\mathrm{d}\phi \int^{\infty}_{0}(f_{\mu,r}\sin\phi+f_{\mu,\phi}\cos\phi)r\mathrm{d}r,\\
\end{split}\end{equation}
with $f_{\mu,x}$, $f_{\mu,y}$, $F_{\mu,x}$ and $F_{\mu,y}$ marking the $x$- and $y$-components of the vectors $\textbf{\textit{f}}_{\mu}$ and $\textbf{\textit{F}}_{\mu}$ in rectangular coordinates. Noting that $\textbf{\textit{e}}_{r}=\cos\phi\textbf{\textit{e}}_{x}+\sin\phi\textbf{\textit{e}}_{y}$ and $\textbf{\textit{e}}_{\phi}=-\sin\phi\textbf{\textit{e}}_{x}+\cos\phi\textbf{\textit{e}}_{y}$, one may reformulate the functions in polar coordinates for simplicity
\begin{equation}\begin{split}\label{eq14}
\textbf{\textit{F}}_{\mu}&=\int^{360^{\circ}}_{0^{\circ}}\bf{\Gamma}_{\mu}\mathrm{d}\phi,\\
\bf{\Gamma}_{\mu}&=\Gamma_{\mu,r}\textbf{\textit{e}}_{r}+\Gamma_{\mu,\phi}\textbf{\textit{e}}_{\phi},\\
\Gamma_{\mu,r}&=\int^{\infty}_{0}f_{\mu,r}r\mathrm{d}r,\\
\Gamma_{\mu,\phi}&=\int^{\infty}_{0}f_{\mu,\phi}r\mathrm{d}r,\\
\end{split}\end{equation}
where $\bf{\Gamma}_{\mu}$ denotes angular density of force $\textbf{\textit{F}}_{\mu}$. $\Gamma_{\mu,r}$ and $\Gamma_{\mu,\phi}$ represent the radial and
tangential components, respectively. As shown in Fig.~\ref{fig5}(a), $\Gamma_{\mu,r}$ makes the different polar angular segments of skyrmion compression or expansion, while $\Gamma_{\mu,\phi}$ pulls them to rotate in clockwise (CW) or counterclockwise (CCW) direction.

Based on Eqs.(\ref{eq9}), (\ref{eq11}), (\ref{eq12}) and (\ref{eq14}), we deduced the formulae for the internal forces and strain force in terms of $\bf{\Gamma}_{\mu}$ by using a semi-analytical approach, with the detailed calculations given in the Appendixes A, B and C,
\begin{equation}\begin{split}\label{eq15}
\textbf{\textit{F}}_\mathrm{dm}=&Dd\int^{360^{\circ}}_{0^{\circ}}\bigg(G_1\textbf{\textit{e}}_{r}
+G_2\frac{\mathrm{d}\rho}{\mathrm{d}\phi}\textbf{\textit{e}}_{\phi}\bigg)\mathrm{d}\phi,\\
\textbf{\textit{F}}_\mathrm{ex}=&\frac{Ad}{w}\int^{360^{\circ}}_{0^{\circ}}
\bigg\{\bigg[H_1+H_2\bigg(\frac{\mathrm{d}\rho}{\mathrm{d}\phi}\bigg)^2\bigg]\textbf{\textit{e}}_{r}\\
&+\bigg[H_3\frac{\mathrm{d}\rho}{\mathrm{d}\phi}+H_4\bigg(\frac{\mathrm{d}\rho}{\mathrm{d}\phi}\bigg)^3
+H_5\frac{\mathrm{d}\rho}{\mathrm{d}\phi}\frac{\mathrm{d}^{2}\rho}{\mathrm{d}\phi^{2}}\bigg]\textbf{\textit{e}}_{\phi}\bigg\}\mathrm{d}\phi,\\
\textbf{\textit{F}}_\mathrm{st}=&-K_{\mathrm{s}}wd
\int^{360^{\circ}}_{0^{\circ}}\bigg[I_1\cos^{2}\beta\textbf{\textit{e}}_{r}\\
&+\bigg(I_2\cos^{2}\beta\frac{\mathrm{d}\rho}{\mathrm{d}\phi}+2I_1\sin\beta\cos\beta\bigg)\textbf{\textit{e}}_{\phi}\bigg]\mathrm{d}\phi,\\
\end{split}\end{equation}
where we introduced the dimensionless quantity $\rho=\rho(\phi)=R(\phi)/w$, and coefficients $G_k=G_k(\rho)$, $H_k=H_k(\rho)$ and $I_k=I_k(\rho)$ are functions of $\rho$. The expressions for the functions $G_k=G_k(\rho)$, $H_k=H_k(\rho)$ and $I_k=I_k(\rho)$ [see Eqs. (\ref{eqa6}), (\ref{eqa7}), (\ref{eqb6}) and (\ref{eqc5})] are given in the Appendixes. From the above functions, we can understand that the forces $\textbf{\textit{F}}_\mathrm{dm}$, $\textbf{\textit{F}}_\mathrm{ex}$, and $\textbf{\textit{F}}_\mathrm{st}$ are related to the deformation of skyrmion structure, which are reflected by the variations of $R$ as a function of $\phi$ in terms of the first and second-order derivatives of $R(\phi)$ (i.e., the terms $\mathrm{d}\rho/\mathrm{d}\phi$, $\mathrm{d}^{2}\rho/\mathrm{d}\phi^2$) as well as the $\phi$-dependent functions $G_k(\rho), H_k(\rho)$, and $I_k(\rho)$. For asymmetric deformation of skyrmion structure, the angular density of force $\bf{\Gamma}_{\mu}$ acting on the whole polar angles segments of skyrmion configuration do not mutually cancel, and thus the integrals in Eq.(\ref{eq15}) always contribute nonzero $\textbf{\textit{F}}_\mathrm{dm}$, $\textbf{\textit{F}}_\mathrm{ex}$ and $\textbf{\textit{F}}_\mathrm{st}$, as schematically illustrated in Fig.~\ref{fig5}(a).

To further examine the relations between the forces $\textbf{\textit{F}}_\mathrm{dm}$, $\textbf{\textit{F}}_\mathrm{ex}$, $\textbf{\textit{F}}_\mathrm{st}$ and the deformation of skyrmion, we first analyzed the deformed skyrmion structures in the motions by measuring the deformation $\Delta R$ defined as
\begin{equation}\begin{split}\label{eq16}
R(\phi)=&R_{0}+\sum_{n=1,2,3,\cdots}a_{n}^{\prime}\cos(n\phi+\phi_n),\\
\Delta R(\phi)=&R(\phi)-R_{0}+\sum_{n=1,2,3,\cdots}a_{n}^{\prime}\\
=&\sum_{n=1,2,3,\cdots}a_{n}^{\prime}[1+\cos(n\phi+\phi_n)],\\
\end{split}\end{equation}
here the Fourier series for function $R(\phi)$ was reformulated from Eq. (\ref{eq10}) with using the new coefficients $a_{n}^{\prime}$, which can be determined with a given skyrmion structure. $\Delta R$ could be used to quantify the shape divergence of deformed skyrmion from the circular one, and it can also be expressed for simplicity as
\begin{equation}\begin{split}\label{eq17}
&\Delta R(\phi)=\sum_{n=1,2,3,\cdots}\Delta R_{n}(\phi),\\
&\Delta R_{n}(\phi)=a_{n}^{\prime}[1+\cos(n\phi+\phi_n)],\\
\end{split}\end{equation}
with $\Delta R_{n}$ resembling the deformation mode with $n$-fold rotational symmetry. The magnitude of coefficients $a_{n}^{\prime}$ represents the contribution of $n$-th mode to the deformation, and $\phi_n$ marks the phase shift.

Next, we numerically calculated the deformation modes of the skyrmion structures that obtained in the simulation, according to the following protocol: First, a set of data of $R(\phi)$ were collected through mearing the distance between the skyrmion core and the points at various polar angle $\phi$ on the $m_z=0$ contour of skyrmion configuration. Second, the Fourier analysis of the data of $R(\phi)$ were carried out to get the coefficient $a_{n}^{\prime}$ and phase shift $\phi_n$ for the skyrmion structure. In this step, the functions $R(\phi)$ and $\Delta R(\phi)$ were obtained in terms of series $\Delta R_{n}(\phi)$, and one may take into account several important terms with large coefficient $a_{n}^{\prime}$ in the series expansion Eqs. (\ref{eq16}) and (\ref{eq17}), which were adequate to reflect the essential deformation in skyrmion structures.

Here, we took the skyrmion structure displayed in Fig.~\ref{fig4}(c) as an example for examining the deformation modes. It was found that the largest coefficients $a_1^{\prime}, a_2^{\prime}$ and $a_3^{\prime}$ in the series expansion of $R(\phi)$, which means that the modes of $\Delta R_1$, $\Delta R_2$ and $\Delta R_3$ have the most important contributions to the deformation $\Delta R$ of the skyrmion, as depicted in Fig.~\ref{fig5}(b). It can be seen that the deformation is actually composed by the three main modes with 1-, 2- and 3-fold rotational symmetries. These results apparently suggest the asymmetric deformation of skyrmion structure, which naturally leads to the appearance of nonzero forces $\textbf{\textit{F}}_\mathrm{dm}$, $\textbf{\textit{F}}_\mathrm{ex}$ and $\textbf{\textit{F}}_\mathrm{st}$  according to Eq. (\ref{eq15}).

To verify the validity of the forces formulae for $\textbf{\textit{F}}_\mathrm{dm}$, $\textbf{\textit{F}}_\mathrm{ex}$ and $\textbf{\textit{F}}_\mathrm{st}$ in Eq. (\ref{eq15}), we used these formulae to calculate analytical results of the three forces appearing in the skyrmion motion, by considering the process displayed in Fig.~\ref{fig2}. In the calculations, we first obtained the series expansion of functions $R(\phi)$ at various simulated time $t$ using the above protocol. Then, the function $R(\phi)$ was substituted into Eq. (\ref{eq15}) to get the forces $\textbf{\textit{F}}_\mathrm{dm}$, $\textbf{\textit{F}}_\mathrm{ex}$ and $\textbf{\textit{F}}_\mathrm{st}$ at each time $t$, as the analytical results for the $F_\mathrm{dm}\mbox{-}t$, $F_\mathrm{ex}\mbox{-}t$ and $F_\mathrm{st}\mbox{-}t$ curves plotted in Fig.~\ref{fig5}(c).\par
We further compared the analytical results with simulation results for time evolutions of $F_\mathrm{dm}$, $F_\mathrm{ex}$ and $F_\mathrm{st}$ in Fig.~\ref{fig5}(c). It was found that analytical results and simulation results for all the three forces are of the same magnitude. In addition, the variation tendency of the curves for analytical results show qualitative consistence with that of the simulation results. We also noted that there are some discrepancies in numerical values between these results. This may be partly because that the geometry effect of nanostructure on $\textbf{\textit{F}}_\mathrm{dm}$ is not considered in our analytical model, and partly because that the expressions Eqs. (\ref{eq8}) and (\ref{eq9}) for describing the deformed skyrmion structure are approximate ansatzes. Despite the complicated sources for the appearance of $\textbf{\textit{F}}_\mathrm{dm}$, $\textbf{\textit{F}}_\mathrm{ex}$ and $\textbf{\textit{F}}_\mathrm{st}$, we may assume that the asymmetric deformation of skyrmion are mainly responsible for the forces, whose underlying physical mechanism has been well captured by our proposed analytical model.

To proceed, we analytically studied the force $\textbf{\textit{F}}_\mathrm{dem}$, as the detailed derivations of formula for $\textbf{\textit{F}}_\mathrm{dem}$ presented in Appendix D. The demagnetization energy in the magnetic nanostructure stems from the interaction of positive and negative magnetic charges in the sample, and it can be calculated using magnetostatic approach \cite{hubert1998magnetic,jackson2012classical,wang2018theory}. It was observed in the simulation results that the non-collinear arrangement of magnetic moments only exists in the regions of skyrmion configuration $S_{\mathrm{sky}}$ and nanostructure edge $S_{\mathrm{edge}}^{\prime}$, and thus the magnetic charges appear in these regions, as indicated by the formula for magnetic charge density [Eq. (\ref{eq20})] below.\par
It is proved in Appendix D that the force $\textbf{\textit{F}}_\mathrm{dem}$ acting on skyrmion is contributed by the the magnetostatic interaction energy $E_{\mathrm{dem,sky\mbox{-}edge}}$ between the magnetic charges in the regions $S_{\mathrm{sky}}$ and $S_{\mathrm{edge}}^{\prime}$ as
\begin{equation}\begin{split}\label{eq18}
E_{\mathrm{dem,sky\mbox{-}edge}}=\frac{\mu_{0}}{4\pi} \iint \limits_{S_{\mathrm{sky}}} \iint \limits_{S_{\mathrm{edge}}^{\prime}}
\frac{\sigma_{\mathrm{m,sky}}\left(\textbf{\textit{r}}\right)\sigma_{\mathrm{m,edge}}\left(\textbf{\textit{r}}^{\prime}\right)}{\left| \textbf{\textit{r}}-\textbf{\textit{r}}^{\prime} \right|}\mathrm{d}S^{\prime}\mathrm{d}S,\\
\end{split}\end{equation}
and
\begin{equation}\begin{split}\label{eq19}
\textbf{\textit{F}}_{\mathrm{dem}}=&-\left(\nabla E_{\mathrm{dem,sky\mbox{-}edge}}\right)_{\mathrm{invariant}\, \sigma_{\mathrm{m,sky}},\,\sigma_{\mathrm{m,edge}}}\\
=&\frac{\mu_{0}}{4\pi} \iint \limits_{S_{\mathrm{sky}}} \iint \limits_{S_{\mathrm{edge}}^{\prime}}
\frac{\sigma_{\mathrm{m,sky}}\left(\textbf{\textit{r}}\right)\sigma_{\mathrm{m,edge}}\left(\textbf{\textit{r}}^{\prime}\right)}{\left| \textbf{\textit{r}}-\textbf{\textit{r}}^{\prime} \right|^{3}}\left(\textbf{\textit{r}}-\textbf{\textit{r}}^{\prime}\right)\mathrm{d}S^{\prime}\mathrm{d}S,\\
\end{split}\end{equation}
where the magnetic charges are considered to distribute on the $xy$-plane with the magnetic surface charge density defined as
\begin{equation}\begin{split}\label{eq20}
\sigma_{\mathrm{m}}\left(\textbf{\textit{r}}\right)=-\mathrm{M}_{\mathrm{s}}d\nabla\cdot\textbf{\textit{m}}\left(\textbf{\textit{r}},z\right)
=-\mathrm{M}_{\mathrm{s}}d\bigg(\frac{\partial \textit{m}_{x}}{\partial x}+\frac{\partial \textit{m}_{y}}{\partial y}\bigg),
\end{split}\end{equation}
with $\textbf{\textit{r}}=\textit{r}\textbf{\textit{e}}_{\textit{r}}=\textit{x}\textbf{\textit{e}}_{\textit{x}}+\textit{y}\textbf{\textit{e}}_{\textit{y}}$ denoting the position vector of the magnetic charges on the $xy$-plane, and $d$ representing the thickness of the thin nanostructure.\par

\begin{figure}[t]
\centering
\includegraphics[width=0.45\textwidth]{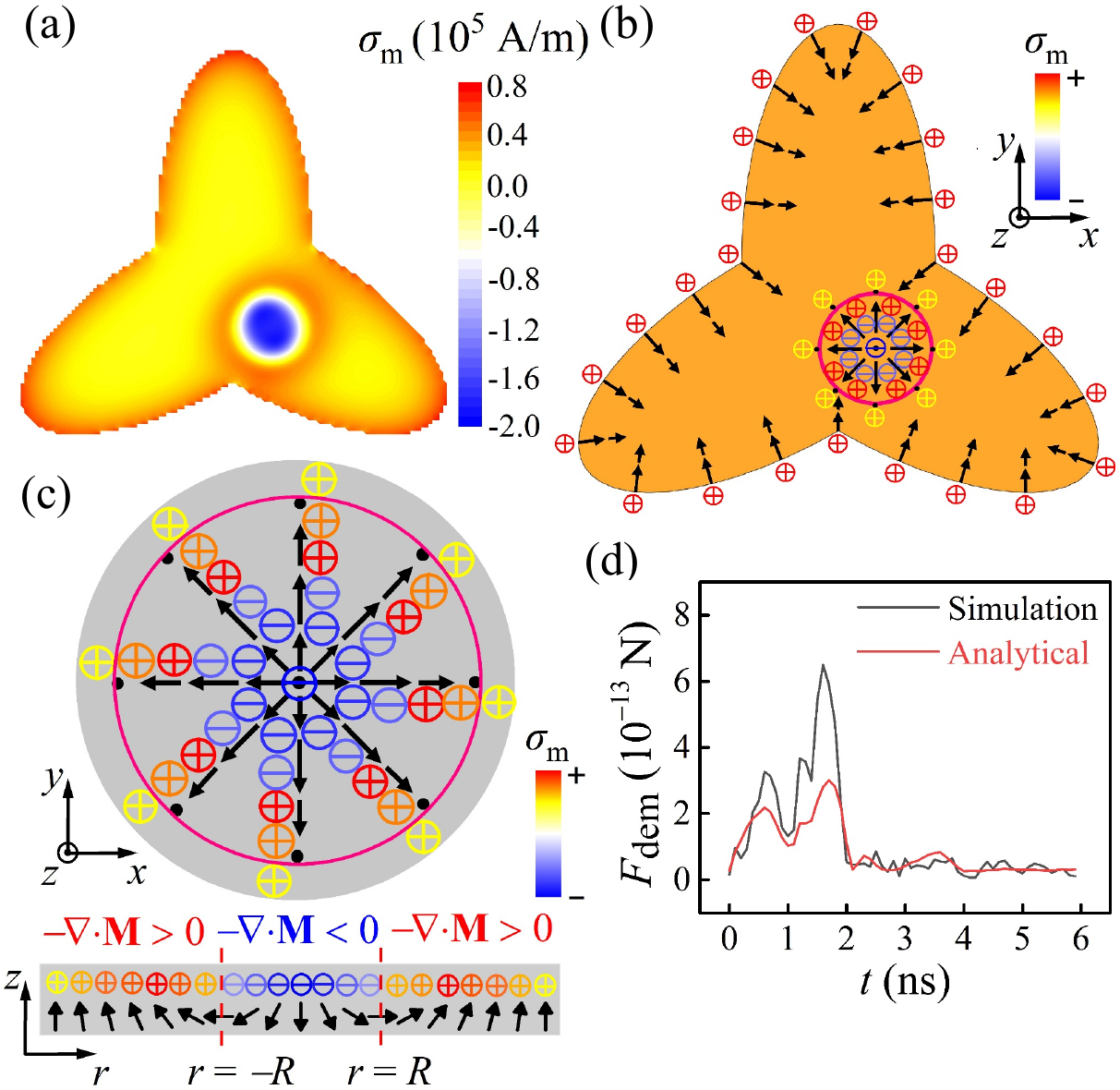}
\caption{(Color online) (a) Plots of the magnetic surface charge density profiles $\sigma_{\mathrm{m}}\left(\textbf{\textit{r}}\right)$ in the nanostructure for the deformed skyrmion shown in Fig.~\ref{fig4}(c). (b) Schematic diagram of the in-plane magnetization components $\textbf{M}_{xy}$ and $\sigma_{\mathrm{m}}\left(\textbf{\textit{r}}\right)$ in the regions of the nanostructure edge and the skyrmion. The positive magnetic charges emerge on the edge of the nanostructure due to the presence of tilted edge magnetization. (c) The enlarged view of $\textbf{M}_{xy}$ and $\sigma_{\mathrm{m}}\left(\textbf{\textit{r}}\right)$ profiles for the skyrmion presented in (b). The insert at the bottom shows the N\'{e}el spiral structure with the distribution of $\sigma_{\mathrm{m}}$ along the radial direction of the skyrmion, in which the presence of  positive and negative $\sigma_{\mathrm{m}}$ are related to the sign of the term $-\nabla\cdot\textbf{M}$. (d) Comparison between the analytical and simulation results of the time evolutions of force $F_\mathrm{dem}$ for the skyrmion motion displayed in Fig.~\ref{fig2}.}
\label{fig6}
\end{figure}

In Fig.~\ref{fig6}(a), the $\sigma_{\mathrm{m}}\left(\textbf{\textit{r}}\right)$  profiles were plotted for the skyrmion structure shown in Fig.~\ref{fig4}(c). As we can see from the schematic diagram in Fig.~\ref{fig6}(b), the in-plane magnetization M$_{xy}$ with $-\nabla\cdot\textbf{{M}}>0$ appears at the nanostructure edge which leads to the presence of positive magnetic charges at the edge. In fact, the in-plane magnetization components at the edge of nanostructure are mainly induced by the DM interaction, which forms the tilted edge magnetization and thus creates magnetic charges at the edge of nanostructure \cite{iwasaki2013current,rohart2013skyrmion,mulkers2018effect}.\par
For the skyrmion, the term $-\nabla\cdot\textbf{{M}}>0$ at the region of $-R < r < R$ while $-\nabla\cdot\textbf{{M}}<0$ at the regions of $r < -R$ and $r > R$, which results in the positive and negative magnetic charges emerging at these regions respectively, as schematically illustrated in Fig.~\ref{fig6}(c). It was also noted that magnetic charges at the periphery of skyrmion structure have the same sign with that at the edge of nanostructure. This means that the repulsive interaction appears between the skyrmion perimeter and the nanostructure edge that restrains the skyrmion from reaching the nanostructure boundary, which clearly illustrates the boundary effects.\par
It is also clearly evident in Eq. (\ref{eq19}) that $\textbf{\textit{F}}_{\mathrm{dem}}$ is dependent on the magnetic charges distributions in the regions $S_{\mathrm{sky}}$ and $S_{\mathrm{edge}}^{\prime}$, as well as the geometric shape and size of the nanostructure. Based on the formula of $\textbf{\textit{F}}_{\mathrm{dem}}$, we numerically calculated the forces $\textit{F}_{\mathrm{dem}}$ for the skyrmion motion displayed in Fig.~\ref{fig2}, as the results shown in Fig.~\ref{fig6}(d). By comparing the analytical and simulated $\textit{F}_{\mathrm{dem}}\mbox{-}t$ curves, we found that the results of $\textit{F}_{\mathrm{dem}}$ are of the same magnitude and the variation tendencies are qualitative consistent in two curves.\par
To further clarify the effects of nanostructure geometry on $\textbf{\textit{F}}_{\mathrm{dem}}$, we rewrite the expression of $E_{\mathrm{dem,sky\mbox{-}edge}}$ in terms of a series of magnetic multipole moments in the skyrmion by employing the multipole expansion technique \cite{jackson2012classical,raab2004multipole},
\begin{equation}\begin{split}\label{eq21}
&E_{\mathrm{dem,sky\mbox{-}edge}}\\
=&\frac{\mu_{0}}{4\pi} \iint \limits_{S_{\mathrm{edge}}^{\prime}}
\bigg(\frac{Q_{\mathrm{m}}}{r^{\prime}}+\frac{\textbf{\textit{P}}_{\mathrm{m}}\cdot\textbf{\textit{r}}^{\prime}}{r^{\prime 3}}+\cdots\bigg)
\sigma_{\mathrm{m,edge}}\left(\textbf{\textit{r}}^{\prime}\right)\mathrm{d}S^{\prime},\\
\end{split}\end{equation}
where $Q_{\mathrm{m}}=\iint \limits_{S_{\mathrm{sky}}} \sigma_{\mathrm{m,sky}}(\textbf{\textit{r}})\mathrm{d}S$ is the total magnetic charge (i.e., magnetic monopole moment) of the skyrmion configuration, and $\textbf{\textit{P}}_{\mathrm{m}}=\iint \limits_{S_{\mathrm{sky}}} \sigma_{\mathrm{m,sky}}(\textbf{\textit{r}})\textbf{\textit{r}}\mathrm{d}S$ is the magnetic dipole moment in the skyrmion configuration.\par
It is proved that the monopole moment $Q_{\mathrm{m}}=0$ in the skyrmion which implies it has none contribution to the force $\textbf{\textit{F}}_{\mathrm{dem}}$. We further derived the formula for force $\textbf{\textit{F}}_{\mathrm{dem,dip}}$ that is contributed by the magnetic dipole moment $\textbf{\textit{P}}_{\mathrm{m}}$ in the skyrmion [see Eqs. (\ref{eqd21}) and (\ref{eqd22}) in Appendix D for more information]. As the leading term in the multipole expansion of force $\textbf{\textit{F}}_{\mathrm{dem}}$, the force $\textbf{\textit{F}}_{\mathrm{dem,dip}}$ is demonstrated to be a consequence of the combined effect of the asymmetric deformation of skyrmion, the geometric shape and size of nanostructure, together with the magnetic charges distributions at the nanostructure edge. This elucidates the critical effect of geometric shape and size of nanostructure on the energy distribution of $E_{\mathrm{dem}}$ in the nanostructure and also the skyrmion dynamics.\par

In the last of this section, we analyzed the effect of force $\textbf{\textit{F}}_\mathrm{pma}$ on the skyrmion motion. The force density $\textbf{\textit{f}}_\mathrm{pma}$ is obtained from Eq. (\ref{eq7}),
\begin{equation}\begin{split}\label{eq22}
\textbf{\textit{f}}_\mathrm{pma}=-\nabla \varepsilon_\mathrm{pma}=2K_{z}m_{z}\nabla m_{z}.\\ \end{split}\end{equation}

Fig.~\ref{fig7} displays the cross section of N\'{e}el skyrmion along the radial direction, which is described by a generalized version of one-dimensional N\'{e}el-type spiral \cite{nagaosa2013topological,wang2018theory,romming2015field}
\begin{equation}\begin{split}\label{eq23}
&\textbf{\textit{m}}=\sin\Theta\textbf{\textit{e}}_{r}+\cos\Theta\textbf{\textit{e}}_{z},\\
&\Theta=\Theta(r)=2\chi\arctan \bigg[ \frac{\sinh(R/w)}{\sinh(r/w)} \bigg]+\Psi,\\
\end{split}\end{equation}
and the constants $\chi=\pm1$ and $\Psi=0$ or $\pi$ classifying the four types of N\'{e}el spirals according to their different chirality and polarity.

The direction of force density $\textbf{\textit{f}}_\mathrm{pma}$ is reflected by the sign of the term $\it{m}_z\nabla\it{m}_z$ from Eq. (\ref{eq22}), which leads to the same features of $\textbf{\textit{f}}_\mathrm{pma}$ alignments in the spirals, regardless of the chirality and polarity of spirals. We can see in Fig.~\ref{fig7} that $\textbf{\textit{f}}_\mathrm{pma}$ points to the spiral core at the region of $-R<r<R$, while it orients to the outer of spiral at the regions of $r<-R$ and $r>R$. In this sense, most results presented in this paper for the effect of $\textbf{\textit{F}}_\mathrm{pma}$ on the skyrmion motion can be generalized to the cases of N\'{e}el skyrmion with various types of chirality and polarity. In addition, it is clearly that $f_\mathrm{pma}$ in the high $K_z$ regions is larger than that in the low $K_z$ regions, since $f_\mathrm{pma}$ is proportional to $K_z$ from Eq. (\ref{eq22}). This means that a non-zero force $\textbf{\textit{F}}_\mathrm{pma}$ naturally appears for the N\'{e}el-type spiral under the nonuniform PMA, and thus the spiral is pulled towards the high $K_z$ region by $\textbf{\textit{F}}_\mathrm{pma}$. Therefore, looking back to the skyrmion motion in Fig.~\ref{fig2}, we can understand that the nonuniform PMA in the capping nanoisland induce the force $\textbf{\textit{F}}_\mathrm{pma}$, which attracts the skyrmion to move towards the capping region at the first stage (see snapshots at $t$ = 0.0 ns $\sim$  1.0 ns). However, the skyrmion finally runs away from the capping region mainly because that the capping region is in fact an energy potential energy barrier for the skyrmion [see Fig.~\ref{fig1}(e)]. Our further simulations found that the skyrmion trajectory in the nanostructure is sensitive to the strength of $K_{\mathrm{c}}$ and the geometry size of capping nanoisland, and the switching of skyrmion state from State I to State II can only been achieved by choosing the suitable parameters of $K_{\mathrm{c}}$ and nanoisland size. All these results demonstrated that the nonuniform PMA and the induced force $\textbf{\textit{F}}_\mathrm{pma}$ have a significant impact on the skyrmion dynamics.
\begin{figure}[t]
\centering
\includegraphics[width=0.50\textwidth]{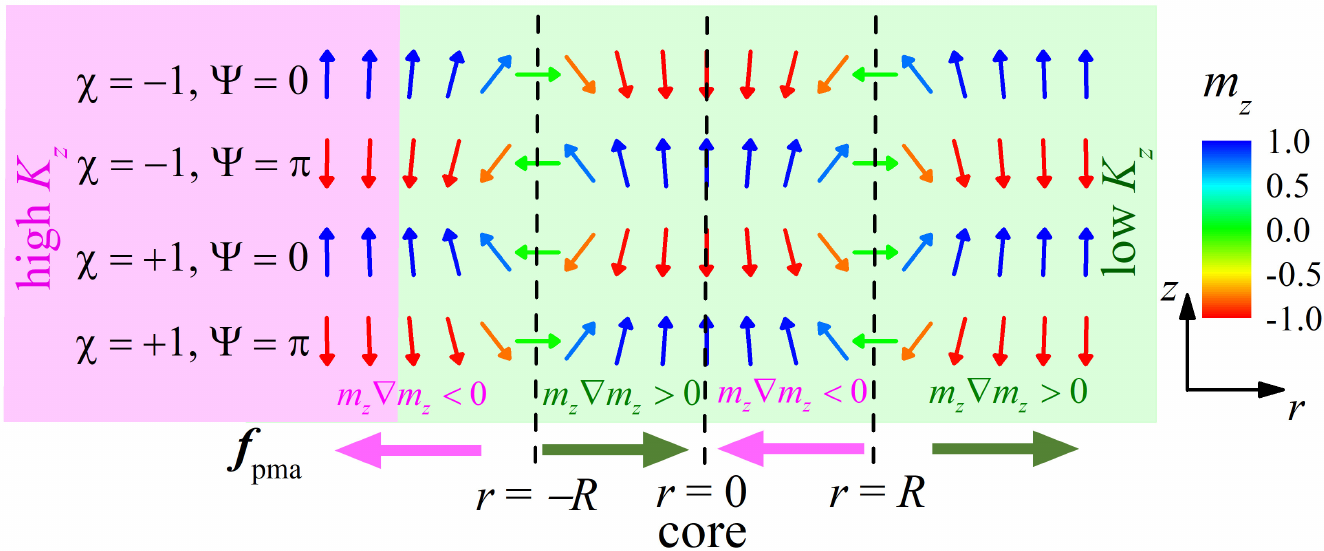}
\caption{(Color online)  Schematic illustrations of the forces density $\textbf{\textit{f}}_\mathrm{pma}$ acting on the one-dimensional N\'{e}el spirals under nonuniform PMA, where the left and right sides of the panel are attributed to the regions of high $K_z$ (colored magenta) and low $K_z$ (colored green), respectively. Here display four types of one-dimensional N\'{e}el spiral with various chirality $\chi$ and polarity $\Psi$. The inserted arrows at the bottom show the directions of $\textbf{\textit{f}}_\mathrm{pma}$ in different regions divided by the dashed lines.}
\label{fig7}
\end{figure}

Before concluding this section, we briefly discuss the strain force $\textbf{\textit{F}}_\mathrm{st}$. It was noted in Eq. (\ref{eqc9}) that $\textbf{\textit{F}}_\mathrm{st}=0$ for the skyrmion in the case of $\rho = R(\phi)/w\geq5.0$ . This implies that $\textbf{\textit{F}}_\mathrm{st}$ disappears for the large size of skyrmion, no matter how asymmetric its deformed shape is. In this sense, one may expect that the automotion of deformed skyrmion driven by internal forces may appear in the case of large-sized skyrmion, in which the strain only distorts the skyrmion structure but not exert force on the skyrmion. Therefore, the conception of automotion of deformed skyrmion may provide a new pathway for development of energy efficient magnetoelectric devices \cite{leutner2022skyrmion,yanes2019skyrmion,du2023strain,nikonov2014automotion,mawass2017switching,yershov2018geometry}, which would bring some insights for future studies.\par
On the other hand, previous studies \cite{liu2023emergent,gong2022skyrmion,iwasaki2013current} demonstrated that the pinning and defects of material system may cause the skyrmion deformation, and thus exert critical impacts on the skyrmion dynamics. Although the effect of pinning and defect on skyrmion were not considered in the present study, to explore the dynamics of deformed skyrmions in the generalized context of various types of pinning and defects is important for extending our understanding the emergent mechanics of magnetic skyrmions, which would be an interesting challenge for our further studies.

\section{Conclusions}
In summary, we have investigated the strain-mediated voltage-controlled motion of deformed skyrmions in multiferroic heterostructures with a flower-shaped magnetic nanostructure, by employing micromagnetic simulation. We first presented three possible states of isolated skyrmion nucleated in the nanostructure with inhomogeneous PMA. Simulations revealed the dynamic behaviors of isolated skyrmion confined in the nanostructure and the mutually switching between different isolated skyrmion states, by applying suitable in-plane strain pulses. Based on these findings, we numerically demonstrated that the skyrmion motions are driven by internal forces and strain force, which mainly originate from the asymmetrically deformed skyrmion with the asymmetric distributions of internal energy density and strain energy density. To confirm this scenario, we constructed an analytical model of deformed skyrmions to elucidate the dependence of the internal forces and strain force on the deformation of skyrmion structures, with a set of formulae derived for these forces in a semi-analytical approach. Besides, our calculations based on the formulae verified the forces appearing in the skyrmion motion, and the resulting forces are in accordance with the simulated data. This suggested the reliability of our semi-analytical model that could capture the main physics responsible for the motion of deformed skyrmion. Therefore, our study extends the understanding of mechanics emerging in deformed skyrmion, and provides an effective approach for deterministically controlling the dynamic behaviors of deformed skyrmion via strain forces and internal forces, which may open new routes to development of skyrmion-based energy efficient spintronic devices.

\acknowledgments
This work was supported by the Natural Science Foundation of China (Grant No. 11604059), the Natural Science Foundation of Guangdong Province, China (Grant No. 2017A030313020), and the Key Discipline of Materials Science and Engineering, Bureau of Education of Guangzhou, China (Grant No. 202255464).

\appendix

\section{Derivation of the formula for force $\textbf{\textit{F}}_\mathrm{dm}$}
In this section, we derived the formula for force $\textbf{\textit{F}}_\mathrm{dm}$ using a semi-analytical approach.
By substituting the ansatz $\Theta(r, \phi)$ [Eq. (\ref{eq9})] into the function $\varepsilon_\mathrm{dm}$ [Eq. (\ref{eq11b})], one may obtain the radial and tangential components of force density $\textbf{\textit{f}}_\mathrm{dm}$ with the definition Eq. (\ref{eq12})
\begin{equation}\begin{split}\label{eqa1}
&f_{\mathrm{dm},r}=-\frac{\partial \varepsilon_{\mathrm{dm}}}{\partial r}=Ddg_1(r,R),\\
&f_{\mathrm{dm},\phi}=-\frac{1}{r}\frac{\partial \varepsilon_{\mathrm{dm}}}{\partial \phi}=Ddg_2(r,R)\frac{\mathrm{d}\rho}{\mathrm{d}\phi},\\
&g_1(r,R)=\frac{\partial^{2}\Theta}{\partial r^{2}}
+\frac{\cos2\Theta}{r}\frac{\partial\Theta}{\partial r}-\frac{\sin2\Theta}{2r^{2}},\\
&g_2(r,R)=\frac{w}{r}\bigg(\frac{\partial^{2}\Theta}{\partial r \partial R}
+\frac{\cos2\Theta}{r}\frac{\partial\Theta}{\partial R}\bigg),\\
\end{split}\end{equation}
where we introduced the dimensionless quantities $\rho=\rho(\phi)=R(\phi)/w$ and $\delta=r/w$. Therefore, the ansatz $\Theta(r, \phi)$ can be rewritten in terms of $\rho$ and $\delta$
\begin{equation}\begin{split}\label{eqa2}
\Theta(\delta, \rho)=-2\arctan\bigg(\frac{\sinh \rho}{\sinh \delta}\bigg).\\
\end{split}\end{equation}

Based on the force formula in Eq. (\ref{eq14}), we obtained
\begin{equation}\begin{split}\label{eqa3}
&\textbf{\textit{F}}_\mathrm{dm}=Dd\int^{360^{\circ}}_{0^{\circ}}\bigg(G_1\textbf{\textit{e}}_{r}
+G_2\frac{\mathrm{d}\rho}{\mathrm{d}\phi}\textbf{\textit{e}}_{\phi}\bigg)\mathrm{d}\phi,\\
&G_1=\int^{\infty}_{0}g_{1}(r,R)r\mathrm{d}r,\\
&G_2=\int^{\infty}_{0}g_{2}(r,R)r\mathrm{d}r,\\
\end{split}\end{equation}
and the integrals $G_1$ and $G_2$ are
\begin{equation}\begin{split}\label{eqa4}
G_1=&\int^{\infty}_{0}\bigg(\frac{\partial^{2}\Theta}{\partial r^{2}}+\frac{\cos2\Theta}{r}\frac{\partial\Theta}{\partial r}-\frac{\sin2\Theta}{2r^{2}}\bigg) r\mathrm{d}r\\
=&\int^{\infty}_{0}\bigg(\frac{\partial^{2}\Theta}{\partial \delta^{2}}+\frac{\cos2\Theta}{\delta}\frac{\partial\Theta}{\partial \delta}-\frac{\sin2\Theta}{2\delta^{2}}\bigg)\delta\mathrm{d}\delta\\
=&G_1(\rho),\\
G_2=&\int^{\infty}_{0}\frac{w}{r}\bigg(\frac{\partial^{2}\Theta}{\partial r \partial R}
+\frac{\cos2\Theta}{r}\frac{\partial\Theta}{\partial R}\bigg)r\mathrm{d}r,\\
=&\int^{\infty}_{0}\bigg(\frac{\partial^{2}\Theta}{\partial \delta \partial \rho}
+\frac{\cos2\Theta}{\delta}\frac{\partial\Theta}{\partial \rho}\bigg)\mathrm{d}\delta\\
=&G_2(\rho),\\
\end{split}\end{equation}
which means that $G_1$ and $G_2$ are dimensionless parameters as a function of $\rho$.

From Eq. (\ref{eqa2}), we had the following relations
\begin{equation}\begin{split}\label{eqa5}
&\frac{\partial\Theta}{\partial \delta}=-\sin\Theta\coth\delta,\\
&\frac{\partial\Theta}{\partial \rho}=\sin\Theta\coth \rho,\\
&\frac{\partial^{2}\Theta}{\partial \delta^{2}}=\sin\Theta\cos\Theta\coth^{2} \delta+\frac{\sin\Theta}{\sinh^{2} \delta},\\
&\frac{\partial^{2}\Theta}{\partial \delta \partial \rho}=-\sin\Theta\cos\Theta\coth\delta\coth \rho,\\
&\sin\Theta=-\frac{2\sinh \delta\sinh \rho}{\sinh^{2} \delta+\sinh^{2} \rho},\\
&\cos\Theta=\frac{\sinh^{2} \delta-\sinh^{2} \rho}{\sinh^{2} \delta+\sinh^{2} \rho}.\\
\end{split}\end{equation}

To further determine the expressions of functions $G_1(\rho)$ and $G_2(\rho)$, we substituted Eq. (\ref{eqa5}) into the integrals Eq. (\ref{eqa4}) for replacing the function $\Theta(\delta, \rho)$ and its derivations, and then numerically computed these integrals for getting a set of data of $G_1(\rho)$ and $G_2(\rho)$ at a broad range of $1.0\leq \rho \leq 20.0$, which is applied to a variety of skyrmion configurations. At last, these data were numerically fitted by a polynomial function for the range of small $\rho$ and an exponential decay function for the range of large $\rho$, as the results presented in Fig.~\ref{figA1}. The resulting fitting formulae for $G_1(\rho)$ and $G_2(\rho)$ are

\begin{equation}\label{eqa6}
G_1=\\ \left\{\!
\begin{aligned}
\!&-3.48605-0.277\rho+0.09949\rho^2,\, \mathrm{if}\, 1.0\leq \rho \leq 2.0 \\
\!&-3.15032-0.57534 e^{-\frac{\rho-1.70376}{2.54134}},\, \mathrm{if}\, 2.0\leq \rho \leq 20.0
\end{aligned},\right.
\end{equation}
\begin{equation}\label{eqa7}
G_2=\\ \left\{\!
\begin{aligned}
\!&-0.76835-0.83849\rho+0.26004\rho^2 \\&- 0.02684\rho^3,\, \mathrm{if}\, 1.0\leq \rho \leq 3.0 \\
\\
\!&-0.1347 e^{-\frac{\rho-3.10118}{2.54333}},\, \mathrm{if}\, 3.0\leq \rho \leq 20.0
\end{aligned}.\right.
\end{equation}
\counterwithin{figure}{section}
\begin{figure}[t]
\centering
\includegraphics[width=0.34\textwidth]{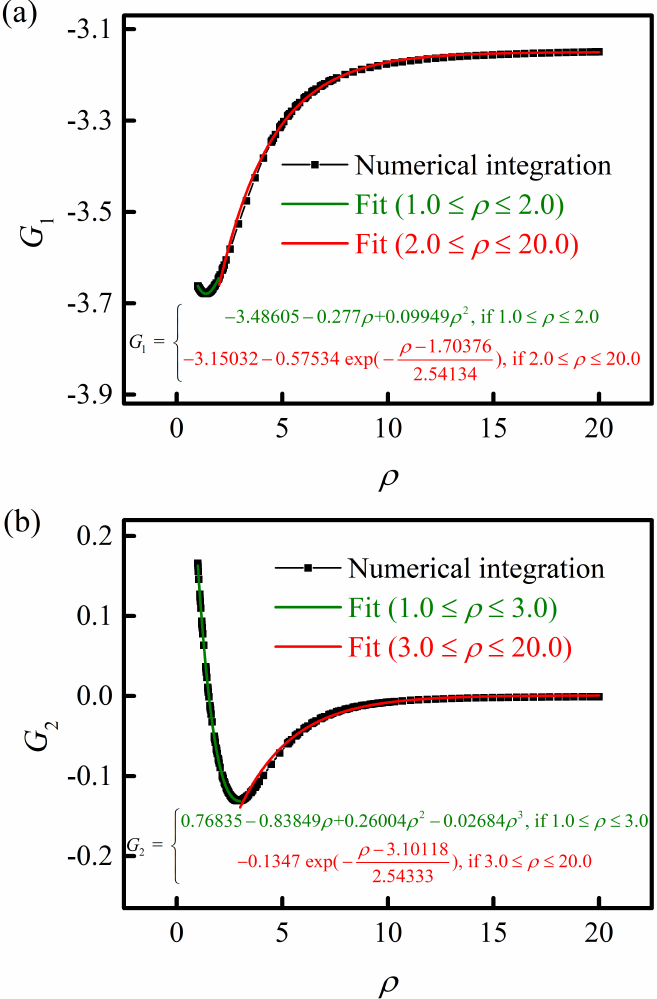}
\setcounter{figure}{0}
\renewcommand{\thefigure}{A\arabic{figure}}
\caption{(Color online) Numerical integration and curve fitting for the functions (a) $G_1(\rho)$ and (b) $G_2(\rho)$ at a broad range of $1.0\leq \rho\leq 20.0$.}
\label{figA1}
\end{figure}
\section{Derivation of the formula for force $\textbf{\textit{F}}_\mathrm{ex}$}
Following the semi-analytical approach in the Appendix A, we derived the formula for force $\textbf{\textit{F}}_\mathrm{ex}$ in this section.
We substituted the ansatz $\Theta(r, \phi)$ [Eq. (\ref{eq9})] into the function $\varepsilon_{\mathrm{ex}}$ [Eq. (\ref{eq11a})], obtaining
\begin{equation}\begin{split}\label{eqb1}
&\varepsilon_\mathrm{ex}=\frac{A\sin^2\Theta}{w^2}\bigg[ \coth^2 \delta+ \frac{\coth^2 \rho}{\delta^2} \bigg(\frac{\mathrm{d}\rho}{\mathrm{d}\phi}\bigg)^{2}+\frac{1}{\delta^{2}} \bigg],\\
\end{split}\end{equation}
where we used the following relations
\begin{equation}\begin{split}\label{eqb2}
&\frac{\partial\sin^{2}\Theta}{\partial r}=-\frac{2}{w}\sin^{2}\Theta\cos\Theta\coth\delta,\\
&\frac{\partial\coth^{2} \delta}{\partial r}=-\frac{2}{w}\frac{\coth\delta}{\sinh^{2} \delta},\\
&\frac{\partial\sin^{2}\Theta}{\partial \phi}=2\sin^{2}\Theta\cos\Theta\coth \rho \frac{\mathrm{d}\rho}{\mathrm{d}\phi},\\
&\frac{\mathrm{d}\coth^{2} \rho}{\mathrm{d}\phi}=\frac{-2\coth \rho}{\sinh^{2} \rho}\frac{\mathrm{d}\rho}{\mathrm{d}\phi}.\\
\end{split}\end{equation}
The force density $\textbf{\textit{f}}_\mathrm{ex}$ with the radial and tangential components is
\begin{equation}\begin{split}\label{eqb3}
&\textbf{\textit{f}}_\mathrm{ex}=-\nabla \varepsilon_\mathrm{ex}=f_{\mathrm{ex},r}\textbf{\textit{e}}_{r}
+f_{\mathrm{ex},\phi}\textbf{\textit{e}}_{\phi},\\
&f_{\mathrm{ex},r}=-\frac{\partial \varepsilon_{\mathrm{ex}}}{\partial r}=\frac{Ad}{w^3}\bigg[h_1+h_2\bigg(\frac{\mathrm{d}\rho}{\mathrm{d}\phi}\bigg)^2\bigg],\\
&f_{\mathrm{ex},\phi}=-\frac{1}{r}\frac{\partial \varepsilon_{\mathrm{ex}}}{\partial \phi}=\frac{Ad}{w^3}\bigg[ h_3\frac{\mathrm{d}\rho}{\mathrm{d}\phi}
+h_4\bigg(\frac{\mathrm{d}\rho}{\mathrm{d}\phi}\bigg)^3
+h_5\frac{\mathrm{d}\rho}{\mathrm{d}\phi} \frac{\mathrm{d}^{2} \rho}{\mathrm{d}\phi^{2}} \bigg],\\
\end{split}\end{equation}
with the functions $h_k\,(k=1,2,3,4,5)$ as

\begin{eqnarray}\begin{split}\label{eqb4}
h_1=&2\sin^{2}\Theta \bigg(\cos\Theta\coth^3 \delta+\frac{\cos\Theta \coth\delta}{\delta^2}\\
&+\frac{\coth\delta}{\sinh^{2} \delta}+\frac{1}{\delta^3} \bigg),\\
h_2=&2\sin^{2}\Theta\coth^2 \rho \bigg(\frac{\cos\Theta \coth\delta}{\delta^2}+\frac{1}{\delta^3} \bigg),\\
h_3=&-2\sin^{2}\Theta\cos\Theta\coth \rho \bigg(\frac{\coth^2 \delta}{\delta}+\frac{1}{\delta^3} \bigg),\\
h_4=&-\frac{2}{\delta^3} \bigg(\sin^{2}\Theta\cos\Theta\coth^3 \rho-\frac{\sin^{2}\Theta\coth \rho}{\sinh^{2} \rho} \bigg),\\
h_5=&-\frac{2}{\delta^3}\sin^{2}\Theta\coth^2 \rho.
\end{split}\end{eqnarray}

From Eq. (\ref{eqb3}) and the force formula in Eq. (\ref{eq14}), one may have
\begin{equation}\begin{split}\label{eqb5}
\textbf{\textit{F}}_\mathrm{ex}&=\frac{Ad}{w}\int^{360^{\circ}}_{0^{\circ}}
\bigg\{\bigg[H_1+H_2\bigg(\frac{\mathrm{d}\rho}{\mathrm{d}\phi}\bigg)^2\bigg]\textbf{\textit{e}}_{r}\\
&+\bigg[H_3\frac{\mathrm{d}\rho}{\mathrm{d}\phi}+H_4\bigg(\frac{\mathrm{d}\rho}{\mathrm{d}\phi}\bigg)^3
+H_5\frac{\mathrm{d}\rho}{\mathrm{d}\phi}\frac{\mathrm{d}^{2} \rho}{\mathrm{d}\phi^{2}}\bigg]\textbf{\textit{e}}_{\phi}\bigg\}\mathrm{d}\phi,\\
H_k&=\frac{1}{w^2}\int^{\infty}_{0}h_k(r,R)r\mathrm{d}r
=\int^{\infty}_{0}h_k(\delta,\rho)\delta\mathrm{d}\delta\\
&=H_k(\rho)\,(k=1,2,3,4,5),\\
\end{split}\end{equation}
where $H_k$ are some dimensionless parameters as a function of $\rho$.

To further determine the functions $H_k(\rho)$, we substituted the expressions of $h_k(\delta,\rho)$ [Eq. (\ref{eqb4})] and $\Theta(\delta, \rho)$ [Eq. (\ref{eqa2})] into the integrals of $H_k(\rho)$ [Eq. (\ref{eqb5})]. Then numerical computations of these integrals yield a set of data of $H_k(\rho)$ at a broad range of $1.0\leq \rho \leq 20.0$, which were further numerically fitted by a double exponential decay function, as the results displayed in Fig.~\ref{figB1}. The resulting fitting formula for $H_k(\rho)$ are

\begin{equation}\begin{split}\label{eqb6}
&H_1(\rho)=0.27422 e^{-\frac{\rho}{3.86992}}+19.60248 e^{-\frac{\rho}{0.46583}},\\
&H_2(\rho)= 0.02229 +4.10751 e^{-\frac{\rho}{1.15609}}+66.64995 e^{-\frac{\rho}{0.30801}},\\
&H_3(\rho)= 0.00967 +72.34246 e^{-\frac{\rho}{0.26742}}+15.9706 e^{-\frac{\rho}{0.7858}},\\
&H_4(\rho)= 0.01386 +718.20378 e^{-\frac{\rho}{0.21647}}+13.26677 e^{-\frac{\rho}{0.75248}},\\
&H_5(\rho)= -0.04421 -129.46738 e^{-\frac{\rho}{0.3095}}-8.11446 e^{-\frac{\rho}{1.16114}},\\
\end{split}\end{equation}
\counterwithin{figure}{section}
\begin{figure*}[t]
\centering
\includegraphics[width=1.0\textwidth]{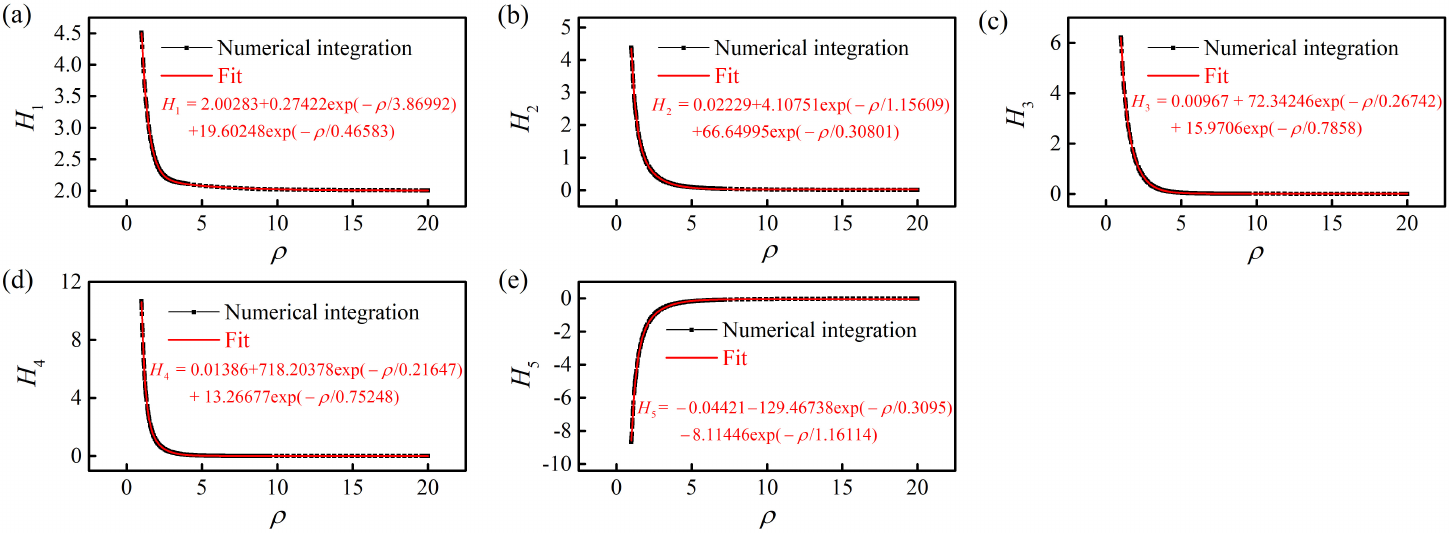}
\setcounter{figure}{0}
\renewcommand{\thefigure}{B\arabic{figure}}
\caption{(Color online) Numerical integration and curve fitting for the functions (a) $H_1(\rho)$, (b) $H_2(\rho)$, (c) $H_3(\rho)$, (d) $H_4(\rho)$ and (e) $H_5(\rho)$ at a broad range of $1.0\leq \rho\leq 20.0$.}
\label{figB1}
\end{figure*}
\section{Derivation of the formula for force $\textbf{\textit{F}}_\mathrm{st}$}
Following the semi-analytical approach in the Appendix A, we derived the formula for force $\textbf{\textit{F}}_\mathrm{st}$ in this section.
We substituted the ansatz $\Theta(r,\phi)$ [Eq. (\ref{eq9})] into the function $\varepsilon_\mathrm{st}$ [Eq. (\ref{eq11c})], obtaining the radial and tangential components of force density $\textbf{\textit{f}}_\mathrm{st}$
\begin{equation}\begin{split}\label{eqc1}
&\textbf{\textit{f}}_\mathrm{st}=-\nabla \varepsilon_\mathrm{st}=f_{\mathrm{st},r}\textbf{\textit{e}}_{r}
+f_{\mathrm{st},\phi}\textbf{\textit{e}}_{\phi},\\
&f_{\mathrm{st},r}=-\frac{\partial \varepsilon_{\mathrm{st}}}{\partial r}=-K_\mathrm{s}dwi_1\cos^{2}\beta,\\
&f_{\mathrm{st},\phi}=-\frac{1}{r}\frac{\partial \varepsilon_{\mathrm{st}}}{\partial \phi}=-K_\mathrm{s}dw\bigg(i_2\cos^{2}\beta\frac{\mathrm{d}\rho}{\mathrm{d}\phi}-i_3\sin\beta\cos\beta\bigg),\\
\end{split}\end{equation}
with the function $i_k (k = 1, 2, 3)$ as
\begin{equation}\begin{split}\label{eqc2}
&i_1=\frac{\sin 2\Theta}{w}\frac{\partial\Theta}{\partial r},\\
&i_2=\frac{\sin 2\Theta}{r}\frac{\partial\Theta}{\partial R},\\
&i_3=\frac{2\sin^{2}\Theta}{wr},\\
\end{split}\end{equation}
From Eq. (\ref{eqc1}) and the force formula Eq. (\ref{eq14}), one may have
\begin{equation}\begin{split}\label{eqc3}
\textbf{\textit{F}}_\mathrm{st}=&-K_{\mathrm{s}}wd
\int^{360^{\circ}}_{0^{\circ}}\bigg[I_1\cos^{2}\beta\textbf{\textit{e}}_{r}\\
&+\bigg(I_2\cos^{2}\beta\frac{\mathrm{d}\rho}{\mathrm{d}\phi}-I_3\sin\beta\cos\beta\bigg)\textbf{\textit{e}}_{\phi}\bigg]\mathrm{d}\phi,\\
\end{split}\end{equation}
and
\begin{equation}\begin{split}\label{eqc4}
&I_1=\int^{\infty}_{0}i_{1}r\mathrm{d}r=\int^{\infty}_{0}\sin2\Theta \frac{\partial\Theta}{\partial \delta}\delta\mathrm{d}\delta=I_1(\rho),\\
&I_2=\int^{\infty}_{0}i_{2}r\mathrm{d}r=\int^{\infty}_{0}\sin2\Theta \frac{\partial\Theta}{\partial \rho}\mathrm{d}\delta=I_2(\rho),\\
&I_3=\int^{\infty}_{0}i_{3}r\mathrm{d}r=\int^{\infty}_{0}2\sin^{2}\Theta\mathrm{d}\delta=I_3(\rho),\\
\end{split}\end{equation}
where $I_1$, $I_2$ and $I_3$ are some dimensionless parameters as a function of $\rho$.

To determine the functions $I_k(\rho)$, we substituted the expressions of $i_k$ [Eq. (\ref{eqc2})] and the relations Eq. (\ref{eqa5}) into the integrals of $I_k(\rho)$ [Eq. (\ref{eqc4})] for replacing the function $\Theta(\delta, \rho)$ and its derivations. Further numerical computations of these integrals yield a set of data of $I_k(\rho)$ at a broad range of $1.0\leq \rho \leq 20.0$. Then, these data were numerically fitted by an exponential decay function, as the results shown in Fig.~\ref{figC1}. The resulting fitting formula for $I_k(\rho)$ are
\begin{equation}\begin{split}\label{eqc5}
&I_1(\rho)=-2.00131+2.1175 e^{-\frac{\rho}{0.78978}},\\
&I_2(\rho)=  2.1098 e^{-\frac{\rho}{0.88304}},\\
&I_3(\rho)= 4.00222 - 4.24172 e^{-\frac{\rho}{0.78875}},\\
\end{split}\end{equation}
and it is noted in Eq. (\ref{eqc5}) that $I_3 \approx -2I_1$ for $1.0\leq \rho \leq 20.0$, thus Eq. (\ref{eqc3}) can be simplified as
\counterwithin{figure}{section}
\begin{figure}[t]
\centering
\includegraphics[width=0.33\textwidth]{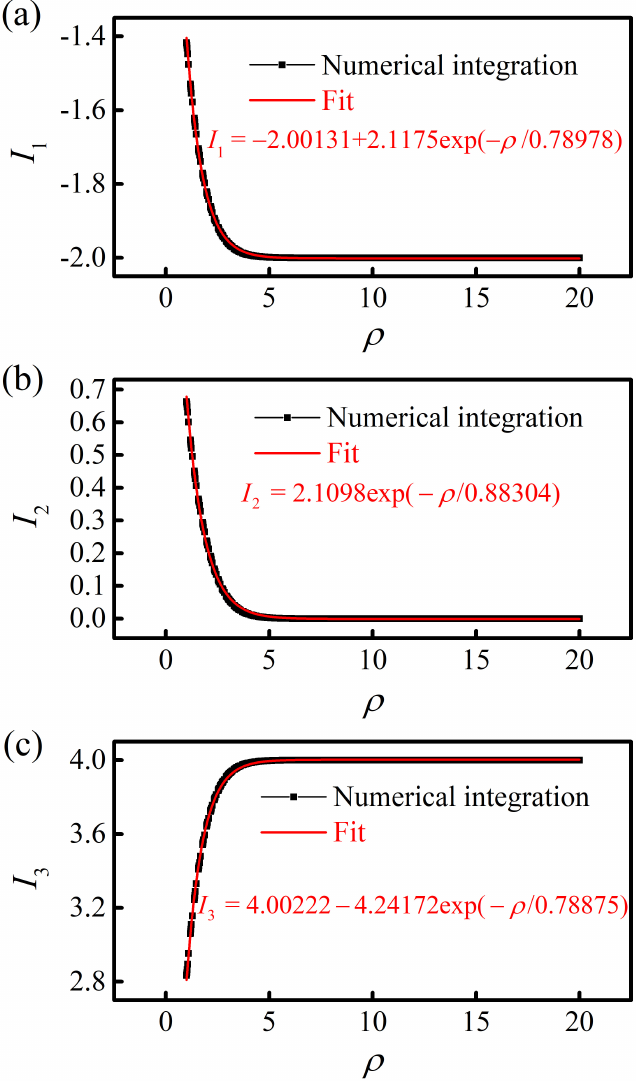}
\setcounter{figure}{0}
\renewcommand{\thefigure}{C\arabic{figure}}
\caption{(Color online) Numerical integration and curve fitting for the functions (a) $I_1(\rho)$, (b) $I_2(\rho)$ and (c) $I_3(\rho)$ at a broad range of $1.0\leq \rho \leq 20.0$.}
\label{figC1}
\end{figure}

\begin{equation}\begin{split}\label{eqc6}
\textbf{\textit{F}}_\mathrm{st}=&-K_{\mathrm{s}}wd
\int^{360^{\circ}}_{0^{\circ}}\bigg[I_1\cos^{2}\beta\textbf{\textit{e}}_{r}\\
&+\bigg(I_2\cos^{2}\beta\frac{\mathrm{d}\rho}{\mathrm{d}\phi}+2I_1\sin\beta\cos\beta\bigg)\textbf{\textit{e}}_{\phi}\bigg]\mathrm{d}\phi,\\
\end{split}\end{equation}

Eq. (\ref{eqc6}) may also be reformulated as
\begin{equation}\begin{split}\label{eqc7}
F_{\mathrm{st},x}=&-K_{\mathrm{s}}wd\bigg[
\int^{360^{\circ}}_{0^{\circ}}I_1(\cos^{2}\beta\cos\phi-2\sin\beta\cos\beta\sin\phi)\mathrm{d}\phi\\
&-\int^{360^{\circ}}_{0^{\circ}}I_2\cos^{2}\beta\frac{\mathrm{d}\rho}{\mathrm{d}\phi}\sin\phi\mathrm{d}\phi\bigg],\\
F_{\mathrm{st},y}=&-K_{\mathrm{s}}wd\bigg[
\int^{360^{\circ}}_{0^{\circ}}I_1(\cos^{2}\beta\sin\phi+2\sin\beta\cos\beta\cos\phi)\mathrm{d}\phi\\
&+\int^{360^{\circ}}_{0^{\circ}}I_2\cos^{2}\beta\frac{\mathrm{d}\rho}{\mathrm{d}\phi}\cos\phi\mathrm{d}\phi\bigg].\\
\end{split}\end{equation}

Noting in Eq. (\ref{eqc7}) that
\begin{equation}\begin{split}\label{eqc8}
&\int^{360^{\circ}}_{0^{\circ}}(\cos^{2}\beta\cos\phi-2\sin\beta\cos\beta\sin\phi)\mathrm{d}\phi=0,\\
&\int^{360^{\circ}}_{0^{\circ}}(\cos^{2}\beta\sin\phi+2\sin\beta\cos\beta\cos\phi)\mathrm{d}\phi=0,\\
\end{split}\end{equation}
and $I_1$ and $I_2$ are nearly constant (i.e., $I_1 \approx -2.0$, $I_2 \approx 0.0$) in the case of $\rho \geq 5.0$ from Eq. (\ref{eqc5}), we have
\begin{equation}\begin{split}\label{eqc9}
F_{\mathrm{st},x}=F_{\mathrm{st},y}=0,\\
\end{split}\end{equation}
which means $\textbf{\textit{F}}_\mathrm{st}=0$ in the case of $\rho \geq 5.0$.

\section{Derivation of the formula for force $\textbf{\textit{F}}_\mathrm{dem}$}
The demagnetization energy in the magnetic nanostructure stems from the interaction of positive and negative magnetic charges in the sample, and it can be calculated using magnetostatic approach \cite{hubert1998magnetic,jackson2012classical,wang2018theory}. The expression for demagnetization energy in cylindrical coordinates is given as
\begin{equation}\begin{split}\label{eqd1}
&E_{\mathrm{dem}}=\frac{\mu_{0}}{2} \iiint \limits_{V_{\mathrm{sample}}} \Phi_{\mathrm{m}}\left(\textbf{\textit{r}},z\right)\rho_{\mathrm{m}}\left(\textbf{\textit{r}},z\right)\mathrm{d}V\\
=&\frac{\mu_{0}}{8\pi} \iint \limits_{S_{\mathrm{sample}}} \mathrm{d}S \,£º\iint \limits_{S_{\mathrm{sample}}^{\prime}}\mathrm{d}S^{\prime}\int^{d}_{0}\mathrm{d}z \int^{d}_{0}\mathrm{d}z^{\prime}
\frac{\rho_{\mathrm{m}}\left(\textbf{\textit{r}},z\right)\rho_{\mathrm{m}}\left(\textbf{\textit{r}}^{\prime},z^{\prime}\right)}{\sqrt{\left| \textbf{\textit{r}}-\textbf{\textit{r}}^{\prime} \right|^{2}+\left(z-z^{\prime}\right)^{2}}},\\
\end{split}\end{equation}
where $\textbf{\textit{r}}=\textit{r}\textbf{\textit{e}}_{\textit{r}}=\textit{x}\textbf{\textit{e}}_{\textit{x}}+\textit{y}\textbf{\textit{e}}_{\textit{y}}$ denotes the position vector of the magnetic charges on the $xy$-plane. The magnetic scalar potential $\Phi_{\mathrm{m}}\left(\textbf{\textit{r}},z\right)$ and magnetic volume charge density $\rho_{\mathrm{m}}\left(\textbf{\textit{r}},z\right)$ are

\begin{equation}\begin{split}\label{eqd2}
\Phi_{\mathrm{m}}\left(\textbf{\textit{r}},z\right) =\frac{1}{4\pi} \iiint \limits_{V^{\prime}_{\mathrm{sample}}}\frac{\rho_{\mathrm{m}}\left(\textbf{\textit{r}}^{\prime},z^{\prime}\right)}{\sqrt{\left| \textbf{\textit{r}}-\textbf{\textit{r}}^{\prime} \right|^{2}+\left(z-z^{\prime}\right)^{2}}}\mathrm{d}V^{\prime},\\
\end{split}\end{equation}
\begin{equation}\begin{split}\label{eqd3}
\rho_{\mathrm{m}}\left(\textbf{\textit{r}},z\right) =-\nabla\cdot\textbf{M}\left(\textbf{\textit{r}},z\right)=-\mathrm{M}_{\mathrm{s}}\nabla\cdot\textbf{\textit{m}}\left(\textbf{\textit{r}},z\right).
\end{split}\end{equation}

For a thin ferromagnetic nanostructure with thickness $d$, with the assumption that the magnetization is uniform in $z$ direction, then $\rho_{\mathrm{m}}\left(\textbf{\textit{r}},z\right)=\rho_{\mathrm{m}}\left(\textbf{\textit{r}},z^{\prime}\right)$  and $z$ part in Eq. (\ref{eqd1}) can be integrated out,

\begin{equation}\begin{split}\label{eqd4}
&\int^{d}_{0}\mathrm{d}z \int^{d}_{0}\mathrm{d}z^{\prime}
\frac{\rho_{\mathrm{m}}\left(\textbf{\textit{r}},z\right)\rho_{\mathrm{m}}\left(\textbf{\textit{r}}^{\prime},z^{\prime}\right)}{\sqrt{\left| \textbf{\textit{r}}-\textbf{\textit{r}}^{\prime} \right|^{2}+\left(z-z^{\prime}\right)^{2}}}\\
=&d^{2}\frac{\rho_{\mathrm{m}}\left(\textbf{\textit{r}}\right)\rho_{\mathrm{m}}\left(\textbf{\textit{r}}^{\prime}\right)}{\left| \textbf{\textit{r}}-\textbf{\textit{r}}^{\prime} \right|}\\
=&\frac{\sigma_{\mathrm{m}}\left(\textbf{\textit{r}}\right)\sigma_{\mathrm{m}}\left(\textbf{\textit{r}}^{\prime}\right)}{\left| \textbf{\textit{r}}-\textbf{\textit{r}}^{\prime} \right|},\\
\end{split}\end{equation}
here the magnetic charges can be considered to distribute on the $xy$-plane with the magnetic surface charge density defined as
\begin{equation}\begin{split}\label{eqd5}
\sigma_{\mathrm{m}}\left(\textbf{\textit{r}}\right)=d\rho_{\mathrm{m}}\left(\textbf{\textit{r}}\right)=-\mathrm{M}_{\mathrm{s}}d\nabla\cdot\textbf{\textit{m}}\left(\textbf{\textit{r}}\right).
\end{split}\end{equation}

With this definition, the integration in Eq. (\ref{eqd1}) can be simplified to
\begin{equation}\begin{split}\label{eqd6}
E_{\mathrm{dem}} =\frac{\mu_{0}}{8\pi} \iint \limits_{S_{\mathrm{sample}}}  \,£º\iint \limits_{S_{\mathrm{sample}}^{\prime}}
\frac{\sigma_{\mathrm{m}}\left(\textbf{\textit{r}}\right)\sigma_{\mathrm{m}}\left(\textbf{\textit{r}}^{\prime}\right)}{\left| \textbf{\textit{r}}-\textbf{\textit{r}}^{\prime} \right|}\mathrm{d}S^{\prime}\mathrm{d}S,\\
\end{split}\end{equation}
or
\begin{equation}\begin{split}\label{eqd7}
&E_{\mathrm{dem}}=\frac{\mu_{0}}{2} \iint \limits_{S_{\mathrm{sample}}} \Phi_{\mathrm{m}}\left(\textbf{\textit{r}}\right)\sigma_{\mathrm{m}}\left(\textbf{\textit{r}}\right)\mathrm{d}S,\\
&\Phi_{\mathrm{m}}\left(\textbf{\textit{r}}\right)=\frac{1}{4\pi} \iint \limits_{S_{\mathrm{sample}}^{\prime}}
\frac{\sigma_{\mathrm{m}}\left(\textbf{\textit{r}}^{\prime}\right)}{\left| \textbf{\textit{r}}-\textbf{\textit{r}}^{\prime} \right|}\mathrm{d}S^{\prime}.\\
\end{split}\end{equation}

It was observed in the simulation results that the non-collinear arrangement of magnetic moments only exists in the regions of skyrmion configuration $S_{\mathrm{sky}}$ and sample edge $S_{\mathrm{edge}}^{\prime}$. Thus the magnetic charges appear in these regions, and the magnetic scalar potential has two contributions coming from the magnetic charges in the regions of skyrmion and sample edge,
\begin{equation}\begin{split}\label{eqd8}
\Phi_{\mathrm{m}}\left(\textbf{\textit{r}}\right)=
\Phi_{\mathrm{m,sky}}\left(\textbf{\textit{r}}\right)+\Phi_{\mathrm{m,edge}}\left(\textbf{\textit{r}}\right).\\
\end{split}\end{equation}

The integration in Eq. (\ref{eqd7}) can be further reformulated according to Eq. (\ref{eqd8}),
\begin{equation}\begin{split}\label{eqd9}
E_{\mathrm{dem}}=&\frac{\mu_{0}}{2} \iint \limits_{S_{\mathrm{sample}}} \left[\Phi_{\mathrm{m,sky}}\left(\textbf{\textit{r}}\right)+\Phi_{\mathrm{m,edge}}\left(\textbf{\textit{r}}\right)\right]
[\sigma_{\mathrm{m,sky}}\left(\textbf{\textit{r}}\right)\\
&\qquad\qquad+\sigma_{\mathrm{m,edge}}\left(\textbf{\textit{r}}\right)]\mathrm{d}S\\
=&E_{\mathrm{dem,sky}}+E_{\mathrm{dem,edge}}+E_{\mathrm{dem,sky\mbox{-}edge}},\\
\end{split}\end{equation}
\begin{equation}\begin{split}\label{eqd10}
E_{\mathrm{dem,sky}}&=\frac{\mu_{0}}{2} \iint \limits_{S_{\mathrm{sky}}} \Phi_{\mathrm{m,sky}}\left(\textbf{\textit{r}}\right)
\sigma_{\mathrm{m,sky}}\left(\textbf{\textit{r}}\right)\mathrm{d}S,\\
&=\frac{\mu_{0}}{8\pi} \iint \limits_{S_{\mathrm{sky}}} \iint \limits_{S_{\mathrm{sky}}}
\frac{\sigma_{\mathrm{m,sky}}\left(\textbf{\textit{r}}_{i}\right)\sigma_{\mathrm{m,sky}}\left(\textbf{\textit{r}}_{j}\right)}{\left| \textbf{\textit{r}}_{i}-\textbf{\textit{r}}_{j} \right|}\mathrm{d}S_{i}\mathrm{d}S_{j},\\
\end{split}\end{equation}

\begin{equation}\begin{split}\label{eqd11}
E_{\mathrm{dem,edge}}&=\frac{\mu_{0}}{2} \iint \limits_{S_{\mathrm{edge}}^{\prime}} \Phi_{\mathrm{m,edge}}\left(\textbf{\textit{r}}^{\prime}\right)
\sigma_{\mathrm{m,edge}}\left(\textbf{\textit{r}}^{\prime}\right)\mathrm{d}S^{\prime},\\
\end{split}\end{equation}

\begin{equation}\begin{split}\label{eqd12}
E_{\mathrm{dem,sky\mbox{-}edge}}&=\mu_{0} \iint \limits_{S_{\mathrm{sky}}} \Phi_{\mathrm{m,edge}}\left(\textbf{\textit{r}}\right)
\sigma_{\mathrm{m,sky}}\left(\textbf{\textit{r}}\right)\mathrm{d}S,\\
&=\mu_{0} \iint \limits_{S_{\mathrm{edge}}^{\prime}} \Phi_{\mathrm{m,sky}}\left(\textbf{\textit{r}}^{\prime}\right)
\sigma_{\mathrm{m,edge}}\left(\textbf{\textit{r}}^{\prime}\right)\mathrm{d}S^{\prime}\\
&=\frac{\mu_{0}}{4\pi} \iint \limits_{S_{\mathrm{sky}}} \iint \limits_{S_{\mathrm{edge}}^{\prime}}
\frac{\sigma_{\mathrm{m,sky}}\left(\textbf{\textit{r}}\right)\sigma_{\mathrm{m,edge}}\left(\textbf{\textit{r}}^{\prime}\right)}{\left| \textbf{\textit{r}}-\textbf{\textit{r}}^{\prime} \right|}\mathrm{d}S^{\prime}\mathrm{d}S,\\
\end{split}\end{equation}
where $E_{\mathrm{dem,sky}}$ and $E_{\mathrm{dem,edge}}$ represent the magnetostatic self-energy of skyrmion configuration and sample edge, respectively. $E_{\mathrm{dem,sky\mbox{-}edge}}$ denotes the magnetostatic interaction energy between the magnetic charges in the regions of skyrmion and sample edge.\par
The force $\textbf{\textit{F}}_\mathrm{dem}$ acting on skyrmion includes two parts,
\begin{equation}\begin{split}\label{eqd13}
\textbf{\textit{F}}_{\mathrm{dem}}=\textbf{\textit{F}}_{\mathrm{dem,sky}}+\textbf{\textit{F}}_{\mathrm{dem,edge}},\\
\end{split}\end{equation}
where $\textbf{\textit{F}}_{\mathrm{dem,sky}}$ is the self-force in the skyrmion and $\textbf{\textit{F}}_{\mathrm{dem,edge}}$ is the force generated by the sample edge, which are related to the energy terms $E_{\mathrm{dem,sky}}$ and $E_{\mathrm{dem,sky\mbox{-}edge}}$ in Eqs. (\ref{eqd10}) and (\ref{eqd12}), respectively.
The self-force $\textbf{\textit{F}}_{\mathrm{dem,sky}}$ is
\begin{equation}\begin{split}\label{eqd14}
&\textbf{\textit{F}}_{\mathrm{dem,sky}}\\=&-\left(\nabla E_{\mathrm{dem,sky}}\right)_{\mathrm{invariant}\, \sigma_{\mathrm{m,sky}}}\\
=&-\left(\frac{\partial E_{\mathrm{dem,sky}}}{\partial \textbf{\textit{r}}_{i}}+\frac{\partial E_{\mathrm{dem,sky}}}{\partial \textbf{\textit{r}}_{j}}\right)_{\mathrm{invariant}\,\sigma_{\mathrm{m,sky}}}\\
=&\frac{\mu_{0}}{8\pi} \iint \limits_{S_{\mathrm{sky}}} \iint \limits_{S_{\mathrm{sky}}}
\frac{\sigma_{\mathrm{m,sky}}\left(\textbf{\textit{r}}_{i}\right)\sigma_{\mathrm{m,sky}}\left(\textbf{\textit{r}}_{j}\right)}{\left| \textbf{\textit{r}}_{i}-\textbf{\textit{r}}_{j} \right|^{3}}[\left(\textbf{\textit{r}}_{i}-\textbf{\textit{r}}_{j}\right)\\
&\qquad\qquad\;\;\;\;+\left(\textbf{\textit{r}}_{j}-\textbf{\textit{r}}_{i}\right)]\mathrm{d}S_{i}\mathrm{d}S_{j}\\
=&0.
\end{split}\end{equation}
This means that the magnetostatic forces between two arbitrary elementary magnetic charges $\sigma_{\mathrm{m,sky}}\left(\textbf{\textit{r}}_{i}\right)\mathrm{d}S_{i}$ and $\sigma_{\mathrm{m,sky}}\left(\textbf{\textit{r}}_{j}\right)\mathrm{d}S_{j}$ counteract each other, and thus the resultant force $\textbf{\textit{F}}_{\mathrm{dem,sky}}$ of all magnetic charges in the skyrmion equals zero.\par
Therefore, the force $\textbf{\textit{F}}_{\mathrm{dem}}$ acting on the skyrmion is only contributed by the magnetic charges at the sample edge,
\begin{equation}\begin{split}\label{eqd15}
&\textbf{\textit{F}}_{\mathrm{dem}}\\
=&\textbf{\textit{F}}_{\mathrm{dem,edge}}\\
=&-\left(\nabla E_{\mathrm{dem,sky\mbox{-}edge}}\right)_{\mathrm{invariant}\, \sigma_{\mathrm{m,sky}},\,\sigma_{\mathrm{m,edge}}}\\
=&\frac{\mu_{0}}{4\pi} \iint \limits_{S_{\mathrm{sky}}} \iint \limits_{S_{\mathrm{edge}}^{\prime}}
\frac{\sigma_{\mathrm{m,sky}}\left(\textbf{\textit{r}}\right)\sigma_{\mathrm{m,edge}}\left(\textbf{\textit{r}}^{\prime}\right)}{\left| \textbf{\textit{r}}-\textbf{\textit{r}}^{\prime} \right|^{3}}\left(\textbf{\textit{r}}-\textbf{\textit{r}}^{\prime}\right)\mathrm{d}S^{\prime}\mathrm{d}S.\\
\end{split}\end{equation}

It is clearly evident in Eq. (\ref{eqd15}) that $\textbf{\textit{F}}_{\mathrm{dem}}$ is dependent on the magnetic charges distributions in the regions $S_{\mathrm{sky}}$ and $S_{\mathrm{edge}}^{\prime}$, as well as the geometric shape and size of the sample. To further clarify this dependence, we rewrite the expression of $E_{\mathrm{dem,sky\mbox{-}edge}}$ using the multipole expansion technique \cite{jackson2012classical,raab2004multipole},
\begin{equation}\begin{split}\label{eqd16}
E_{\mathrm{dem,sky\mbox{-}edge}}&=\mu_{0} \iint \limits_{S_{\mathrm{edge}}^{\prime}} \Phi_{\mathrm{m,sky}}\left(\textbf{\textit{r}}^{\prime}\right)
\sigma_{\mathrm{m,edge}}\left(\textbf{\textit{r}}^{\prime}\right)\mathrm{d}S^{\prime},\\
\Phi_{\mathrm{m,sky}}\left(\textbf{\textit{r}}^{\prime}\right)&=\frac{1}{4\pi} \iint \limits_{S_{\mathrm{sky}}}
\frac{\sigma_{\mathrm{m,sky}}\left(\textbf{\textit{r}}\right)}{\left| \textbf{\textit{r}}-\textbf{\textit{r}}^{\prime} \right|}\mathrm{d}S\\
&=\frac{1}{4\pi} \iint \limits_{S_{\mathrm{sky}}}
\sum_{l=0}^{\infty}\frac{{\textit{r}^{l}}P_{l}(\cos \alpha)}{\textit{r}^{\prime\,l+1}}\sigma_{\mathrm{m,sky}}\left(\textbf{\textit{r}}\right)\mathrm{d}S,
\end{split}\end{equation}
where the Legendre polynomials $P_{l}(\cos \alpha)$ are introduced, with angle $\alpha=\phi-\phi^{\prime} $ representing the included angle between the position vectors $\textbf{\textit{r}}$ and $\textbf{\textit{r}}^{\prime}$. $r$ and $r^{\prime}$ are the radial distance of magnetic charges in the regions of skyrmion and sample edge, respectively.\par
Thus the multipole expansion of the energy expression of $E_{\mathrm{dem,sky\mbox{-}edge}}$ is obtained,
\begin{equation}\begin{split}\label{eqd17}
&E_{\mathrm{dem,sky\mbox{-}edge}}\\
=&\frac{\mu_{0}}{4\pi} \iint \limits_{S_{\mathrm{edge}}^{\prime}}
\bigg(\frac{Q_{\mathrm{m}}}{r^{\prime}}+\frac{\textbf{\textit{P}}_{\mathrm{m}}\cdot\textbf{\textit{r}}^{\prime}}{r^{\prime 3}}+\cdots\bigg)
\sigma_{\mathrm{m,edge}}\left(\textbf{\textit{r}}^{\prime}\right)\mathrm{d}S^{\prime},\\
\end{split}\end{equation}
where $Q_{\mathrm{m}}=\iint \limits_{S_{\mathrm{sky}}} \sigma_{\mathrm{m,sky}}(\textbf{\textit{r}})\mathrm{d}S$ is the total magnetic charge (i.e., magnetic monopole moment) of the skyrmion configuration, and $\textbf{\textit{P}}_{\mathrm{m}}=\iint \limits_{S_{\mathrm{sky}}} \sigma_{\mathrm{m,sky}}(\textbf{\textit{r}})\textbf{\textit{r}}\mathrm{d}S$ is the magnetic dipole moment in the skyrmion configuration.\par
For convenience of calculation, we computed the total magnetic charge $Q_{\mathrm{m}}$ by considering the magnetic moments $\textbf{M}(\textbf{\textit{r}},z)$  of skyrmion configuration distributed in a three-dimensional cylindrical-like space,
\begin{equation}\begin{split}\label{eqd18}
Q_{\mathrm{m}}
=&\iiint \limits_{V_{\mathrm{sky}}} \rho_{\mathrm{m,sky}}(\textbf{\textit{r}})\mathrm{d}V\\
=&-\iiint \limits_{V_{\mathrm{sky}}} \nabla\cdot\textbf{M}(\textbf{\textit{r}},z)\mathrm{d}V\\
=&-\mathop{{\int\!\!\!\!\!\int}\mkern-21mu \bigcirc}\limits_{S_{\mathrm{sky}}}\textbf{M}(\textbf{\textit{r}},z)\cdot\mathrm{d}\textbf{\textit{S}}\\
=&-\iint \limits_{S_{\mathrm{sky}}} [\textbf{M}(\textbf{\textit{r}},z_{\mathrm{top}})\cdot\mathrm{d}\textbf{\textit{S}}_{\mathrm{top}}
+\textbf{M}(\textbf{\textit{r}},z_{\mathrm{bot}})\cdot\mathrm{d}\textbf{\textit{S}}_{\mathrm{bot}}\\
&\qquad\quad+\textbf{M}(\textbf{\textit{r}},z_{\mathrm{lat}})\cdot\mathrm{d}\textbf{\textit{S}}_{\mathrm{lat}}] \\
=&-\iint \limits_{S_{\mathrm{sky}}} [\mathrm{M}_{z}(\textbf{\textit{r}},z_{\mathrm{top}})\mathrm{d}\textit{S}_{\mathrm{top}}
-\mathrm{M}_{z}(\textbf{\textit{r}},z_{\mathrm{bot}})\mathrm{d}\textit{S}_{\mathrm{bot}}\\
&\qquad\quad+\textbf{M}(\textbf{\textit{r}},z_{\mathrm{lat}})\cdot\mathrm{d}\textbf{\textit{S}}_{\mathrm{lat}}], \\
\end{split}\end{equation}
where the divergence theorem is used for transforming volumetric integrals of $\nabla\cdot\textbf{M}(\textbf{\textit{r}},z)$ into surface integrals of $\textbf{M}(\textbf{\textit{r}},z)$. The envelope surface for the cylindrical-like skyrmion configuration is composed of the near-surface of top magnetic layer $\textbf{\textit{S}}_{\mathrm{top}}$, the near-surface of bottom magnetic layer $\textbf{\textit{S}}_{\mathrm{bot}}$, and the lateral surface on the periphery of skyrmion configuration $\textbf{\textit{S}}_{\mathrm{lat}}$.\\

For the thin ferromagnetic layers,
\begin{equation}\begin{split}\label{eqd19}
&\mathrm{M}_{z}(\textbf{\textit{r}},z_{\mathrm{top}})=\mathrm{M}_{z}(\textbf{\textit{r}},z_{\mathrm{bot}}),
\mathrm{d}\textit{S}_{\mathrm{top}}=\mathrm{d}\textit{S}_{\mathrm{bot}},\\
&\textbf{M}(\textbf{\textit{r}},z_{\mathrm{lat}})=\mathrm{M}_{z}(\textbf{\textit{r}},z_{\mathrm{lat}})\hat{z},
\mathrm{d}\textbf{\textit{S}}_{\mathrm{lat}}\bot\hat{z},\textbf{M}(\textbf{\textit{r}},z_{\mathrm{lat}})\cdot\mathrm{d}\textbf{\textit{S}}_{\mathrm{lat}}=0,\\
\end{split}\end{equation}
which yields
\begin{equation}\begin{split}\label{eqd20}
Q_{\mathrm{m}}=0.
\end{split}\end{equation}

This means that the total magnetic charge $Q_{\mathrm{m}}$ keeps zero even the skyrmion configuration deforms. Therefore, the first expansion energy term in Eq. (\ref{eqd17}) has none contribution to the force $\textbf{\textit{F}}_{\mathrm{dem}}$.
We further considered the second expansion energy term (i.e., the magnetic dipole moment energy term) in Eq. (\ref{eqd17}), which can be expressed as
\begin{equation}\begin{split}\label{eqd21}
&E_{\mathrm{dem,dip}}=-\frac{\mu_{0}}{4\pi} \textbf{\textit{P}}_{\mathrm{m}}\cdot\textbf{\textit{H}}_{\mathrm{m,edge}},\\
&\textbf{\textit{H}}_{\mathrm{m,edge}}=-\iint \limits_{S_{\mathrm{edge}}^{\prime}}
\frac{\textbf{\textit{r}}^{\prime}}{r^{\prime 3}}
\sigma_{\mathrm{m,edge}}\left(\textbf{\textit{r}}^{\prime}\right)\mathrm{d}S^{\prime},\\
\end{split}\end{equation}
where $\textbf{\textit{H}}_{\mathrm{m,edge}}$ represents the effective magnetic field acting on the magnetic dipole moment $\textbf{\textit{P}}_{\mathrm{m}}$. This field is generated by the magnetic charges at the sample edge, and it is strongly dependent on the distributions of magnetic charges at the sample edge $\sigma_{\mathrm{m,edge}}\left(\textbf{\textit{r}}^{\prime}\right)$, and also the geometric shape and size of the sample (noting that the field strength is proportional to $1/r^{\prime 2}$, and here $r^{\prime}$ is the radial distance of magnetic charges at the sample edge).\par
The corresponding force for the energy $E_{\mathrm{dem,dip}}$ is
\begin{equation}\begin{split}\label{eqd22}
\textbf{\textit{F}}_{\mathrm{dem,dip}}=&-\nabla E_{\mathrm{dem,dip}}\\
=&\frac{\mu_{0}}{4\pi} \nabla(\textbf{\textit{P}}_{\mathrm{m}}\cdot\textbf{\textit{H}}_{\mathrm{m,edge}})\\
=&\frac{\mu_{0}}{4\pi}(\textbf{\textit{P}}_{\mathrm{m}}\cdot\nabla)\textbf{\textit{H}}_{\mathrm{m,edge}}.\\
\end{split}\end{equation}

Further numerical calculations of $\textbf{\textit{P}}_{\mathrm{m}}$ based on the simulated data showed that $\textbf{\textit{P}}_{\mathrm{m}}$ appears in the motion of the deformed skyrmion. In fact, the appearance of $\textbf{\textit{P}}_{\mathrm{m}}$ is related to the inhomogeneous distributions of magnetic charges in the skyrmion configuration, and it is sourced mainly from the asymmetric deformation of skyrmion configuration. Therefore, one may consider $\textbf{\textit{F}}_{\mathrm{dem,dip}}$ to be a consequence of the combined effect of the asymmetric deformation of skyrmion, the geometric shape and size of sample, together with the magnetic charges distributions on the sample edge, as demonstrated in Eqs. (\ref{eqd21}) and (\ref{eqd22}). In addition to $\textbf{\textit{F}}_{\mathrm{dem,dip}}$, the other magnetic multipole moments such as the quadrupole moment in skyrmion may also have contributions to $\textbf{\textit{F}}_{\mathrm{dem}}$, because these multipole moments may also appear in such an asymmetrically deformed skyrmion, as suggested by the energy expression $E_{\mathrm{dem,sky\mbox{-}edge}}$ in Eq. (\ref{eqd17}).\par

\bibliography{reference}

\begin{thebibliography}{65}
\expandafter\ifx\csname natexlab\endcsname\relax\def\natexlab#1{#1}\fi
\expandafter\ifx\csname bibnamefont\endcsname\relax
  \def\bibnamefont#1{#1}\fi
\expandafter\ifx\csname bibfnamefont\endcsname\relax
  \def\bibfnamefont#1{#1}\fi
\expandafter\ifx\csname citenamefont\endcsname\relax
  \def\citenamefont#1{#1}\fi
\expandafter\ifx\csname url\endcsname\relax
  \def\url#1{\texttt{#1}}\fi
\expandafter\ifx\csname urlprefix\endcsname\relax\def\urlprefix{URL }\fi
\providecommand{\bibinfo}[2]{#2}
\providecommand{\eprint}[2][]{\url{#2}}

\bibitem[{\citenamefont{Nagaosa and Tokura}(2013)}]{nagaosa2013topological}
\bibinfo{author}{\bibfnamefont{N.}~\bibnamefont{Nagaosa}} \bibnamefont{and}
  \bibinfo{author}{\bibfnamefont{Y.}~\bibnamefont{Tokura}},
  \bibinfo{journal}{Nat. Nanotech.} \textbf{\bibinfo{volume}{8}},
  \bibinfo{pages}{899} (\bibinfo{year}{2013}).

\bibitem[{\citenamefont{Fert et~al.}(2017)\citenamefont{Fert, Reyren, and
  Cros}}]{fert2017magnetic}
\bibinfo{author}{\bibfnamefont{A.}~\bibnamefont{Fert}},
  \bibinfo{author}{\bibfnamefont{N.}~\bibnamefont{Reyren}}, \bibnamefont{and}
  \bibinfo{author}{\bibfnamefont{V.}~\bibnamefont{Cros}},
  \bibinfo{journal}{Nat. Rev. Mater.} \textbf{\bibinfo{volume}{2}},
  \bibinfo{pages}{1} (\bibinfo{year}{2017}).

\bibitem[{\citenamefont{Wiesendanger}(2016)}]{wiesendanger2016nanoscale}
\bibinfo{author}{\bibfnamefont{R.}~\bibnamefont{Wiesendanger}},
  \bibinfo{journal}{Nat. Rev. Mater.} \textbf{\bibinfo{volume}{1}},
  \bibinfo{pages}{1} (\bibinfo{year}{2016}).

\bibitem[{\citenamefont{Jiang et~al.}(2015{\natexlab{a}})\citenamefont{Jiang,
  Upadhyaya, Zhang, Yu, Jungfleisch, Fradin, Pearson, Tserkovnyak, Wang, and
  Heinonen}}]{jiang2015sge}
\bibinfo{author}{\bibfnamefont{W.}~\bibnamefont{Jiang}},
  \bibinfo{author}{\bibfnamefont{P.}~\bibnamefont{Upadhyaya}},
  \bibinfo{author}{\bibfnamefont{W.}~\bibnamefont{Zhang}},
  \bibinfo{author}{\bibfnamefont{G.}~\bibnamefont{Yu}},
  \bibinfo{author}{\bibfnamefont{M.}~\bibnamefont{Jungfleisch}},
  \bibinfo{author}{\bibfnamefont{F.}~\bibnamefont{Fradin}},
  \bibinfo{author}{\bibfnamefont{J.}~\bibnamefont{Pearson}},
  \bibinfo{author}{\bibfnamefont{Y.}~\bibnamefont{Tserkovnyak}},
  \bibinfo{author}{\bibfnamefont{K.}~\bibnamefont{Wang}}, \bibnamefont{and}
  \bibinfo{author}{\bibfnamefont{O.}~\bibnamefont{Heinonen}},
  \bibinfo{journal}{Science} \textbf{\bibinfo{volume}{349}},
  \bibinfo{pages}{283} (\bibinfo{year}{2015}{\natexlab{a}}).

\bibitem[{\citenamefont{Kang et~al.}(2016)\citenamefont{Kang, Huang, Zhang,
  Zhou, and Zhao}}]{kang2016skyrmion}
\bibinfo{author}{\bibfnamefont{W.}~\bibnamefont{Kang}},
  \bibinfo{author}{\bibfnamefont{Y.}~\bibnamefont{Huang}},
  \bibinfo{author}{\bibfnamefont{X.}~\bibnamefont{Zhang}},
  \bibinfo{author}{\bibfnamefont{Y.}~\bibnamefont{Zhou}}, \bibnamefont{and}
  \bibinfo{author}{\bibfnamefont{W.}~\bibnamefont{Zhao}},
  \bibinfo{journal}{Proc. IEEE} \textbf{\bibinfo{volume}{104}},
  \bibinfo{pages}{2040} (\bibinfo{year}{2016}).

\bibitem[{\citenamefont{Zhang et~al.}(2020{\natexlab{a}})\citenamefont{Zhang,
  Zhou, Song, Park, Xia, Ezawa, Liu, Zhao, Zhao, and Woo}}]{zhang2020skyrmion}
\bibinfo{author}{\bibfnamefont{X.}~\bibnamefont{Zhang}},
  \bibinfo{author}{\bibfnamefont{Y.}~\bibnamefont{Zhou}},
  \bibinfo{author}{\bibfnamefont{K.~M.} \bibnamefont{Song}},
  \bibinfo{author}{\bibfnamefont{T.-E.} \bibnamefont{Park}},
  \bibinfo{author}{\bibfnamefont{J.}~\bibnamefont{Xia}},
  \bibinfo{author}{\bibfnamefont{M.}~\bibnamefont{Ezawa}},
  \bibinfo{author}{\bibfnamefont{X.}~\bibnamefont{Liu}},
  \bibinfo{author}{\bibfnamefont{W.}~\bibnamefont{Zhao}},
  \bibinfo{author}{\bibfnamefont{G.}~\bibnamefont{Zhao}}, \bibnamefont{and}
  \bibinfo{author}{\bibfnamefont{S.}~\bibnamefont{Woo}}, \bibinfo{journal}{J.
  Phys. Condens. Matter.} \textbf{\bibinfo{volume}{32}},
  \bibinfo{pages}{143001} (\bibinfo{year}{2020}{\natexlab{a}}).

\bibitem[{\citenamefont{Romming et~al.}(2013)\citenamefont{Romming, Hanneken,
  Menzel, Bickel, Wolter, von Bergmann, Kubetzka, and
  Wiesendanger}}]{romming2013writing}
\bibinfo{author}{\bibfnamefont{N.}~\bibnamefont{Romming}},
  \bibinfo{author}{\bibfnamefont{C.}~\bibnamefont{Hanneken}},
  \bibinfo{author}{\bibfnamefont{M.}~\bibnamefont{Menzel}},
  \bibinfo{author}{\bibfnamefont{J.~E.} \bibnamefont{Bickel}},
  \bibinfo{author}{\bibfnamefont{B.}~\bibnamefont{Wolter}},
  \bibinfo{author}{\bibfnamefont{K.}~\bibnamefont{von Bergmann}},
  \bibinfo{author}{\bibfnamefont{A.}~\bibnamefont{Kubetzka}}, \bibnamefont{and}
  \bibinfo{author}{\bibfnamefont{R.}~\bibnamefont{Wiesendanger}},
  \bibinfo{journal}{Science} \textbf{\bibinfo{volume}{341}},
  \bibinfo{pages}{636} (\bibinfo{year}{2013}).

\bibitem[{\citenamefont{Sampaio et~al.}(2013)\citenamefont{Sampaio, Cros,
  Rohart, Thiaville, and Fert}}]{sampaio2013nucleation}
\bibinfo{author}{\bibfnamefont{J.}~\bibnamefont{Sampaio}},
  \bibinfo{author}{\bibfnamefont{V.}~\bibnamefont{Cros}},
  \bibinfo{author}{\bibfnamefont{S.}~\bibnamefont{Rohart}},
  \bibinfo{author}{\bibfnamefont{A.}~\bibnamefont{Thiaville}},
  \bibnamefont{and} \bibinfo{author}{\bibfnamefont{A.}~\bibnamefont{Fert}},
  \bibinfo{journal}{Nat. Nanotech.} \textbf{\bibinfo{volume}{8}},
  \bibinfo{pages}{839} (\bibinfo{year}{2013}).

\bibitem[{\citenamefont{Fert and Cros}(2013)}]{fert2013v}
\bibinfo{author}{\bibfnamefont{A.}~\bibnamefont{Fert}} \bibnamefont{and}
  \bibinfo{author}{\bibfnamefont{V.}~\bibnamefont{Cros}},
  \bibinfo{journal}{Nanotechnology} \textbf{\bibinfo{volume}{8}},
  \bibinfo{pages}{152} (\bibinfo{year}{2013}).

\bibitem[{\citenamefont{Iwasaki et~al.}(2013)\citenamefont{Iwasaki, Mochizuki,
  and Nagaosa}}]{iwasaki2013current}
\bibinfo{author}{\bibfnamefont{J.}~\bibnamefont{Iwasaki}},
  \bibinfo{author}{\bibfnamefont{M.}~\bibnamefont{Mochizuki}},
  \bibnamefont{and} \bibinfo{author}{\bibfnamefont{N.}~\bibnamefont{Nagaosa}},
  \bibinfo{journal}{Nat. Nanotech.} \textbf{\bibinfo{volume}{8}},
  \bibinfo{pages}{742} (\bibinfo{year}{2013}).

\bibitem[{\citenamefont{Hsu et~al.}(2017)\citenamefont{Hsu, Kubetzka, Finco,
  Romming, Von~Bergmann, and Wiesendanger}}]{hsu2017electric}
\bibinfo{author}{\bibfnamefont{P.-J.} \bibnamefont{Hsu}},
  \bibinfo{author}{\bibfnamefont{A.}~\bibnamefont{Kubetzka}},
  \bibinfo{author}{\bibfnamefont{A.}~\bibnamefont{Finco}},
  \bibinfo{author}{\bibfnamefont{N.}~\bibnamefont{Romming}},
  \bibinfo{author}{\bibfnamefont{K.}~\bibnamefont{Von~Bergmann}},
  \bibnamefont{and}
  \bibinfo{author}{\bibfnamefont{R.}~\bibnamefont{Wiesendanger}},
  \bibinfo{journal}{Nat. Nanotech.} \textbf{\bibinfo{volume}{12}},
  \bibinfo{pages}{123} (\bibinfo{year}{2017}).

\bibitem[{\citenamefont{Jiang et~al.}(2015{\natexlab{b}})\citenamefont{Jiang,
  Upadhyaya, Zhang, Yu, Jungfleisch, Fradin, Pearson, Tserkovnyak, Wang,
  Heinonen et~al.}}]{jiang2015blowing}
\bibinfo{author}{\bibfnamefont{W.}~\bibnamefont{Jiang}},
  \bibinfo{author}{\bibfnamefont{P.}~\bibnamefont{Upadhyaya}},
  \bibinfo{author}{\bibfnamefont{W.}~\bibnamefont{Zhang}},
  \bibinfo{author}{\bibfnamefont{G.}~\bibnamefont{Yu}},
  \bibinfo{author}{\bibfnamefont{M.~B.} \bibnamefont{Jungfleisch}},
  \bibinfo{author}{\bibfnamefont{F.~Y.} \bibnamefont{Fradin}},
  \bibinfo{author}{\bibfnamefont{J.~E.} \bibnamefont{Pearson}},
  \bibinfo{author}{\bibfnamefont{Y.}~\bibnamefont{Tserkovnyak}},
  \bibinfo{author}{\bibfnamefont{K.~L.} \bibnamefont{Wang}},
  \bibinfo{author}{\bibfnamefont{O.}~\bibnamefont{Heinonen}},
  \bibnamefont{et~al.}, \bibinfo{journal}{Science}
  \textbf{\bibinfo{volume}{349}}, \bibinfo{pages}{283}
  (\bibinfo{year}{2015}{\natexlab{b}}).

\bibitem[{\citenamefont{Wang et~al.}(2020)\citenamefont{Wang, Wang, Xia, Lai,
  Tian, Zhang, Hou, Gao, Mi, Feng et~al.}}]{wang2020electric}
\bibinfo{author}{\bibfnamefont{Y.}~\bibnamefont{Wang}},
  \bibinfo{author}{\bibfnamefont{L.}~\bibnamefont{Wang}},
  \bibinfo{author}{\bibfnamefont{J.}~\bibnamefont{Xia}},
  \bibinfo{author}{\bibfnamefont{Z.}~\bibnamefont{Lai}},
  \bibinfo{author}{\bibfnamefont{G.}~\bibnamefont{Tian}},
  \bibinfo{author}{\bibfnamefont{X.}~\bibnamefont{Zhang}},
  \bibinfo{author}{\bibfnamefont{Z.}~\bibnamefont{Hou}},
  \bibinfo{author}{\bibfnamefont{X.}~\bibnamefont{Gao}},
  \bibinfo{author}{\bibfnamefont{W.}~\bibnamefont{Mi}},
  \bibinfo{author}{\bibfnamefont{C.}~\bibnamefont{Feng}}, \bibnamefont{et~al.},
  \bibinfo{journal}{Nat. Commun.} \textbf{\bibinfo{volume}{11}},
  \bibinfo{pages}{3577} (\bibinfo{year}{2020}).

\bibitem[{\citenamefont{Ba et~al.}(2021)\citenamefont{Ba, Zhuang, Zhang, Wang,
  Gao, Zhou, Chen, Sun, Liu, Chai et~al.}}]{ba2021electric}
\bibinfo{author}{\bibfnamefont{Y.}~\bibnamefont{Ba}},
  \bibinfo{author}{\bibfnamefont{S.}~\bibnamefont{Zhuang}},
  \bibinfo{author}{\bibfnamefont{Y.}~\bibnamefont{Zhang}},
  \bibinfo{author}{\bibfnamefont{Y.}~\bibnamefont{Wang}},
  \bibinfo{author}{\bibfnamefont{Y.}~\bibnamefont{Gao}},
  \bibinfo{author}{\bibfnamefont{H.}~\bibnamefont{Zhou}},
  \bibinfo{author}{\bibfnamefont{M.}~\bibnamefont{Chen}},
  \bibinfo{author}{\bibfnamefont{W.}~\bibnamefont{Sun}},
  \bibinfo{author}{\bibfnamefont{Q.}~\bibnamefont{Liu}},
  \bibinfo{author}{\bibfnamefont{G.}~\bibnamefont{Chai}}, \bibnamefont{et~al.},
  \bibinfo{journal}{Nat. Commun.} \textbf{\bibinfo{volume}{12}},
  \bibinfo{pages}{322} (\bibinfo{year}{2021}).

\bibitem[{\citenamefont{Geirhos et~al.}(2020)\citenamefont{Geirhos, Gross,
  Szigeti, Mehlin, Philipp, White, Cubitt, Widmann, Ghara, Lunkenheimer
  et~al.}}]{geirhos2020macroscopic}
\bibinfo{author}{\bibfnamefont{K.}~\bibnamefont{Geirhos}},
  \bibinfo{author}{\bibfnamefont{B.}~\bibnamefont{Gross}},
  \bibinfo{author}{\bibfnamefont{B.~G.} \bibnamefont{Szigeti}},
  \bibinfo{author}{\bibfnamefont{A.}~\bibnamefont{Mehlin}},
  \bibinfo{author}{\bibfnamefont{S.}~\bibnamefont{Philipp}},
  \bibinfo{author}{\bibfnamefont{J.~S.} \bibnamefont{White}},
  \bibinfo{author}{\bibfnamefont{R.}~\bibnamefont{Cubitt}},
  \bibinfo{author}{\bibfnamefont{S.}~\bibnamefont{Widmann}},
  \bibinfo{author}{\bibfnamefont{S.}~\bibnamefont{Ghara}},
  \bibinfo{author}{\bibfnamefont{P.}~\bibnamefont{Lunkenheimer}},
  \bibnamefont{et~al.}, \bibinfo{journal}{npj Quantum Mater.}
  \textbf{\bibinfo{volume}{5}}, \bibinfo{pages}{44} (\bibinfo{year}{2020}).

\bibitem[{\citenamefont{Chen et~al.}(2016)\citenamefont{Chen, Zhang, and
  Liu}}]{chen2016exotic}
\bibinfo{author}{\bibfnamefont{J.-P.} \bibnamefont{Chen}},
  \bibinfo{author}{\bibfnamefont{D.-W.} \bibnamefont{Zhang}}, \bibnamefont{and}
  \bibinfo{author}{\bibfnamefont{J.-M.} \bibnamefont{Liu}},
  \bibinfo{journal}{Sci. Rep.} \textbf{\bibinfo{volume}{6}},
  \bibinfo{pages}{29126} (\bibinfo{year}{2016}).

\bibitem[{\citenamefont{Lin et~al.}(2022)\citenamefont{Lin, Chen, Tan, Chen,
  Chen, Li, Gao, and Liu}}]{lin2022manipulation}
\bibinfo{author}{\bibfnamefont{J.-Q.} \bibnamefont{Lin}},
  \bibinfo{author}{\bibfnamefont{J.-P.} \bibnamefont{Chen}},
  \bibinfo{author}{\bibfnamefont{Z.-Y.} \bibnamefont{Tan}},
  \bibinfo{author}{\bibfnamefont{Y.}~\bibnamefont{Chen}},
  \bibinfo{author}{\bibfnamefont{Z.-F.} \bibnamefont{Chen}},
  \bibinfo{author}{\bibfnamefont{W.-A.} \bibnamefont{Li}},
  \bibinfo{author}{\bibfnamefont{X.-S.} \bibnamefont{Gao}}, \bibnamefont{and}
  \bibinfo{author}{\bibfnamefont{J.-M.} \bibnamefont{Liu}},
  \bibinfo{journal}{Nanomaterials} \textbf{\bibinfo{volume}{12}},
  \bibinfo{pages}{278} (\bibinfo{year}{2022}).

\bibitem[{\citenamefont{Yao and Dong}(2022)}]{yao2022vector}
\bibinfo{author}{\bibfnamefont{X.}~\bibnamefont{Yao}} \bibnamefont{and}
  \bibinfo{author}{\bibfnamefont{S.}~\bibnamefont{Dong}},
  \bibinfo{journal}{Phys. Rev. B} \textbf{\bibinfo{volume}{105}},
  \bibinfo{pages}{014444} (\bibinfo{year}{2022}).

\bibitem[{\citenamefont{Sch{\"u}tte et~al.}(2014)\citenamefont{Sch{\"u}tte,
  Iwasaki, Rosch, and Nagaosa}}]{schutte2014inertia}
\bibinfo{author}{\bibfnamefont{C.}~\bibnamefont{Sch{\"u}tte}},
  \bibinfo{author}{\bibfnamefont{J.}~\bibnamefont{Iwasaki}},
  \bibinfo{author}{\bibfnamefont{A.}~\bibnamefont{Rosch}}, \bibnamefont{and}
  \bibinfo{author}{\bibfnamefont{N.}~\bibnamefont{Nagaosa}},
  \bibinfo{journal}{Phys. Rev. B} \textbf{\bibinfo{volume}{90}},
  \bibinfo{pages}{174434} (\bibinfo{year}{2014}).

\bibitem[{\citenamefont{Lin et~al.}(2013{\natexlab{a}})\citenamefont{Lin,
  Reichhardt, Batista, and Saxena}}]{lin2013particle}
\bibinfo{author}{\bibfnamefont{S.-Z.} \bibnamefont{Lin}},
  \bibinfo{author}{\bibfnamefont{C.}~\bibnamefont{Reichhardt}},
  \bibinfo{author}{\bibfnamefont{C.~D.} \bibnamefont{Batista}},
  \bibnamefont{and} \bibinfo{author}{\bibfnamefont{A.}~\bibnamefont{Saxena}},
  \bibinfo{journal}{Phys. Rev. B} \textbf{\bibinfo{volume}{87}},
  \bibinfo{pages}{214419} (\bibinfo{year}{2013}{\natexlab{a}}).

\bibitem[{\citenamefont{Yasin et~al.}(2022)\citenamefont{Yasin, Masell, Karube,
  Kikkawa, Taguchi, Tokura, and Yu}}]{yasin2022real}
\bibinfo{author}{\bibfnamefont{F.~S.} \bibnamefont{Yasin}},
  \bibinfo{author}{\bibfnamefont{J.}~\bibnamefont{Masell}},
  \bibinfo{author}{\bibfnamefont{K.}~\bibnamefont{Karube}},
  \bibinfo{author}{\bibfnamefont{A.}~\bibnamefont{Kikkawa}},
  \bibinfo{author}{\bibfnamefont{Y.}~\bibnamefont{Taguchi}},
  \bibinfo{author}{\bibfnamefont{Y.}~\bibnamefont{Tokura}}, \bibnamefont{and}
  \bibinfo{author}{\bibfnamefont{X.}~\bibnamefont{Yu}}, \bibinfo{journal}{Proc.
  Natl. Acad. Sci.} \textbf{\bibinfo{volume}{119}},
  \bibinfo{pages}{e2200958119} (\bibinfo{year}{2022}).

\bibitem[{\citenamefont{Liu et~al.}(2022)\citenamefont{Liu, Chen, and
  Zheng}}]{liu2022flexoresponses}
\bibinfo{author}{\bibfnamefont{L.}~\bibnamefont{Liu}},
  \bibinfo{author}{\bibfnamefont{W.}~\bibnamefont{Chen}}, \bibnamefont{and}
  \bibinfo{author}{\bibfnamefont{Y.}~\bibnamefont{Zheng}},
  \bibinfo{journal}{Phys. Rev. Lett.} \textbf{\bibinfo{volume}{128}},
  \bibinfo{pages}{257201} (\bibinfo{year}{2022}).

\bibitem[{\citenamefont{Liu et~al.}(2023)\citenamefont{Liu, Chen, and
  Zheng}}]{liu2023emergent}
\bibinfo{author}{\bibfnamefont{L.}~\bibnamefont{Liu}},
  \bibinfo{author}{\bibfnamefont{W.}~\bibnamefont{Chen}}, \bibnamefont{and}
  \bibinfo{author}{\bibfnamefont{Y.}~\bibnamefont{Zheng}},
  \bibinfo{journal}{Phys. Rev. Lett.} \textbf{\bibinfo{volume}{131}},
  \bibinfo{pages}{246701} (\bibinfo{year}{2023}).

\bibitem[{\citenamefont{Cheng et~al.}(2021)\citenamefont{Cheng, Yan, Dong, Liu,
  Xia, Li, and Han}}]{cheng2021elliptical}
\bibinfo{author}{\bibfnamefont{C.}~\bibnamefont{Cheng}},
  \bibinfo{author}{\bibfnamefont{Z.}~\bibnamefont{Yan}},
  \bibinfo{author}{\bibfnamefont{J.}~\bibnamefont{Dong}},
  \bibinfo{author}{\bibfnamefont{Y.}~\bibnamefont{Liu}},
  \bibinfo{author}{\bibfnamefont{Z.}~\bibnamefont{Xia}},
  \bibinfo{author}{\bibfnamefont{L.}~\bibnamefont{Li}}, \bibnamefont{and}
  \bibinfo{author}{\bibfnamefont{X.}~\bibnamefont{Han}},
  \bibinfo{journal}{Phys. Rev. B} \textbf{\bibinfo{volume}{104}},
  \bibinfo{pages}{174409} (\bibinfo{year}{2021}).

\bibitem[{\citenamefont{Thiele}(1973)}]{thiele1973steady}
\bibinfo{author}{\bibfnamefont{A.}~\bibnamefont{Thiele}},
  \bibinfo{journal}{Phys. Rev. Lett.} \textbf{\bibinfo{volume}{30}},
  \bibinfo{pages}{230} (\bibinfo{year}{1973}).

\bibitem[{\citenamefont{Koshibae and Nagaosa}(2016)}]{koshibae2016berry}
\bibinfo{author}{\bibfnamefont{W.}~\bibnamefont{Koshibae}} \bibnamefont{and}
  \bibinfo{author}{\bibfnamefont{N.}~\bibnamefont{Nagaosa}},
  \bibinfo{journal}{New J. Phys.} \textbf{\bibinfo{volume}{18}},
  \bibinfo{pages}{045007} (\bibinfo{year}{2016}).

\bibitem[{\citenamefont{Koshibae and Nagaosa}(2017)}]{koshibae2017theory}
\bibinfo{author}{\bibfnamefont{W.}~\bibnamefont{Koshibae}} \bibnamefont{and}
  \bibinfo{author}{\bibfnamefont{N.}~\bibnamefont{Nagaosa}},
  \bibinfo{journal}{Sci. Rep.} \textbf{\bibinfo{volume}{7}},
  \bibinfo{pages}{42645} (\bibinfo{year}{2017}).

\bibitem[{\citenamefont{Leutner et~al.}(2022)\citenamefont{Leutner, Winkler,
  G{\"u}ttinger, Fangohr, and Kl{\"a}ui}}]{leutner2022skyrmion}
\bibinfo{author}{\bibfnamefont{K.}~\bibnamefont{Leutner}},
  \bibinfo{author}{\bibfnamefont{T.~B.} \bibnamefont{Winkler}},
  \bibinfo{author}{\bibfnamefont{J.}~\bibnamefont{G{\"u}ttinger}},
  \bibinfo{author}{\bibfnamefont{H.}~\bibnamefont{Fangohr}}, \bibnamefont{and}
  \bibinfo{author}{\bibfnamefont{M.}~\bibnamefont{Kl{\"a}ui}},
  \bibinfo{journal}{arXiv preprint arXiv:2211.05711}  (\bibinfo{year}{2022}).

\bibitem[{\citenamefont{Yanes et~al.}(2019)\citenamefont{Yanes, Garcia-Sanchez,
  Luis, Martinez, Raposo, Torres, and Lopez-Diaz}}]{yanes2019skyrmion}
\bibinfo{author}{\bibfnamefont{R.}~\bibnamefont{Yanes}},
  \bibinfo{author}{\bibfnamefont{F.}~\bibnamefont{Garcia-Sanchez}},
  \bibinfo{author}{\bibfnamefont{R.}~\bibnamefont{Luis}},
  \bibinfo{author}{\bibfnamefont{E.}~\bibnamefont{Martinez}},
  \bibinfo{author}{\bibfnamefont{V.}~\bibnamefont{Raposo}},
  \bibinfo{author}{\bibfnamefont{L.}~\bibnamefont{Torres}}, \bibnamefont{and}
  \bibinfo{author}{\bibfnamefont{L.}~\bibnamefont{Lopez-Diaz}},
  \bibinfo{journal}{Appl. Phys. Lett.} \textbf{\bibinfo{volume}{115}},
  \bibinfo{pages}{132401} (\bibinfo{year}{2019}).

\bibitem[{\citenamefont{Du et~al.}(2023)\citenamefont{Du, Hu, and
  Kawasaki}}]{du2023strain}
\bibinfo{author}{\bibfnamefont{D.}~\bibnamefont{Du}},
  \bibinfo{author}{\bibfnamefont{J.}~\bibnamefont{Hu}}, \bibnamefont{and}
  \bibinfo{author}{\bibfnamefont{J.~K.} \bibnamefont{Kawasaki}},
  \bibinfo{journal}{Appl. Phys. Lett.} \textbf{\bibinfo{volume}{122}},
  \bibinfo{pages}{170501} (\bibinfo{year}{2023}).

\bibitem[{\citenamefont{Nikonov et~al.}(2014)\citenamefont{Nikonov,
  Manipatruni, and Young}}]{nikonov2014automotion}
\bibinfo{author}{\bibfnamefont{D.~E.} \bibnamefont{Nikonov}},
  \bibinfo{author}{\bibfnamefont{S.}~\bibnamefont{Manipatruni}},
  \bibnamefont{and} \bibinfo{author}{\bibfnamefont{I.~A.} \bibnamefont{Young}},
  \bibinfo{journal}{J. Appl. Phys.} \textbf{\bibinfo{volume}{115}},
  \bibinfo{pages}{213902} (\bibinfo{year}{2014}).

\bibitem[{\citenamefont{Mawass et~al.}(2017)\citenamefont{Mawass, Richter,
  Bisig, Reeve, Kr{\"u}ger, Weigand, Stoll, Krone, Kronast, Sch{\"u}tz
  et~al.}}]{mawass2017switching}
\bibinfo{author}{\bibfnamefont{M.-A.} \bibnamefont{Mawass}},
  \bibinfo{author}{\bibfnamefont{K.}~\bibnamefont{Richter}},
  \bibinfo{author}{\bibfnamefont{A.}~\bibnamefont{Bisig}},
  \bibinfo{author}{\bibfnamefont{R.~M.} \bibnamefont{Reeve}},
  \bibinfo{author}{\bibfnamefont{B.}~\bibnamefont{Kr{\"u}ger}},
  \bibinfo{author}{\bibfnamefont{M.}~\bibnamefont{Weigand}},
  \bibinfo{author}{\bibfnamefont{H.}~\bibnamefont{Stoll}},
  \bibinfo{author}{\bibfnamefont{A.}~\bibnamefont{Krone}},
  \bibinfo{author}{\bibfnamefont{F.}~\bibnamefont{Kronast}},
  \bibinfo{author}{\bibfnamefont{G.}~\bibnamefont{Sch{\"u}tz}},
  \bibnamefont{et~al.}, \bibinfo{journal}{Phys. Rev. Appl.}
  \textbf{\bibinfo{volume}{7}}, \bibinfo{pages}{044009} (\bibinfo{year}{2017}).

\bibitem[{\citenamefont{Yershov et~al.}(2018)\citenamefont{Yershov, Kravchuk,
  Sheka, Pylypovskyi, Makarov, and Gaididei}}]{yershov2018geometry}
\bibinfo{author}{\bibfnamefont{K.~V.} \bibnamefont{Yershov}},
  \bibinfo{author}{\bibfnamefont{V.~P.} \bibnamefont{Kravchuk}},
  \bibinfo{author}{\bibfnamefont{D.~D.} \bibnamefont{Sheka}},
  \bibinfo{author}{\bibfnamefont{O.~V.} \bibnamefont{Pylypovskyi}},
  \bibinfo{author}{\bibfnamefont{D.}~\bibnamefont{Makarov}}, \bibnamefont{and}
  \bibinfo{author}{\bibfnamefont{Y.}~\bibnamefont{Gaididei}},
  \bibinfo{journal}{Phys. Rev. B} \textbf{\bibinfo{volume}{98}},
  \bibinfo{pages}{060409} (\bibinfo{year}{2018}).

\bibitem[{\citenamefont{Gao et~al.}(2016)\citenamefont{Gao, Hu, Nelson, Yang,
  Shen, Chen, Ramesh, and Nan}}]{gao2016dynamic}
\bibinfo{author}{\bibfnamefont{Y.}~\bibnamefont{Gao}},
  \bibinfo{author}{\bibfnamefont{J.-M.} \bibnamefont{Hu}},
  \bibinfo{author}{\bibfnamefont{C.}~\bibnamefont{Nelson}},
  \bibinfo{author}{\bibfnamefont{T.}~\bibnamefont{Yang}},
  \bibinfo{author}{\bibfnamefont{Y.}~\bibnamefont{Shen}},
  \bibinfo{author}{\bibfnamefont{L.}~\bibnamefont{Chen}},
  \bibinfo{author}{\bibfnamefont{R.}~\bibnamefont{Ramesh}}, \bibnamefont{and}
  \bibinfo{author}{\bibfnamefont{C.}~\bibnamefont{Nan}}, \bibinfo{journal}{Sci.
  Rep.} \textbf{\bibinfo{volume}{6}}, \bibinfo{pages}{23696}
  (\bibinfo{year}{2016}).

\bibitem[{\citenamefont{Yao et~al.}(2018)\citenamefont{Yao, Song, Gao, Tian,
  Li, Fan, Huang, Yang, Chen, Fan et~al.}}]{yao2018electrically}
\bibinfo{author}{\bibfnamefont{J.}~\bibnamefont{Yao}},
  \bibinfo{author}{\bibfnamefont{X.}~\bibnamefont{Song}},
  \bibinfo{author}{\bibfnamefont{X.}~\bibnamefont{Gao}},
  \bibinfo{author}{\bibfnamefont{G.}~\bibnamefont{Tian}},
  \bibinfo{author}{\bibfnamefont{P.}~\bibnamefont{Li}},
  \bibinfo{author}{\bibfnamefont{H.}~\bibnamefont{Fan}},
  \bibinfo{author}{\bibfnamefont{Z.}~\bibnamefont{Huang}},
  \bibinfo{author}{\bibfnamefont{W.}~\bibnamefont{Yang}},
  \bibinfo{author}{\bibfnamefont{D.}~\bibnamefont{Chen}},
  \bibinfo{author}{\bibfnamefont{Z.}~\bibnamefont{Fan}}, \bibnamefont{et~al.},
  \bibinfo{journal}{ACS Nano} \textbf{\bibinfo{volume}{12}},
  \bibinfo{pages}{6767} (\bibinfo{year}{2018}).

\bibitem[{\citenamefont{Cui et~al.}(2013)\citenamefont{Cui, Hockel, Nordeen,
  Pisani, Liang, Carman, and Lynch}}]{cui2013method}
\bibinfo{author}{\bibfnamefont{J.}~\bibnamefont{Cui}},
  \bibinfo{author}{\bibfnamefont{J.~L.} \bibnamefont{Hockel}},
  \bibinfo{author}{\bibfnamefont{P.~K.} \bibnamefont{Nordeen}},
  \bibinfo{author}{\bibfnamefont{D.~M.} \bibnamefont{Pisani}},
  \bibinfo{author}{\bibfnamefont{C.-y.} \bibnamefont{Liang}},
  \bibinfo{author}{\bibfnamefont{G.~P.} \bibnamefont{Carman}},
  \bibnamefont{and} \bibinfo{author}{\bibfnamefont{C.~S.} \bibnamefont{Lynch}},
  \bibinfo{journal}{Appl. Phys. Lett.} \textbf{\bibinfo{volume}{103}},
  \bibinfo{pages}{232905} (\bibinfo{year}{2013}).

\bibitem[{\citenamefont{Hu and Nan}(2009)}]{hu2009electric}
\bibinfo{author}{\bibfnamefont{J.-M.} \bibnamefont{Hu}} \bibnamefont{and}
  \bibinfo{author}{\bibfnamefont{C.}~\bibnamefont{Nan}},
  \bibinfo{journal}{Phys. Rev. B} \textbf{\bibinfo{volume}{80}},
  \bibinfo{pages}{224416} (\bibinfo{year}{2009}).

\bibitem[{\citenamefont{Song et~al.}(2022)\citenamefont{Song, Chen, Hou, Qin,
  Gao, and Liu}}]{song2022strain}
\bibinfo{author}{\bibfnamefont{X.}~\bibnamefont{Song}},
  \bibinfo{author}{\bibfnamefont{J.-P.} \bibnamefont{Chen}},
  \bibinfo{author}{\bibfnamefont{Z.-P.} \bibnamefont{Hou}},
  \bibinfo{author}{\bibfnamefont{M.-H.} \bibnamefont{Qin}},
  \bibinfo{author}{\bibfnamefont{X.-S.} \bibnamefont{Gao}}, \bibnamefont{and}
  \bibinfo{author}{\bibfnamefont{J.-M.} \bibnamefont{Liu}},
  \bibinfo{journal}{J. Magn. Magn. Mater.} \textbf{\bibinfo{volume}{547}},
  \bibinfo{pages}{168729} (\bibinfo{year}{2022}).

\bibitem[{\citenamefont{Cui et~al.}(2015)\citenamefont{Cui, Liang, Paisley,
  Sepulveda, Ihlefeld, Carman, and Lynch}}]{cui2015generation}
\bibinfo{author}{\bibfnamefont{J.}~\bibnamefont{Cui}},
  \bibinfo{author}{\bibfnamefont{C.-Y.} \bibnamefont{Liang}},
  \bibinfo{author}{\bibfnamefont{E.~A.} \bibnamefont{Paisley}},
  \bibinfo{author}{\bibfnamefont{A.}~\bibnamefont{Sepulveda}},
  \bibinfo{author}{\bibfnamefont{J.~F.} \bibnamefont{Ihlefeld}},
  \bibinfo{author}{\bibfnamefont{G.~P.} \bibnamefont{Carman}},
  \bibnamefont{and} \bibinfo{author}{\bibfnamefont{C.~S.} \bibnamefont{Lynch}},
  \bibinfo{journal}{Appl. Phys. Lett.} \textbf{\bibinfo{volume}{107}},
  \bibinfo{pages}{092903} (\bibinfo{year}{2015}).

\bibitem[{\citenamefont{Chen et~al.}(2020)\citenamefont{Chen, Sablik, Khojah,
  Domann, Dyro, Hu, Mehta, Xiao, Candler, Carman et~al.}}]{chen2020voltage}
\bibinfo{author}{\bibfnamefont{C.}~\bibnamefont{Chen}},
  \bibinfo{author}{\bibfnamefont{J.}~\bibnamefont{Sablik}},
  \bibinfo{author}{\bibfnamefont{R.}~\bibnamefont{Khojah}},
  \bibinfo{author}{\bibfnamefont{J.}~\bibnamefont{Domann}},
  \bibinfo{author}{\bibfnamefont{R.}~\bibnamefont{Dyro}},
  \bibinfo{author}{\bibfnamefont{J.}~\bibnamefont{Hu}},
  \bibinfo{author}{\bibfnamefont{S.}~\bibnamefont{Mehta}},
  \bibinfo{author}{\bibfnamefont{Z.~M.} \bibnamefont{Xiao}},
  \bibinfo{author}{\bibfnamefont{R.}~\bibnamefont{Candler}},
  \bibinfo{author}{\bibfnamefont{G.}~\bibnamefont{Carman}},
  \bibnamefont{et~al.}, \bibinfo{journal}{J. Phys. D: Appl. Phys.}
  \textbf{\bibinfo{volume}{53}}, \bibinfo{pages}{174002}
  (\bibinfo{year}{2020}).

\bibitem[{\citenamefont{Vansteenkiste et~al.}(2014)\citenamefont{Vansteenkiste,
  Leliaert, Dvornik, Helsen, Garcia-Sanchez, and
  Van~Waeyenberge}}]{vansteenkiste2014design}
\bibinfo{author}{\bibfnamefont{A.}~\bibnamefont{Vansteenkiste}},
  \bibinfo{author}{\bibfnamefont{J.}~\bibnamefont{Leliaert}},
  \bibinfo{author}{\bibfnamefont{M.}~\bibnamefont{Dvornik}},
  \bibinfo{author}{\bibfnamefont{M.}~\bibnamefont{Helsen}},
  \bibinfo{author}{\bibfnamefont{F.}~\bibnamefont{Garcia-Sanchez}},
  \bibnamefont{and}
  \bibinfo{author}{\bibfnamefont{B.}~\bibnamefont{Van~Waeyenberge}},
  \bibinfo{journal}{AIP Adv.} \textbf{\bibinfo{volume}{4}},
  \bibinfo{pages}{107133} (\bibinfo{year}{2014}).

\bibitem[{\citenamefont{Landau and Lifshitz}(1992)}]{landau1992theory}
\bibinfo{author}{\bibfnamefont{L.}~\bibnamefont{Landau}} \bibnamefont{and}
  \bibinfo{author}{\bibfnamefont{E.}~\bibnamefont{Lifshitz}}, in
  \emph{\bibinfo{booktitle}{Perspectives in Theoretical Physics}}
  (\bibinfo{publisher}{Elsevier}, \bibinfo{year}{1992}), pp.
  \bibinfo{pages}{51--65}.

\bibitem[{\citenamefont{Gilbert}(1955)}]{gilbert1955lagrangian}
\bibinfo{author}{\bibfnamefont{T.~L.} \bibnamefont{Gilbert}},
  \bibinfo{journal}{Phys. Rev.} \textbf{\bibinfo{volume}{100}},
  \bibinfo{pages}{1243} (\bibinfo{year}{1955}).

\bibitem[{\citenamefont{Neilinger et~al.}(2018)\citenamefont{Neilinger,
  {\v{S}}{\v{c}}epka, Mruczkiewicz, D{\'e}rer, Manca, Dobro{\v{c}}ka, Samardak,
  Grajcar, and Cambel}}]{neilinger2018ferromagnetic}
\bibinfo{author}{\bibfnamefont{P.}~\bibnamefont{Neilinger}},
  \bibinfo{author}{\bibfnamefont{T.}~\bibnamefont{{\v{S}}{\v{c}}epka}},
  \bibinfo{author}{\bibfnamefont{M.}~\bibnamefont{Mruczkiewicz}},
  \bibinfo{author}{\bibfnamefont{J.}~\bibnamefont{D{\'e}rer}},
  \bibinfo{author}{\bibfnamefont{D.}~\bibnamefont{Manca}},
  \bibinfo{author}{\bibfnamefont{E.}~\bibnamefont{Dobro{\v{c}}ka}},
  \bibinfo{author}{\bibfnamefont{A.}~\bibnamefont{Samardak}},
  \bibinfo{author}{\bibfnamefont{M.}~\bibnamefont{Grajcar}}, \bibnamefont{and}
  \bibinfo{author}{\bibfnamefont{V.}~\bibnamefont{Cambel}},
  \bibinfo{journal}{Appl. Surf. Sci.} \textbf{\bibinfo{volume}{461}},
  \bibinfo{pages}{202} (\bibinfo{year}{2018}).

\bibitem[{\citenamefont{Mizukami et~al.}(2010)\citenamefont{Mizukami, Sajitha,
  Watanabe, Wu, Miyazaki, Naganuma, Oogane, and Ando}}]{mizukami2010gilbert}
\bibinfo{author}{\bibfnamefont{S.}~\bibnamefont{Mizukami}},
  \bibinfo{author}{\bibfnamefont{E.}~\bibnamefont{Sajitha}},
  \bibinfo{author}{\bibfnamefont{D.}~\bibnamefont{Watanabe}},
  \bibinfo{author}{\bibfnamefont{F.}~\bibnamefont{Wu}},
  \bibinfo{author}{\bibfnamefont{T.}~\bibnamefont{Miyazaki}},
  \bibinfo{author}{\bibfnamefont{H.}~\bibnamefont{Naganuma}},
  \bibinfo{author}{\bibfnamefont{M.}~\bibnamefont{Oogane}}, \bibnamefont{and}
  \bibinfo{author}{\bibfnamefont{Y.}~\bibnamefont{Ando}},
  \bibinfo{journal}{Appl. Phys. Lett.} \textbf{\bibinfo{volume}{96}},
  \bibinfo{pages}{152502} (\bibinfo{year}{2010}).

\bibitem[{\citenamefont{Barman et~al.}(2007)\citenamefont{Barman, Wang,
  Hellwig, Berger, Fullerton, and Schmidt}}]{barman2007ultrafast}
\bibinfo{author}{\bibfnamefont{A.}~\bibnamefont{Barman}},
  \bibinfo{author}{\bibfnamefont{S.}~\bibnamefont{Wang}},
  \bibinfo{author}{\bibfnamefont{O.}~\bibnamefont{Hellwig}},
  \bibinfo{author}{\bibfnamefont{A.}~\bibnamefont{Berger}},
  \bibinfo{author}{\bibfnamefont{E.~E.} \bibnamefont{Fullerton}},
  \bibnamefont{and} \bibinfo{author}{\bibfnamefont{H.}~\bibnamefont{Schmidt}},
  \bibinfo{journal}{J. Appl. Phys.} \textbf{\bibinfo{volume}{101}},
  \bibinfo{pages}{09D102} (\bibinfo{year}{2007}).

\bibitem[{\citenamefont{Barati et~al.}(2014)\citenamefont{Barati, Cinal,
  Edwards, and Umerski}}]{barati2014gilbert}
\bibinfo{author}{\bibfnamefont{E.}~\bibnamefont{Barati}},
  \bibinfo{author}{\bibfnamefont{M.}~\bibnamefont{Cinal}},
  \bibinfo{author}{\bibfnamefont{D.}~\bibnamefont{Edwards}}, \bibnamefont{and}
  \bibinfo{author}{\bibfnamefont{A.}~\bibnamefont{Umerski}},
  \bibinfo{journal}{Phys. Rev. B} \textbf{\bibinfo{volume}{90}},
  \bibinfo{pages}{014420} (\bibinfo{year}{2014}).

\bibitem[{\citenamefont{Chauwin et~al.}(2019)\citenamefont{Chauwin, Hu,
  Garcia-Sanchez, Betrabet, Paler, Moutafis, and
  Friedman}}]{chauwin2019skyrmion}
\bibinfo{author}{\bibfnamefont{M.}~\bibnamefont{Chauwin}},
  \bibinfo{author}{\bibfnamefont{X.}~\bibnamefont{Hu}},
  \bibinfo{author}{\bibfnamefont{F.}~\bibnamefont{Garcia-Sanchez}},
  \bibinfo{author}{\bibfnamefont{N.}~\bibnamefont{Betrabet}},
  \bibinfo{author}{\bibfnamefont{A.}~\bibnamefont{Paler}},
  \bibinfo{author}{\bibfnamefont{C.}~\bibnamefont{Moutafis}}, \bibnamefont{and}
  \bibinfo{author}{\bibfnamefont{J.~S.} \bibnamefont{Friedman}},
  \bibinfo{journal}{Phys. Rev. Appl.} \textbf{\bibinfo{volume}{12}},
  \bibinfo{pages}{064053} (\bibinfo{year}{2019}).

\bibitem[{\citenamefont{Zhang et~al.}(2020{\natexlab{b}})\citenamefont{Zhang,
  Zhu, Kang, Zhang, and Zhao}}]{zhang2020stochastic}
\bibinfo{author}{\bibfnamefont{H.}~\bibnamefont{Zhang}},
  \bibinfo{author}{\bibfnamefont{D.}~\bibnamefont{Zhu}},
  \bibinfo{author}{\bibfnamefont{W.}~\bibnamefont{Kang}},
  \bibinfo{author}{\bibfnamefont{Y.}~\bibnamefont{Zhang}}, \bibnamefont{and}
  \bibinfo{author}{\bibfnamefont{W.}~\bibnamefont{Zhao}},
  \bibinfo{journal}{Phys. Rev. Appl.} \textbf{\bibinfo{volume}{13}},
  \bibinfo{pages}{054049} (\bibinfo{year}{2020}{\natexlab{b}}).

\bibitem[{\citenamefont{Fattouhi et~al.}(2021)\citenamefont{Fattouhi, Mak,
  Zhou, Zhang, Liu, and El~Hafidi}}]{fattouhi2021logic}
\bibinfo{author}{\bibfnamefont{M.}~\bibnamefont{Fattouhi}},
  \bibinfo{author}{\bibfnamefont{K.~Y.} \bibnamefont{Mak}},
  \bibinfo{author}{\bibfnamefont{Y.}~\bibnamefont{Zhou}},
  \bibinfo{author}{\bibfnamefont{X.}~\bibnamefont{Zhang}},
  \bibinfo{author}{\bibfnamefont{X.}~\bibnamefont{Liu}}, \bibnamefont{and}
  \bibinfo{author}{\bibfnamefont{M.}~\bibnamefont{El~Hafidi}},
  \bibinfo{journal}{Phys. Rev. Appl.} \textbf{\bibinfo{volume}{16}},
  \bibinfo{pages}{014040} (\bibinfo{year}{2021}).

\bibitem[{\citenamefont{Purnama et~al.}(2015)\citenamefont{Purnama, Gan, Wong,
  and Lew}}]{purnama2015guided}
\bibinfo{author}{\bibfnamefont{I.}~\bibnamefont{Purnama}},
  \bibinfo{author}{\bibfnamefont{W.~L.} \bibnamefont{Gan}},
  \bibinfo{author}{\bibfnamefont{D.~W.} \bibnamefont{Wong}}, \bibnamefont{and}
  \bibinfo{author}{\bibfnamefont{W.~S.} \bibnamefont{Lew}},
  \bibinfo{journal}{Sci. Rep.} \textbf{\bibinfo{volume}{5}},
  \bibinfo{pages}{10620} (\bibinfo{year}{2015}).

\bibitem[{\citenamefont{Hao et~al.}(2021)\citenamefont{Hao, Zhuo, Manchon,
  Wang, Li, and Cheng}}]{hao2021skyrmion}
\bibinfo{author}{\bibfnamefont{X.}~\bibnamefont{Hao}},
  \bibinfo{author}{\bibfnamefont{F.}~\bibnamefont{Zhuo}},
  \bibinfo{author}{\bibfnamefont{A.}~\bibnamefont{Manchon}},
  \bibinfo{author}{\bibfnamefont{X.}~\bibnamefont{Wang}},
  \bibinfo{author}{\bibfnamefont{H.}~\bibnamefont{Li}}, \bibnamefont{and}
  \bibinfo{author}{\bibfnamefont{Z.}~\bibnamefont{Cheng}},
  \bibinfo{journal}{Appl. Phys. Rev.} \textbf{\bibinfo{volume}{8}},
  \bibinfo{pages}{021402} (\bibinfo{year}{2021}).

\bibitem[{\citenamefont{Lin et~al.}(2013{\natexlab{b}})\citenamefont{Lin,
  Reichhardt, Batista, and Saxena}}]{lin2013driven}
\bibinfo{author}{\bibfnamefont{S.-Z.} \bibnamefont{Lin}},
  \bibinfo{author}{\bibfnamefont{C.}~\bibnamefont{Reichhardt}},
  \bibinfo{author}{\bibfnamefont{C.~D.} \bibnamefont{Batista}},
  \bibnamefont{and} \bibinfo{author}{\bibfnamefont{A.}~\bibnamefont{Saxena}},
  \bibinfo{journal}{Phys. Rev. Lett.} \textbf{\bibinfo{volume}{110}},
  \bibinfo{pages}{207202} (\bibinfo{year}{2013}{\natexlab{b}}).

\bibitem[{\citenamefont{Streubel et~al.}(2016)\citenamefont{Streubel, Fischer,
  Kronast, Kravchuk, Sheka, Gaididei, Schmidt, and
  Makarov}}]{streubel2016magnetism}
\bibinfo{author}{\bibfnamefont{R.}~\bibnamefont{Streubel}},
  \bibinfo{author}{\bibfnamefont{P.}~\bibnamefont{Fischer}},
  \bibinfo{author}{\bibfnamefont{F.}~\bibnamefont{Kronast}},
  \bibinfo{author}{\bibfnamefont{V.~P.} \bibnamefont{Kravchuk}},
  \bibinfo{author}{\bibfnamefont{D.~D.} \bibnamefont{Sheka}},
  \bibinfo{author}{\bibfnamefont{Y.}~\bibnamefont{Gaididei}},
  \bibinfo{author}{\bibfnamefont{O.~G.} \bibnamefont{Schmidt}},
  \bibnamefont{and} \bibinfo{author}{\bibfnamefont{D.}~\bibnamefont{Makarov}},
  \bibinfo{journal}{J. Phys. D: Appl. Phys.} \textbf{\bibinfo{volume}{49}},
  \bibinfo{pages}{363001} (\bibinfo{year}{2016}).

\bibitem[{\citenamefont{Zheng and Chen}(2017)}]{zheng2017characteristics}
\bibinfo{author}{\bibfnamefont{Y.}~\bibnamefont{Zheng}} \bibnamefont{and}
  \bibinfo{author}{\bibfnamefont{W.}~\bibnamefont{Chen}},
  \bibinfo{journal}{Prog. Phys.} \textbf{\bibinfo{volume}{80}},
  \bibinfo{pages}{086501} (\bibinfo{year}{2017}).

\bibitem[{\citenamefont{Cubukcu et~al.}(2016)\citenamefont{Cubukcu, Sampaio,
  Bouzehouane, Apalkov, Khvalkovskiy, Cros, and
  Reyren}}]{cubukcu2016dzyaloshinskii}
\bibinfo{author}{\bibfnamefont{M.}~\bibnamefont{Cubukcu}},
  \bibinfo{author}{\bibfnamefont{J.}~\bibnamefont{Sampaio}},
  \bibinfo{author}{\bibfnamefont{K.}~\bibnamefont{Bouzehouane}},
  \bibinfo{author}{\bibfnamefont{D.}~\bibnamefont{Apalkov}},
  \bibinfo{author}{\bibfnamefont{A.}~\bibnamefont{Khvalkovskiy}},
  \bibinfo{author}{\bibfnamefont{V.}~\bibnamefont{Cros}}, \bibnamefont{and}
  \bibinfo{author}{\bibfnamefont{N.}~\bibnamefont{Reyren}},
  \bibinfo{journal}{Phys. Rev. B} \textbf{\bibinfo{volume}{93}},
  \bibinfo{pages}{020401} (\bibinfo{year}{2016}).

\bibitem[{\citenamefont{Chen et~al.}(2021)\citenamefont{Chen, Lin, Song, Chen,
  Chen, Li, Qin, Hou, Gao, and Liu}}]{chen2021control}
\bibinfo{author}{\bibfnamefont{J.-P.} \bibnamefont{Chen}},
  \bibinfo{author}{\bibfnamefont{J.-Q.} \bibnamefont{Lin}},
  \bibinfo{author}{\bibfnamefont{X.}~\bibnamefont{Song}},
  \bibinfo{author}{\bibfnamefont{Y.}~\bibnamefont{Chen}},
  \bibinfo{author}{\bibfnamefont{Z.-F.} \bibnamefont{Chen}},
  \bibinfo{author}{\bibfnamefont{W.-A.} \bibnamefont{Li}},
  \bibinfo{author}{\bibfnamefont{M.-H.} \bibnamefont{Qin}},
  \bibinfo{author}{\bibfnamefont{Z.-P.} \bibnamefont{Hou}},
  \bibinfo{author}{\bibfnamefont{X.-S.} \bibnamefont{Gao}}, \bibnamefont{and}
  \bibinfo{author}{\bibfnamefont{J.-M.} \bibnamefont{Liu}},
  \bibinfo{journal}{Front. Phys.} \textbf{\bibinfo{volume}{9}},
  \bibinfo{pages}{680698} (\bibinfo{year}{2021}).

\bibitem[{\citenamefont{Wang et~al.}(2018)\citenamefont{Wang, Yuan, and
  Wang}}]{wang2018theory}
\bibinfo{author}{\bibfnamefont{X.}~\bibnamefont{Wang}},
  \bibinfo{author}{\bibfnamefont{H.}~\bibnamefont{Yuan}}, \bibnamefont{and}
  \bibinfo{author}{\bibfnamefont{X.}~\bibnamefont{Wang}},
  \bibinfo{journal}{Commun. Phys.} \textbf{\bibinfo{volume}{1}},
  \bibinfo{pages}{31} (\bibinfo{year}{2018}).

\bibitem[{\citenamefont{Romming et~al.}(2015)\citenamefont{Romming, Kubetzka,
  Hanneken, von Bergmann, and Wiesendanger}}]{romming2015field}
\bibinfo{author}{\bibfnamefont{N.}~\bibnamefont{Romming}},
  \bibinfo{author}{\bibfnamefont{A.}~\bibnamefont{Kubetzka}},
  \bibinfo{author}{\bibfnamefont{C.}~\bibnamefont{Hanneken}},
  \bibinfo{author}{\bibfnamefont{K.}~\bibnamefont{von Bergmann}},
  \bibnamefont{and}
  \bibinfo{author}{\bibfnamefont{R.}~\bibnamefont{Wiesendanger}},
  \bibinfo{journal}{Phys. Rev. Lett.} \textbf{\bibinfo{volume}{114}},
  \bibinfo{pages}{177203} (\bibinfo{year}{2015}).

\bibitem[{\citenamefont{Hubert and Sch{\"a}fer}(1998)}]{hubert1998magnetic}
\bibinfo{author}{\bibfnamefont{A.}~\bibnamefont{Hubert}} \bibnamefont{and}
  \bibinfo{author}{\bibfnamefont{R.}~\bibnamefont{Sch{\"a}fer}},
  \emph{\bibinfo{title}{Magnetic domains: the analysis of magnetic
  microstructures}} (\bibinfo{publisher}{Springer Science \& Business Media},
  \bibinfo{year}{1998}).

\bibitem[{\citenamefont{Jackson}(2012)}]{jackson2012classical}
\bibinfo{author}{\bibfnamefont{J.~D.} \bibnamefont{Jackson}},
  \emph{\bibinfo{title}{Classical electrodynamics}} (\bibinfo{publisher}{John
  Wiley \& Sons}, \bibinfo{year}{2012}).

\bibitem[{\citenamefont{Rohart and Thiaville}(2013)}]{rohart2013skyrmion}
\bibinfo{author}{\bibfnamefont{S.}~\bibnamefont{Rohart}} \bibnamefont{and}
  \bibinfo{author}{\bibfnamefont{A.}~\bibnamefont{Thiaville}},
  \bibinfo{journal}{Phys. Rev. B} \textbf{\bibinfo{volume}{88}},
  \bibinfo{pages}{184422} (\bibinfo{year}{2013}).

\bibitem[{\citenamefont{Mulkers et~al.}(2018)\citenamefont{Mulkers, Hals,
  Leliaert, Milo{\v{s}}evi{\'c}, Van~Waeyenberge, and
  Everschor-Sitte}}]{mulkers2018effect}
\bibinfo{author}{\bibfnamefont{J.}~\bibnamefont{Mulkers}},
  \bibinfo{author}{\bibfnamefont{K.~M.} \bibnamefont{Hals}},
  \bibinfo{author}{\bibfnamefont{J.}~\bibnamefont{Leliaert}},
  \bibinfo{author}{\bibfnamefont{M.~V.} \bibnamefont{Milo{\v{s}}evi{\'c}}},
  \bibinfo{author}{\bibfnamefont{B.}~\bibnamefont{Van~Waeyenberge}},
  \bibnamefont{and}
  \bibinfo{author}{\bibfnamefont{K.}~\bibnamefont{Everschor-Sitte}},
  \bibinfo{journal}{Phys. Rev. B} \textbf{\bibinfo{volume}{98}},
  \bibinfo{pages}{064429} (\bibinfo{year}{2018}).

\bibitem[{\citenamefont{Raab and De~Lange}(2004)}]{raab2004multipole}
\bibinfo{author}{\bibfnamefont{R.~E.} \bibnamefont{Raab}} \bibnamefont{and}
  \bibinfo{author}{\bibfnamefont{O.~L.} \bibnamefont{De~Lange}},
  \emph{\bibinfo{title}{Multipole theory in electromagnetism}}, vol.
  \bibinfo{volume}{128} (\bibinfo{publisher}{Oxford University Press},
  \bibinfo{year}{2004}).

\bibitem[{\citenamefont{Gong et~al.}(2022)\citenamefont{Gong, Jing, Lu, and
  Wang}}]{gong2022skyrmion}
\bibinfo{author}{\bibfnamefont{X.}~\bibnamefont{Gong}},
  \bibinfo{author}{\bibfnamefont{K.}~\bibnamefont{Jing}},
  \bibinfo{author}{\bibfnamefont{J.}~\bibnamefont{Lu}}, \bibnamefont{and}
  \bibinfo{author}{\bibfnamefont{X.}~\bibnamefont{Wang}},
  \bibinfo{journal}{Phys. Rev. B} \textbf{\bibinfo{volume}{105}},
  \bibinfo{pages}{094437} (\bibinfo{year}{2022}).

\end{thebibliography}

\end{document}